\documentclass{aa}  

\usepackage{graphicx}
\usepackage{placeins}
\usepackage{txfonts}
\usepackage{natbib}
\usepackage[colorlinks=true, linkcolor=blue, urlcolor=blue, citecolor=blue]{hyperref}
\newcommand{\iram}{IRAM-30\,m}

\newcommand{\lsim}{\raisebox{-.4ex}{$\stackrel{<}{\scriptstyle \sim}$}}
\newcommand{\gsim}{\raisebox{-.4ex}{$\stackrel{>}{\scriptstyle \sim}$}}
\newcommand{\farc}{\mbox{$.\!\!^{\prime\prime}$}}
\newcommand{\mydeg}{$^{\circ }$}
\newcommand{\s}{\mbox{$''$}}
\newcommand{\mloss}{\mbox{$\dot{M}$}}

\newcommand{\my}{\mbox{$M_{\odot}$~yr$^{-1}$}}
\newcommand{\ls}{\mbox{$L_{\odot}$}}
\newcommand{\msun}{\mbox{$M_{\odot}$}}
\newcommand{\md}{\mbox{$M_{\rm d}$}}
\newcommand{\mini}{\mbox{$M_{\rm i}$}}

\newcommand{\rs}{\mbox{$R_{\star}$}}
\newcommand{\rd}{\mbox{$R_{\rm d}$}}

\newcommand{\kms}{\mbox{km\,s$^{-1}$}}

\newcommand{\vorb}{\mbox{$\upsilon_{\rm orb}$}}

\newcommand{\vexp}{\mbox{$V_{\rm exp}$}}
\newcommand{\vtrb}{\mbox{$V_{\rm turb}$}}

\newcommand{\vrot}{\mbox{$V_{\rm rot}$}}
 
\newcommand{\vsys}{\mbox{$V_{\rm sys}$}} 
\newcommand{\vlsr}{\mbox{$V_{\rm LSR}$}}

\newcommand{\h}{$^{\rm h}$}
\newcommand{\m}{$^{\rm m}$}

\newcommand{\td}{\mbox{$T_{\rm d}$}}
\newcommand{\teff}{\mbox{$T_{\rm eff}$}}
\newcommand{\trot}{\mbox{$T_{\rm rot}$}}
\newcommand{\tvib}{\mbox{$T_{\rm vib}$}}
\newcommand{\tkin}{\mbox{$T_{\rm kin}$}}      
\newcommand{\tdyn}{\mbox{$t_{\rm kin}$}}      
\newcommand{\zrot}{\mbox{$Z_{\rm rot}$}}
\newcommand{\zvib}{\mbox{$Z_{\rm vib}$}}
\newcommand{\dens}{\mbox{$n_{\rm H_2}$}}
\newcommand{\mtot}{\mbox{$M_{\rm H_2}$}}

\newcommand{\ncol}{\mbox{$N^{\rm col}$}}

\newcommand{\ntot}{\mbox{$N^{\rm tot}$}}
\newcommand{\nc}{\mbox{$n_{\rm crit}$}}

\newcommand{\jb}{\mbox{Jy\,beam$^{-1}$}}

\newcommand{\eu}{\mbox{E$_{\rm u}$}}

\newcommand{\tmb}{\mbox{$T_{\rm MB}$}}

\newcommand{\qx}{\mbox{QX\,Pup}}
\newcommand{\cs}{\mbox{clump {\sl S}}}
\newcommand{\sso}{\mbox{SS-outflow}}
\newcommand{\ohs}{\mbox{OH\,231.8}}
\newcommand{\oh}{\mbox{OH\,231.8$+$4.2}}
\newcommand{\lbol}{\mbox{$L_{\rm bol}$}}

\newcommand{\san}{\mbox{SCetal18}}
\newcommand{\oha}{\mbox{OH231.8\_a\_06\_TM1}}
\newcommand{\ohb}{\mbox{OH231.8\_b\_06\_TM1}}

\newcommand{\water}{\mbox{H$_2$O}}

\newcommand{\naclts}{Na$^{37}$Cl}

\newcommand{\nacldo}{NaCl\,($\upsilonup$=0, $J$=18-17)}
\newcommand{\naclvd}{NaCl\,($\upsilonup$=0, $J$=20-19)}
\newcommand{\waterv}{\mbox{o-H$_2$O\,($\nu_2$=1,5$_{5,0}$-6$_{4,3}$)}}

\newcommand{\alo}{\mbox{Al$_2$O$_3$}}
\newcommand{\sisdo}{\mbox{SiS\,$\upsilonup$=0, $J$=12-11}}

\newcommand{\siov}{\mbox{SiO\,$\upsilonup$=1, $J$=6-5}}
\newcommand{\metanol}{CH$_3$OH}

\def\snu#1{\ifmmode {S_\nu\,\propto\,\nu^{#1}}
          \else \hbox{$S_\nu$\,$\propto$\,$\nu^{#1}$}\fi}
\def\cm#1{\ifmmode {\,{\rm cm^{-#1}}}                  
        \else \hbox{$\,${\rm cm$^{\rm -#1}$}}\fi}
\def\raw {\ifmmode\rightarrow\else$\rightarrow$\fi}
\def\ex#1{\ifmmode {\times 10^{#1}}         
  \else \hbox{{$\times 10^{\rm #1}$}}\fi}
\def\dex#1{\ifmmode {10^{#1}}         
        \else \hbox{{10$^{\rm #1}$}}\fi}

\def\app#1#2{%
  \mathrel{%
    \setbox0=\hbox{$#1\sim$}%
    \setbox2=\hbox{%
      \rlap{\hbox{$#1\propto$}}%
      \lower1.1\ht0\box0%
    }%
    \raise0.25\ht2\box2%
  }%
}
\def\approxprop{\mathpalette\app\relax}

\begin{document}

   \title{Dissecting the central regions of \oh\ with ALMA: a salty rotating disk at the base of a young bipolar outflow}


      \author{ C.~S\'anchez Contreras\inst{1}
          \and
           J.~Alcolea\inst{2} 
             \and
           R.~Rodr{\'i}guez Cardoso\inst{1}
              \and
           V.~Bujarrabal\inst{3}
              \and 
           A.~Castro-Carrizo\inst{4}
             \and 
           G.~Quintana-Lacaci\inst{5}
              \and
           L.~Velilla-Prieto\inst{5}
              \and
           M.~Santander-Garc\'ia\inst{2}
   }

      \institute{Centro de Astrobiolog{\'i}a (CSIC-INTA), Postal address:
  ESAC, Camino Bajo del Castillo s/n, Urb. Villafranca del Castillo,
  E-28691 Villanueva de la Ca\~nada, Madrid, Spain\\
  \email{csanchez@cab.inta-csic.es}
  \and
  Observatorio Astron\'omico Nacional (IGN), Alfonso XII
  No 3, 28014 Madrid, Spain
  \and
  Observatorio Astron\'omico Nacional
  (IGN), Ap 112, 28803 Alcal\'a de Henares, Madrid, Spain
  \and
  Institut de Radioastronomie Millimetrique, 300 rue de la Piscine, 38406 Saint Martin d’Heres, France
  \and
  Instituto de Fisica Fundamental (CSIC), C/ Serrano, 123, E-28006, Madrid, Spain  
       }


 
      \abstract{We present Atacama Large Millimeter/submillimeter Array (ALMA)
    continuum and molecular line emission maps
    at $\sim$1\,mm wavelength of \oh, a well studied bipolar nebula around an
    Asymptotic Giant Branch (AGB) star, which is key to investigate the origin of 
    the remarkable changes in nebular morphology and kinematics during the short
    transition from the AGB to the Planetary Nebula (PN) phase.
    The excellent angular resolution of
    our maps ($\sim$20\,mas\,$\approx$\,30\,au) allows us to scrutinise the
    central nebular regions of \oh, which hold the clues to
unravel
how this iconic object assembled its complex nebular architecture. We
report, for the first time in this object and others of its kind
(i.e.\,pre-PN with massive bipolar outflows), the discovery of a
rotating circumbinary disk selectively traced by NaCl, KCl, and H$_2$O
emission lines. This represents the first detection of KCl in an
oxygen-rich (O-rich) AGB circumstellar envelope (CSE).  The rotating disk, of radius
$\sim$30\,au, lies at the base of a young bipolar wind traced by SiO
and SiS emission (referred to as the \sso), which also presents signs
of rotation at its base.  The NaCl equatorial structure is
characterised by a mean rotation velocity of $\vrot$$\sim$4\,\kms\ and
extremely low expansion speeds, \vexp$\sim$3\,\kms.
    The \sso\ has a predominantly expansive kinematics characterised by
    a constant radial velocity gradient of
    $\sim$65\,\kms\,arcsec$^{-1}$ at its base. Beyond
    $r$$\sim$350\,au, the gas in the \sso\ continues radially flowing
    at a constant terminal speed of \vexp$\sim$16\,\kms.  Our
    continuum maps reveal a spatially resolved dust disk-like
    structure perpendicular to the \sso, with the 
    NaCl, KCl and \water\ emission arising from the disk's surface layers. 
    Within the disk, we also identify an unresolved point continuum
    source, which likely represents the central Mira-type star \qx\ enshrouded
    by a $\sim$3\,\rs\ component
    of hot ($\sim$1400\,K) freshly formed dust. 
    The point source is slightly off-centered (by $\sim$6.6\,mas) from
    the disk centroid, enabling us for the first time to place
    constraints to the orbital separation and period of the central
    binary system, $a$$\sim$20\,au and $P_{\rm orb}$$\sim$55\,yr,
    respectively.  The formation of the dense rotating equatorial
    structure at the core of \oh\ is most likely the result of wind
    Roche lobe overflow (WRLOF) mass transfer from \qx\ to the
    main-sequence companion; this scenario is greatly favored by the
    extremely low AGB wind velocity, the relatively high mass of the
    companion, and the comparable sizes of the dust condensation
    radius and the Roche lobe radius deduced from our data. The
    \vexp$\propto$$r$ kinematic pattern observed within the
    $r$$\lsim$350\,au inner regions of the \sso\ suggest that we are
    witnessing the active acceleration of the companion-perturbed wind
    from \qx\ as it flows through low-density polar regions.}

\titlerunning{A salty rotating disk at the core of OH\,231.8+4.2}
\authorrunning{S\'anchez Contreras et al.}

   \maketitle
%

\section{Introduction}
\label{intro}

The majority of stars (with initial masses $\sim$1-8\,\msun) will
become planetary nebulae (PNe) near the end of their lives.
At these late stages, the slow ($\sim$10-20\,\kms), roughly spherical stellar
winds blown during the preceding asymptotic giant branch (AGB) phase
transform into shining PNe with high-speed
($\approx$100\,\kms) outflows and a puzzling variety of aspherical morphologies \citep[see e.g.][for a review]{bal02}. 
The rapid ($\approx$1000\,yr) evolution from (quasi-) spherical to
bipolar or multipolar ejections is initiated prior to the PN stage,
during the so-called pre-PN (pPN) or post-AGB (pAGB) phase, in which
the most spectacular and extreme aspherical geometries are indeed
observed \citep[e.g.][]{mei99,uet00,sah07}.  Binarity is a widely
accepted mechanism for producing fast collimated winds (jets) in dying
stars and, in turn, these jets are potentially the primary agents for the breaking of the
spherical symmetry during the AGB-to-PN transition
\citep[e.g.][]{sah98,DeM09,taf20}.

Rotating equatorial structures are expected to form associated with the presence
of stellar or substellar companions to mass-losing stars
\citep[e.g.][]{elm20,chen17,zou20}, however, direct empirical
confirmation and characterisation of such structures has proven
to be difficult. To date, rotating circumbinary disks have been
found in a population of binary post-AGB stars with near-infrared
(NIR) excess, referred to as disk-prominent post-AGB (dpAGB) stars
\citep{vW17}. The vast majority of dpAGB stars curiously lack of
extended, prominent nebulosities and in all cases lack of massive fast
outflows like those present in the majority of well-studied pPNe, which we then
refer to as wind-prominent pPNe (wpPNe) \citep[see e.g.][and
  references therein]{buj13,buj01,san12}. Rotating circumbinary disks
were spatially and kinematically resolved for the first time in the
Red Rectangle, the best studied dpAGB star, by means of
interferometric observations of the CO molecular emission
\citep{buj03}. In recent years, molecular rotating disks have been
identified and mapped in a few more dpAGB objects
\citep{buj15,buj17,buj18,gall21} and also in the semiregular AGB stars
L$_2$Pup \citep{ker16,hom17} and, tentatively, R Dor and EP Aquarii
\citep{hom18a,hom18b}.

Detection of rotating structures in wpPNe has remained elusive to
date, probably impeded by the large amount of gas and dust material in
the central nebular regions, arranged in different, partially
overlapping, structural components that are difficult to disentangle
observationally. In this work, we report on the first confirmed
detection of a rotating disk in a pPN with massive bipolar outflows.

\oh\ (hereafter \ohs) is a well known bipolar nebula around a mass-losing AGB star,
\qx.  \qx\ is a Mira-type variable that has prematurely developed a
massive ($\sim$1\,\msun) pPN-like nebula with a spectacular bipolar
morphology and very fast outflows, with velocities of up to a few
hundred \kms\ that are reached at the tips of large-scale
($\approx$0.1-0.2\,pc-sized) bipolar lobes \citep{alc01}. As for most
pPNe, the linear momentum of the fast, large-scale bipolar lobes of
\ohs\ notably exceeds that provided by radiation pressure on dust
particles \citep{san97,alc01}, which is believed to be the mechanism
driving the winds of AGB stars \citep{hof18}. This discrepancy, which is observed in most pPNe \citep{buj01}, 
indicates that a different mechanism for the release of kinetic
momentum by the star must be at work. \ohs\ is the best dying star's
example demonstrating that the onset of asymmetry and vigorous dynamics
can begin while the central star is still on the AGB.  \qx, with an M\,9-10\,III spectral
type, is part of a binary system with (at least) one companion, an A0
main-sequence star \citep{san04}, whose presence is probably at the
root of the seemingly premature evolution of this object to the next
pPN phase.

The structure and kinematics of the molecular envelope of \ohs\ has
been recently characterized with unprecedented detail based on
$\sim$0\farc2-0\farc3-angular resolution continuum and molecular line
maps obtained with the Atacama Large Millimeter/submillimeter Array
\citep[ALMA;][hereafter \san]{san18}.  These observations unveiled a
nebular structure much more complex than previously thought, marking a
before and after in our understanding of the mass-loss history and
nebular shaping of this object. \san\ discovered an extravagant array of
nested (but not always co-axial) small-to-large scale structures
indicative of multiple non-spherical mass ejections. The main nebular
components traced by the ALMA observations studied in \san\ are
schematically depicted and listed in their Fig.\,2 and Table\,4,
respectively. Below we summarize the main properties of those nebular
components that are most relevant to this work.

The high-velocity, bipolar lobes ({\sl HV lobes}) and the slowly-expanding, 
equatorially dense waist ({\sl large waist}) from which the lobes
emerge, both known from previous works, are highly structured,
specially the former. In the large-waist the expansion
velocity ranges from \vexp$\sim$\,3\,\kms\ in the inner edge (at a
radius of $r$$\sim$250\,au) to \vexp$\sim$\,25\kms\ in the outer boundary
(at $r$$\sim$2700\,au). The spatio-kinematics of the large waist and
the HV lobes indicate that they were both shaped nearly simultaneously
about 800-900\,yr ago.
 
In the central regions of the nebula, \san\ uncovered two main
small-scale structures: $i$) a compact parcel of gas and dust that
surrounds the mass-losing AGB star, referred to as \cs, which is
selectively traced by certain species, including NaCl; and $ii$) a
compact bipolar outflow that emanates from \cs, which is selectively
traced by SiO. The SiO-outflow is oriented similarly to the
large-scale nebula but is significantly more symmetric, slower (\vexp$\lsim$20\,\kms), and younger (\tdyn$<$500\,yr).
Unexpectedly, \san\ found that \cs\ does not lie on the equatorial plane of the
large-scale nebula but is off-centered by $\sim$0\farc6 to the south
along the nebular axis, perhaps due to a combination of orbital-motion
and recoil of the binary system after strong asymmetrical mass
ejections.

\san\ also reported first detection of methanol (\metanol) and
sodium chloride (in the form of \naclts) in \ohs, adding to the long list of 
$>$30 species already detected toward this object
\citep[e.g.][]{mor87,san15,vel15}, which is the chemically richest envelope
amongst O-rich AGB and post-AGB stars.

The pulsation layers of \qx\  (within a few au) were mapped with $\sim$1\,milliarcsecond (mas) resolution in SiO maser
emission by \cite{san02} using the Very Long Baseline Array (VLBA). In
contrast to normal AGB stars, with SiO maser spots distributed
(nearly) spherically around the star, in \ohs\ the SiO masers are found
to trace an equatorial torus-like structure around \qx. The velocity
gradient measured in the torus, with a radius of $\sim$6\,au,
indicates a composite velocity field with infall motions (attributed
to the stellar pulsation) as well as rotation in these inner
regions. Comparable infall and rotation velocities are found, both in
the range $\sim$7-10\,\kms.  The precise origin of the elongated, equatorially
dense rotating structure traced by the SiO masers is unclear but it is
probably rooted in the gravitational interaction of \qx\ with a
companion.

\ohs\ is a member of the $\sim$0.25\,Gyr old cluster M\,46
\citep{jur85,dav13} located at a distance of
$d$$\sim$1500\,pc. 
The distance to \ohs\ has been most accurately
determined from a trigonometric parallax measurement of the
\water\ masers, leading to $d$=1.54$^{+0.02}$$^{(-0.01)}$\,kpc \citep{choi12}.
\ohs\ has a total luminosity of
\lbol$\sim$7000\,\ls\ that, considering an average stellar temperature
of \teff$\sim$2500\,K, implies a stellar radius of \rs$\sim$2.1\,au.

In this work, we present a new study of the central nebular regions of
\ohs\ (down to linear scales of $\sim$30\,au) based on high-angular
resolution continuum and molecular line observations with ALMA.  We
focus on line emission maps of NaCl, KCl, \water, SiS and SiO, which
are some of the molecules that selectively trace the central \cs\ and
the compact SiO-outflow (now, referred to as the \sso). This paper is
organized as follows.  The observations and the data reduction are
described in \S\,\ref{obs}. The continuum and the molecular line
emission maps are presented in \S\,\ref{res-cont} and
\S\,\ref{res-mol}, respectively. The analysis of the NaCl line
emission, including a rotational diagram analysis and a local thermal
equilibrium (LTE) radiative transfer model, is presented in
\S\,\ref{res-anal}. In \S\,\ref{binary}, we estimate the central
binary's orbital separation, and the central dynamical mass and
angular momentum of the rotating disk.  In \S\,\ref{discussion}, we
discuss our main findings and present our conclusions on the formation
process of the two principal structures here under study, the rotating
disk and the young \sso.  Finally, a summary of the points addressed
in this work is provided in \S\,\ref{summ}.

\section{Observations and data reduction}
\label{obs}
%
\begin{table}
  \caption{Properties of the spectral windows in project 2017.1.00706.S and continuum flux measurements (last column).}
  \label{t-obs}
  \small
\centering          
\begin{tabular}{c r c c }
\hline\hline       
Center       &     Bandwidth    &    Velocity resolution  & Continuum Flux ($\pm$$\sigma$) \\ 
 (GHz)         &     (km/s)          &    (\kms)              &  (mJy) \\    
\hline                    
\multicolumn{4}{c}{\sl OH231.8\_a\_06\_TM1 (2017-Oct-15:20)}\\ 
261.240 & 537.9  & 0.562          &  34.0($\pm$4.1)  \\  
260.336 & 539.8  & 0.562    	 &  33.0($\pm$4.1)  \\  
258.988 & 135.7  & 0.283    	 &  32.0($\pm$5.1)  \\   
258.684 & 135.8  & 0.283    	 &  33.0($\pm$4.1)  \\   
258.272 & 136.0  & 0.283    	 &  33.0($\pm$4.1)  \\  
257.232 & 136.6  & 0.285    	 &  31.0($\pm$4.0)  \\  
244.913  & 2294.4 & 1.195    	 & 29.1($\pm$3.1)  \\   
242.617  & 144.8  & 0.302    	 & 28.2($\pm$3.0)  \\   
241.935  & 290.4  & 0.303    	 & 28.4($\pm$3.1)  \\   
241.752  & 145.3  & 0.303    	 & 28.2($\pm$3.1)  \\  
\multicolumn{4}{c}{\sl OH231.8\_b\_06\_TM1 (2017-Oct-16:18)}\\ 
234.219  & 299.5      &	0.312 & 26.7($\pm$2.0) \\   
232.665  & 301.4      &	0.314 & 26.1($\pm$2.0) \\   
231.880  & 606.2      &	1.263 & 26.5($\pm$2.0) \\   
231.484  & 607.2      &	1.265 & 26.3($\pm$2.0) \\   
231.199  & 607.7      &	1.266 & 26.0($\pm$2.0) \\    
230.517  & 609.6      &	1.270 & 25.8($\pm$2.0) \\   
219.778  & 319.7      &	0.666 & 23.7($\pm$2.0)  \\    
219.540  & 319.7      &	0.666 & 24.2($\pm$2.0)  \\   
218.961  & 320.6      &	0.668 & 23.5($\pm$3.0)  \\    
218.202  & 321.6      &	0.670 & 23.4($\pm$2.0)  \\   
217.921  & 160.8      &	0.335 & 23.5($\pm$2.0)  \\   
217.798  & 161.3      &	0.336 & 23.2($\pm$3.0)  \\   
217.085  & 161.8      &	0.337 & 23.4($\pm$4.0)  \\   
216.718  & 161.8      &	0.337 & 32.0($\pm$12)   \\   
\hline                  
\end{tabular}
\end{table}
%


This work is based on observations obtained with the ALMA 12\,m
interferometric array during cycle 5 as part of project 2017.1.00706.S
(PI: S\'anchez Contreras). Two frequency settings (\oha\ and \ohb)
within band 6 ($\sim$242-261\,GHz and $\sim$217-234\,GHz,
respectively) were used to map the molecular line and continuum
emission toward \ohs.  In Table\,\ref{t-obs}, we list the different
spectral windows (SPWs) in our data set, a total of 24,
with their central frequencies, bandwidths, and velocity
resolutions (three first columns) as well as the observing dates of each frequency setting.
Observations of \oha\ and \ohb\ were done in three and two $\la$1.7\,hr-long different sessions
or blocks, respectively, executed within a window of a few days.  The
data were obtained with 50-52 antennas with baselines ranging from
41.4\,m to 16.2\,km, resulting in a highest angular resolution of
about $\sim$0\farc02.
The maximum recoverable scale (MRS) is $\sim$0\farc3 and 0\farc4 for
\oha\ and \ohb, respectively.  Following the standard calibration
procedure, a number of sources (J0750+123, J0730$-$1141, and
J0746$-$1555) were observed as bandpass, complex gain, and flux
calibrators.  The flux density adopted for J0750+1231 is
0.793$\pm$0.08\,Jy at $\sim$217\,GHz.
We estimate the absolute flux calibration to be accurate to within 7\%-15\%, depending on the spectral window.

\begin{figure}[ht]
     \centering
      \includegraphics[width=0.9\hsize]{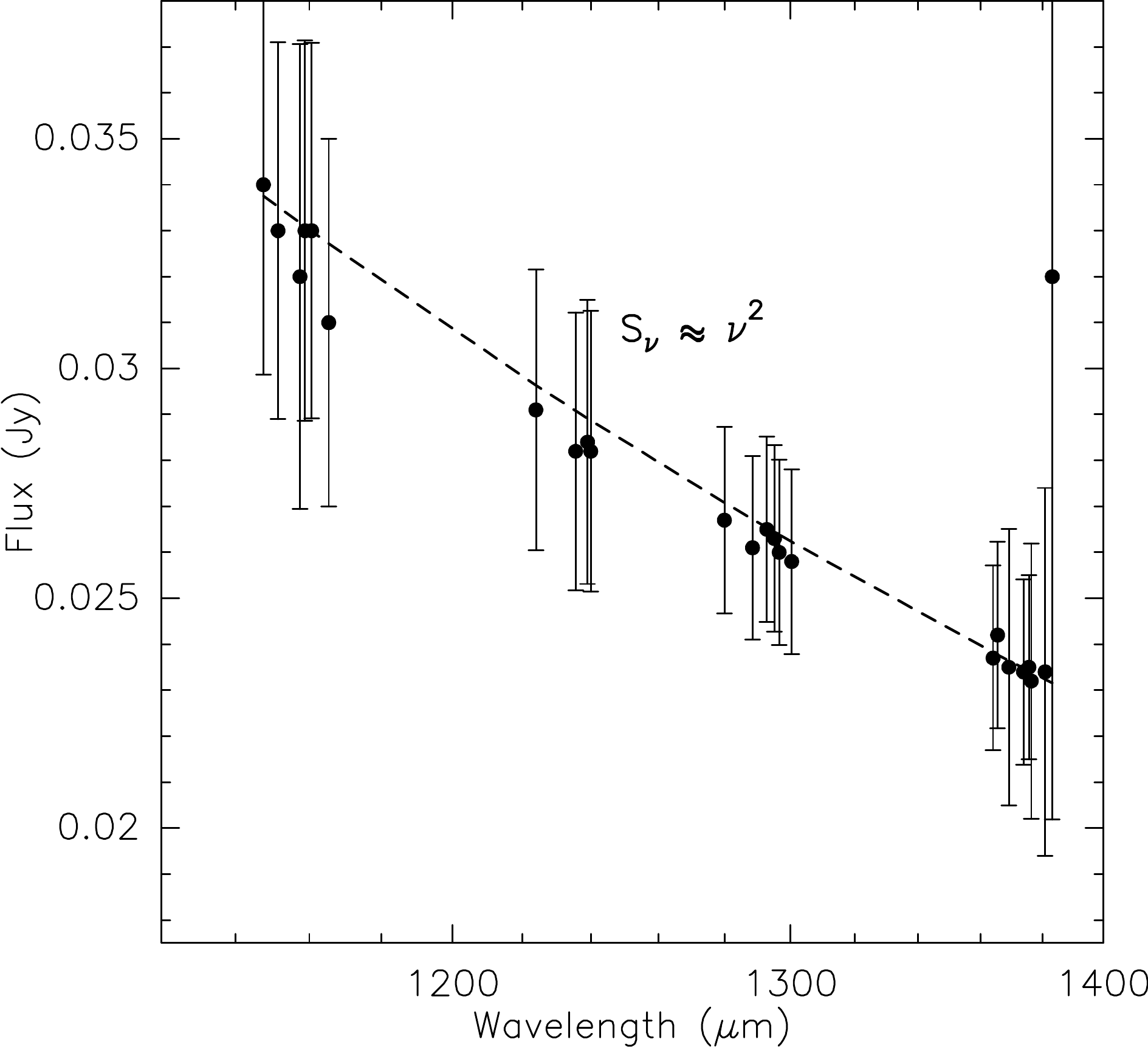}     
       \caption{Spectral energy distribution (SED) of \ohs\ showing
         the continuum emission flux measurements from the individual
         SPWs observed in this project (Table\,\ref{t-obs}). The
         dashed line represents the best-fit power law of the
         continuum flux. The outlier at 216.7\,GHz
         (nearest to 1400\,$\mu$m) has very large errorbars possibly
         reflecting a calibration problem and/or some contamination by
         H$_2$S line emission.
       }
         \label{f-sed}
   \end{figure}

The calibration of the data was performed using the automated ALMA
pipeline of the Common Astronomy Software Applications package
(CASA\footnote{\tt https://casa.nrao.edu}; versions 5.1.1 and
5.4.0). We created continuum images for each of the 24 SPWs using
line-free channels.
These continuum images were used to measure the continuum fluxes from
\cs\ for each of the SPWs observed (Table\,\ref{t-obs} and
Fig.\,\ref{f-sed}).  Subsequently, self-calibration was performed on
the strong continuum using the initial model of the source derived
from the standard calibration to improve the fidelity of the continuum
images.  The line emission cubes, produced after subtracting the
underlying continuum emission (using the SPW containing each line),
have been obtained from the data with the initial standard calibration
because, in contrast to the continuum, the noise in the line maps was
not dominated by residual calibration errors and no improvement was
achieved after applying self-calibration.  Self-calibration of the
continuum as well as the final image restoration and deconvolution was done
using the GILDAS\footnote{http://www.iram.fr/IRAMFR/GILDAS} software
MAPPING.

\begin{figure*}[ht]
     \centering
     \includegraphics*[bb= 5 1 734 590, width=0.33\hsize]{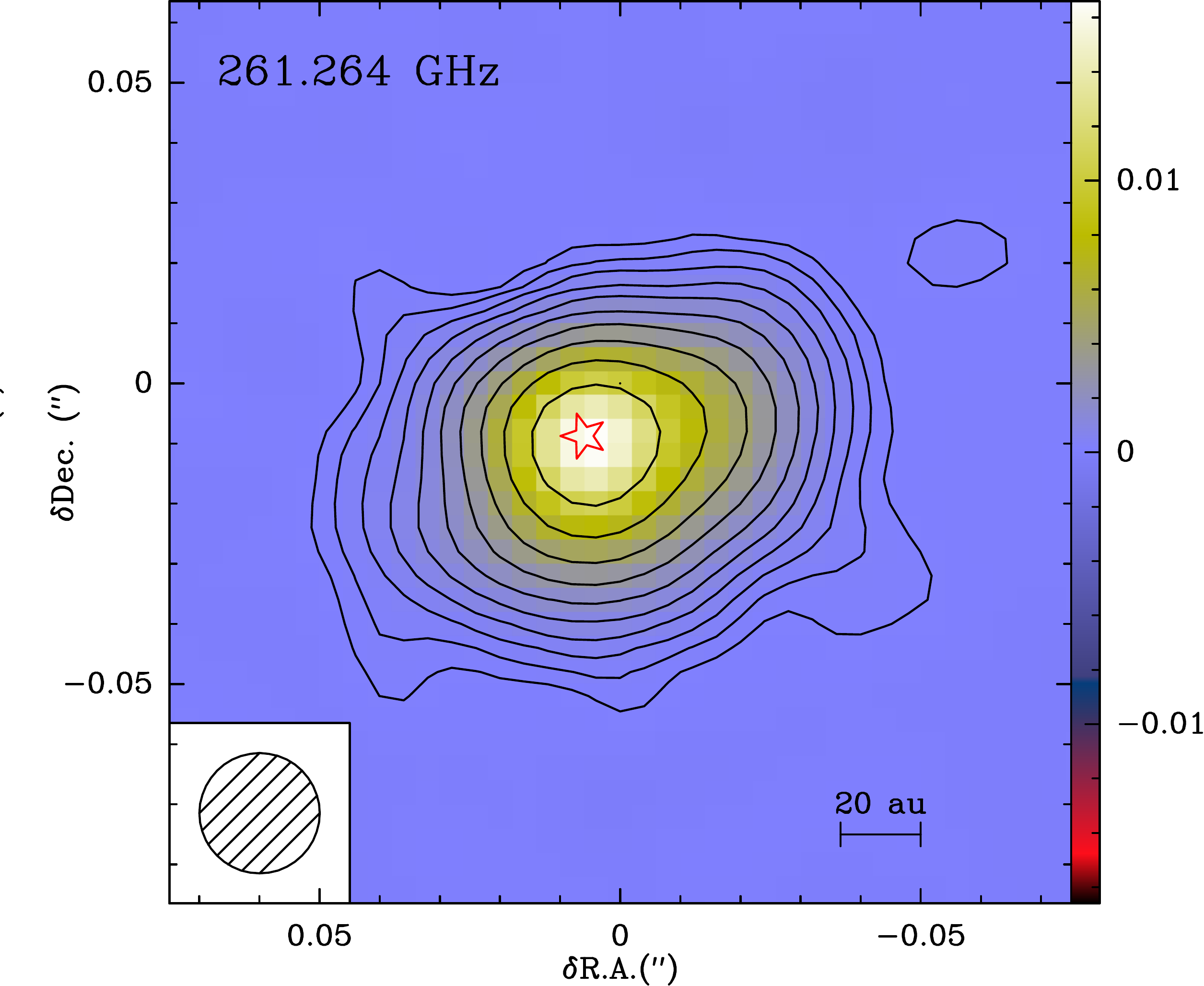}
     \includegraphics*[bb= 5 1 734 590, width=0.33\hsize]{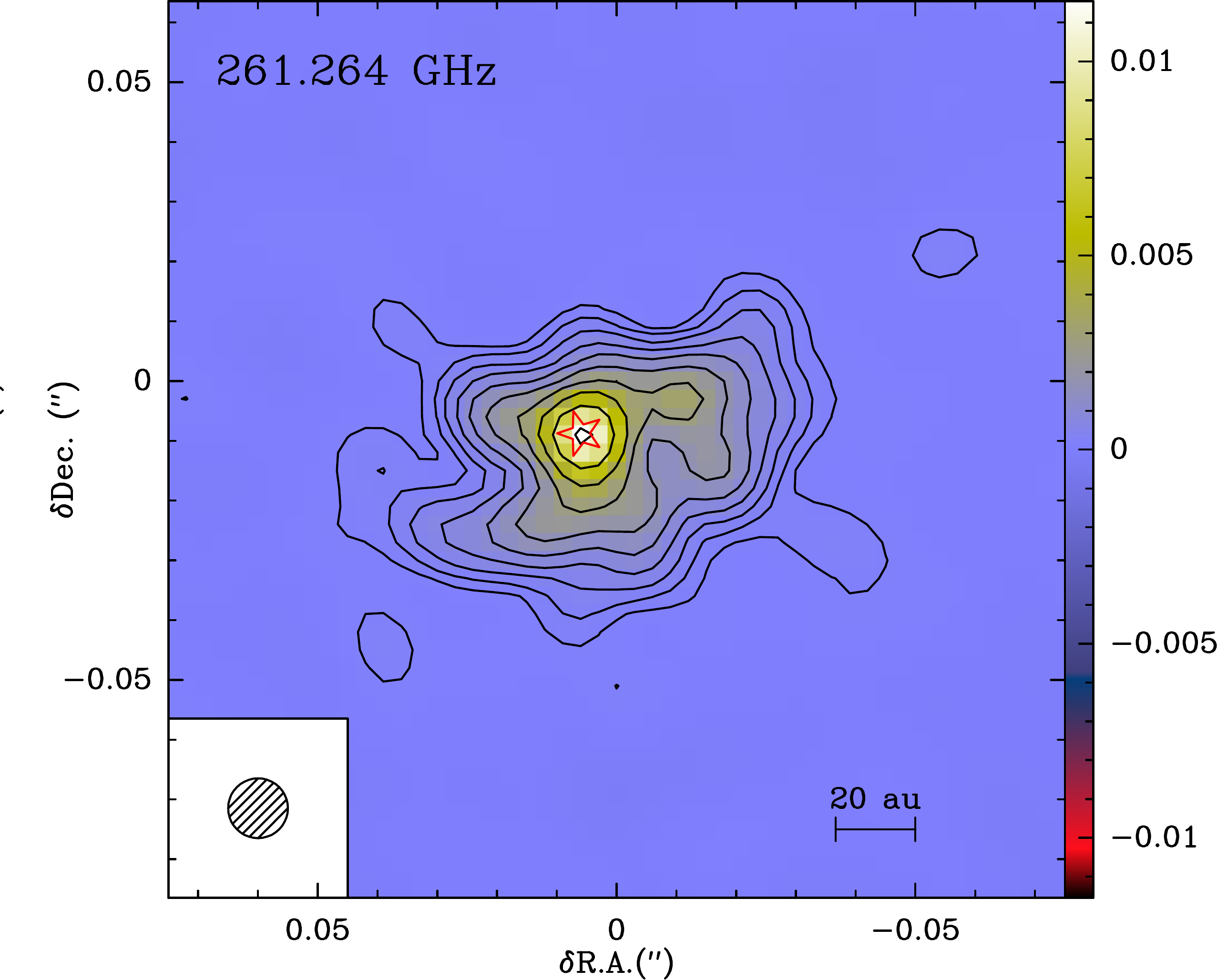}
     \includegraphics*[bb= 5 1 734 590, width=0.33\hsize]{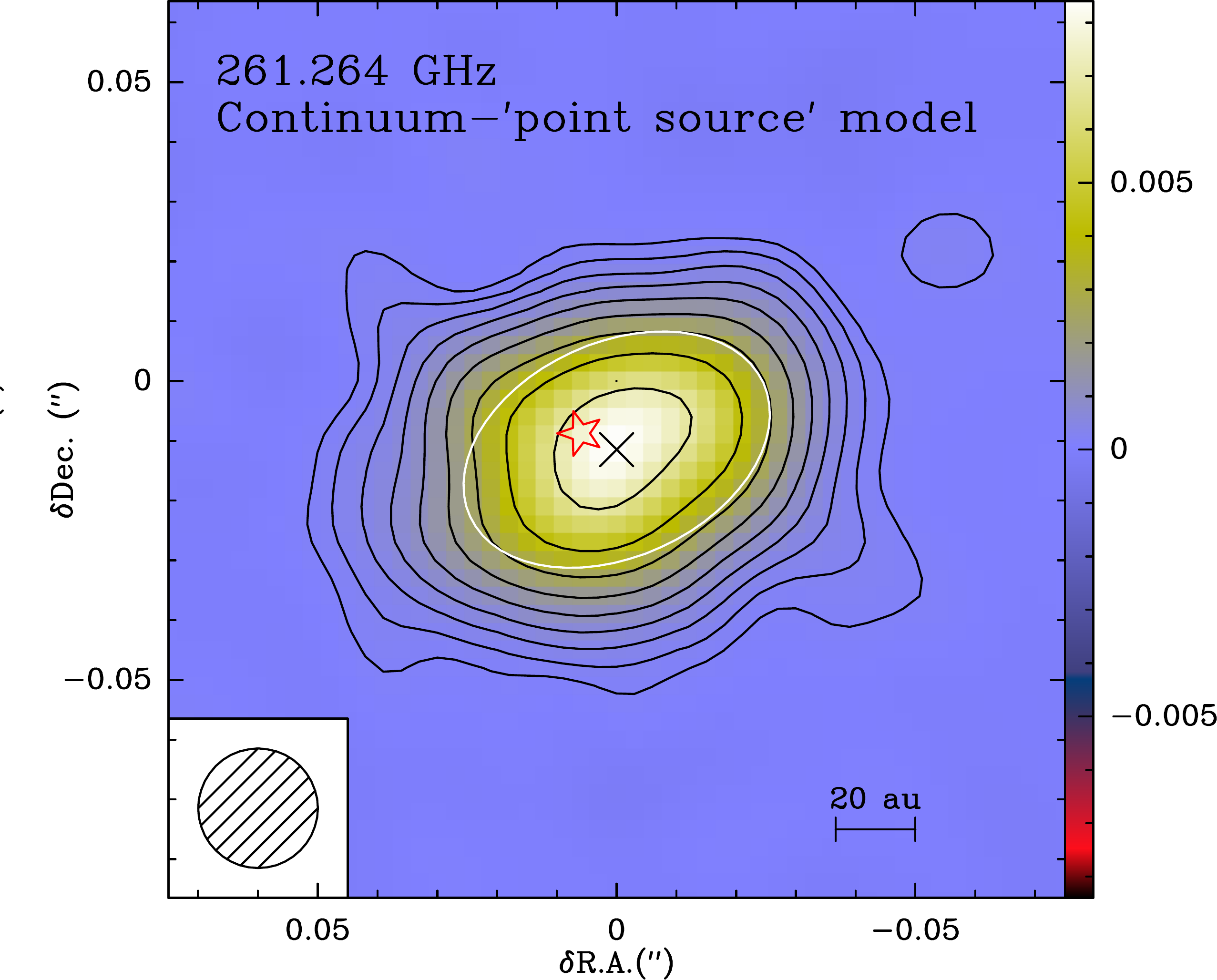}
     \caption{ALMA continuum emission maps at 261.264\,GHz. {\bf Left)} and {\bf Middle)}: continuum maps with 20 and 10\,mas
       restoring beams, respectively (\S\,\ref{obs}). {\bf Right)}: continuum map with 20\,mas restoring beam after
       fitting and subtracting a point source model that is located
       at the position marked by the star-like symbol (see
       Sect.\,\ref{res-cont}).  The white ellipse represents the
       size and orientation of the extended disk model
       that best fits the emission in this map. Note the offset
       between the position of the point-like continuum source
       (starlike symbol at J2000, R.A.=07\h42\m16\fs91543
       Dec.=$-$14\degr42\arcmin50\farc0691) and the center of the disk
       (cross at J2000, R.A.=07\h42\m16\fs91500 and
       Dec.=$-$14\degr42\arcmin50\farc0716). In all three panels, the level contours are 10$^{(-3.77+(i-1)*0.2)}$\,Jy\,beam$^{-1}$, $i$=1 to 11 by 1 and the shaded circular areas
       at the bottom-left corner of the maps represent the
       half-power-beam-width (HPBW).}
         \label{f-cont}
   \end{figure*}

The final line cubes and self-calibrated continuum images here
presented were created using by default the Hogbom deconvolution
method with a robust weighting scheme\footnote{We used a value of 0.56 for the threshold of the robust weighting in MAPPING.}, 
resulting in angular resolutions of $\sim$0\farc016-0\farc019, with a
nearly circular beam. For an optimal comparison of all the maps with
the same angular resolution, the final cubes have been created by
imposing a circular restoring beam of 20\,mas.  Additionally, SiS maps
have been also restored using natural weighting and tapering (with a
tapering distance of 5700 m), resulting in a $\sim$60\,mas beam, which
offers a good compromise between sensitivity to smooth medium-size
structures and angular resolution.  For the continuum, we also
analysed the distribution of clean components and used it to produce a 
higher angular resolution version of the continuum images by imposing
a circular restoring beam of 10\,mas, i.e.\,smaller than the nominal
$\sim$20\,mas beam. This 10\,mas-resolution clean-components map has been
used exclusively to better discern the two different emission
components already inferred from the continuum maps with a nominal
angular resolution of 20\,mas and for a better determination of their
relative positions (\S\,\ref{res-cont}, Fig.\,\ref{f-cont}).

The typical rms noise level per channel of our 20\,mas-resolution
spectral cubes is $\sigma$$\sim$0.5-0.6\,m\jb\ at 2\,\kms\ resolution.
The rms noise level range in the 20\,mas-continuum maps is
$\sigma$$\sim$0.05-0.09\,m\jb\ for the individual SPWs.

\section{Continuum emission}
\label{res-cont}
%
%

\subsection{Surface brightness distribution}
\label{cmaps}
We have made ALMA continuum emission maps of \ohs\ for the 24
SPWs observed in this project (within 
$\sim$217-232\,GHz and $\sim$241-261\,GHz -- see Table\,\ref{t-obs}). The
surface brightness distribution of the continuum is very similar 
(within $\pm$3$\sigma$ errors) at all these frequencies. In
Fig.\,\ref{f-cont}, we show a representative continuum emission
map at 261.3\,GHz, the highest frequency observed in this work, with
20\,mas and 10\,mas restoring beams (left and middle panel, respectively).

As shown in \san, the continuum emission of \ohs\ at (sub)mm wavelengths
is due to dust thermal emission from two major components: (1)
a $\sim$8\arcsec$\times$4\arcsec\ hourglass-like structure with
optically thin cold ($\sim$75\,K) dust and (2) a bright compact
($<$0\farc1) condensation with optically thin hotter ($\sim$300-400\,K) dust - referred to as \cs.
For the ALMA configurations used in this project, the angular size of
the largest smooth structure to which our observations are sensitive 
(or MRS) is $\sim$0\farc3-0\farc4.
Therefore, as expected, the continuum emission from the
$\sim$8\arcsec$\times$4\arcsec\ hourglass-like structure is now filtered
out almost completely. We note that this is
not problematic since the focus of our high-angular resolution
observations here presented is to dissect the small-scale structures
discovered in the central regions of \ohs\ (including \,\cs), which 
should offer clues as to how this iconic object assembled its
complex nebular architecture at larger scales.


We spatially resolve the bright compact condensation of dust at the
center of \ohs\ referred to as \cs\ (\san). Our maps reveal an
elongated structure with its long axis oriented roughly perpendicular
to the main symmetry axis of the bipolar lobes of \ohs. This suggests that 
this structure is an inclined disk.
The emission peak of the continuum (marked with a starlike symbol in
Fig.\,\ref{f-cont}) appears to be slightly offset (toward the
northeast) from the centroid of the extended, disk-like continuum emitting
region. The offset becomes more clear in the
10\,mas maps, where the strong continuum emission stands
out very clearly as a point-like source on top of a more diffuse and
extended emission component.

To confirm the presence of the point source and the disk-like emission
components inferred above and to better constrain their positions and
dimensions, we have fitted a double component model to the $uv$
continuum data. To do that we used the GILDAS/MAPPING task {\tt
  uv\_fit}. The fitting has been done in two steps: first, a
point-source model is fitted and subtracted to the $uv$ continuum data
and, after that, the residual emission is fitted again using a uniform
elliptical disk model.  To fit the point-source model we have 
used only the longest baselines (length $>$6000\,m), which enables us to
filter out partially the emission by the extended component and to
better isolate the emission from the point source. 
The point-source model has been subtracted from the original $uv$ data, after which a cleaned image of
the residual continuum has been created, shown in the right panel of Fig.\,\ref{f-cont}. 
The continuum $uv$ data and the two fits (of the point-source and the extended disk) are shown in the Appendix in
Fig.\,\ref{f-uvfit} together with a final residual map of the continuum after subtraction of the (point-source+extended-disk) model.

The position of the point-source deduced from the fit is
R.A.=07\h42\m16\fs91543 Dec.=$-$14\degr42\arcmin50\farc0691 (J2000),
with formal errors from the fit of 0.18 and 0.14\,mas, respectively,
and is marked by a starlike symbol in the continuum maps shown in
Fig.\,\ref{f-cont}. The continuum flux of the point-like source
deduced from the fit is $\sim$9.1$\pm$0.1\,mJy.
As we show later in \S\,\ref{sed} and \S\,\ref{res-mol}, the
point-like continuum emission source marks the position of the central
AGB star \qx.

By fitting a disk source model to the residual extended
261\,GHz-continuum emission ($\sim$25.6$\pm$0.2\,mJy) we find that the
dimensions and orientation of this structure are
52.6\,mas$\times$34.0\,mas ($\pm$0.5\,mas) and PA=115$\pm$1.6\degr,
respectively. The major-to-minor axis ratio found is consistent with a
circular disk of radius $\sim$40\,au with its plane inclined by
$i$$\la$40\degr\ with respect to the line of sight, consistent with
the inclination of the large-scale bipolar lobes of \ohs\ with respect
to the plane of the sky
\citep[$i$$\sim$35\degr,][]{bow84,kas92,shu95}. This suggests that the
dust continuum disk and the bipolar lobes are indeed orthogonal as
already suspected given their relative orientation (mutually
perpendicular) in the plane of the sky.
The upper limit to $i$ derived from the dimensions of the continuum
disk arises because the disk may have a nonzero vertical thickness. 
In this case the minor axis of the projected geometrically thick disk
would increase by $h\times\cos(i)$ with respect to the 
infinitesimally thin ($h$=0) disk case. Adopting an inclination $i$=30\degr, the
dimensions of the continuum suggest a disk vertical thickness of
$h$$\sim$12-13\,au. 

As already guessed from a quick eye inspection of the continuum maps, the center of the disk (R.A.=07\h42\m16\fs91500 and
Dec.=$-$14\degr42\arcmin50\farc0716, J20000) is offset by  
6.1$\pm$0.3\,mas and  2.5$\pm$0.2\,mas toward the west and south, respectively, 
from the position of the central mass-losing star \qx.
The offset between \qx\ and the centroid of the disk can be easily
explained if the disk is circumbinary, in which case the disk centroid
is expected to coincide with the center of mass of the binary system.  We
further develop this idea in \S\,\ref{binary}, where we use it to
place constraints on the orbital separation based on the `\qx'-`disk
center' relative offset observed (6.6\,mas\,$\sim$\,10\,au at $d$=1500\,pc).



\subsection{Spectral energy distribution}
\label{sed}

The continuum flux measurements (Table\,\ref{t-obs}), obtained
integrating the surface brightness over the emitting region
(i.e.\ within \cs), at the different frequencies observed in this
project are shown in Fig.\,\ref{f-sed}. The $\sim$242-261\,GHz (band
6) continuum flux from \cs\ follows a \snu2 frequency dependence, in
agreement (within errors) with what was observed with ALMA at somewhat
higher frequencies, $\sim$294-344\,GHz (band 7), and with more compact
configurations (\san). The absolute continuum flux in band 6 measured
in this work is $\sim$15\%-20\% lower than that expected by
extrapolating the observed \snu2\ power-law of the continuum flux in
band 7. Since absolute flux errors of up to $\sim$15\% are possible in
both bands, we confirm small (less than 15\%-20\%)
interferometric continuum flux losses in our high-angular resolution
continuum maps from \cs.


  From our $\sim$242-261\,GHz continuum maps obtained using the
  data for the individual observed spectral windows, we find no signs of significant
  deviations from a \snu2 power-law (within errors) either for the
  point-like source at the location of \qx\ or for the extended
  disk-like component.  However, the uncertainty in the spectral index
  determination for the individual (point-like and extended) components is very high and no
  further attempt to perform a separate analysis will be performed.  For a
  reliable characterization of the spectral index distribution across
  the continuum-emitting region more sensitive and higher angular
  resolution multifrequency continuum maps are needed.)
As discussed in \san, the \snu2
continuum distribution and the total flux measured from \cs\ suggests
a dominant component of optically thin thermal-emission produced by
large ($\ga$100$\mu$m-sized) dust grains with a flat emissivity
law. In the following paragraphs, we show (1) that the photosphere of
the AGB star \qx, also expected to follow a frequency dependence
approximately consistent with black-body emission, is responsible for
part ($\sim$10\%-20\%) of the continuum emission from the point-like
source, and (2) that the rest of the unresolved continuum source is
probably hot ($\sim$1400\,K), freshly formed dust in the close
vicinity (within $\sim$3-4\rs) of \qx.

The contribution to the observed 261\,GHz-continuum by the photosphere
of \qx\ (\teff$\sim$2500\,K and \rs$\sim$2.1\,au, \S\,\ref{intro}) in
main-beam brightness temperature units is
\tmb=2500\,K$\times$($\frac{\rs=2.1\,au}{(beam/2)=7.5\,au}$)$^2$=196\,K,
which is equivalent to $\sim$1\,mJy (considering the ALMA
beam=0\farc01$\times$0\farc01 and mJy-to-K=5.5\ex{-3} conversion
factor at this frequency). This is notably smaller than the flux of
the point-like continuum emission source ($\sim$9\,mJy, as derived
from the $uv$-continuum data analysis, \S\ref{cmaps}).
The contribution to the continuum flux of the main-sequence
companion (a factor 4 hotter but over 200 times smaller than \qx) is
10$^4$ times smaller than that by \qx\ and, thus, totally negligible.

Long-period variables, including Mira-type stars such as \qx, are
thought to have radio photospheres near 2\rs\ (i.e.\,just inside the
SiO-maser shell and dust formation zone) that can also in principle
contribute to the mm-to-cm wavelength continuum emission
\citep{reid97,reid07}. In the case of \qx, however, the radio
photospheric emission at 261\,GHz is expected to be only of
$\sim$1\,mJy according to equation (7) in \cite{reid97}, i.e.\,still
far below the emission level of the point-like continuum source
observed by us.

We conclude then that the excess continuum flux of the unresolved
source ($\sim$7-8\,mJy, after subtracting the stellar photo-/radio-sphere) is
dominated by thermal emission from a source other than the star, most
likely due to hot dust within beam/2$\sim$7.5\,au of \qx.

\subsection{Continuum emission components}
\label{dcomponents}
\subsubsection{Emission from a compact hot-dust shell and a circumbinary warm-dust disk}
\label{hot&warm-dust}
The spectral index of the point-like continuum indicates that the dust
emission is either optically thick or optically thin due to grains with emissivity index
$\alpha$$\sim$0 (i.e.\, produced by large solid particles).  In case
of optically thick dust, the 261\,GHz-continuum excess flux from the
point-like source
\tmb$\sim$[7-8]\,mJy/5.5\ex{-3}\,K\,mJy$^{-1}$$\sim$1350($\pm$150)\,K
would imply a dust temperature of
\td$\sim$1350($\pm$150)$\times$($\frac{(beam/2)=7.5\,au}{r_{\rm
    d}[au]}$)$^2$\,K.  (Errorbars in \tmb\ include absolute flux
calibration uncertainties, which could be up to $\sim$15\%, formal
errors from the two-component $uv$-continuum fit, and errors in the
estimate of \qx's contribution to the observed continuum.)  We believe
that the radius of the compact, hot dust region around \qx\ is
unlikely to be significantly smaller than $\sim$7.5\,au because in
that case the \td\ implied would exceed the grain
condensation temperatures ($\lsim$1500\,K, expected for
condensates typical of O-rich CSEs), above which solid particles
cannot form or survive (see \S\,\ref{alo}).
%
%

We have computed the mass of the compact hot-dust component in case of
optically thin and thick thermal dust emission following the procedure
and assumptions described in detail in \san. For optically thin dust
(with flat emissivity) at \td$\sim$1400\,K, the mass of dust inside
the compact region around \qx\ is
\md$\sim$10$^{-6}$-10$^{-5}$\,\msun\ for a grain radius of
$r_{\rm g}$=100-1000\,$\mu$m.
For optically thick dust (and adopting more standard values for the
grain radius and emissivity, Li \& Draine 2001), we deduce a rather
high value for the lower bound to the mass of
\md$\ga$9\ex{-5}($\frac{1400\,K}{\td}$)\,\msun, which would imply a
high amount of material of $\ga$\,0.01\msun, using the
canonical value of the gas-to-dust mass ratio, g/d$\sim$160, for O-rich AGB stars \citep{kna85}, or even up to one order of magnitude more 
if we use the larger values of g/d from most recent works \citep{ram08,bla19}.
Assuming an expansion velocity \vexp$\sim$3\,\kms\ (see
\S\,\ref{res-mol}) for the $\sim$7.5\,au inner layers of \qx's wind,
resulting in kinematical ages of only $\sim$10\,yr, the previous value
of the mass would imply an unrealistically high mass-loss rate of
$\ga$\dex{-3}-\dex{-2}\,\my, never observed before in AGB stars
\citep[typically \mloss$\approx$\dex{-7}-\dex{-4}\,\my,
  e.g.][]{hof18}.

According to this, we believe that the point-like hot dust emission is
probably due to the emission of large $\sim$100-1000$\mu$m grains, like the extended disk-like warm dust
component. If this is the case, the observed value of
\tmb$\sim$1350($\pm$150)\,K is a lower limit to the dust
temperature. Since \tmb\ is already close to the largest temperatures
that solid particles can survive to, the dust emission is probably not
very optically thin.  In the optically thin dust scenario, the
mass-loss rate inferred for $r_{\rm g}$$\sim$100$\mu$m grains,
\mloss$\sim$\dex{-5}\,\my, is in good agreement with the present-day
mass-loss rate of \ohs\ estimated from the analysis of SiO maser
emission \citep{san02} and from a molecular line study of mid- to
high-$J$ CO transitions observed with $Herschel$ \citep{ram18}.


We note that our estimates of the present-day mass-loss rate presented
above, which are in any case uncertain, should be considered as
average or mean values since we have assumed for simplicity that the
hot dust is distributed in a (nearly) spherical expanding shell, which remains
to be validated with higher angular resolution observations.

Across the extended disk-like component (which accounts for
$\sim$80\% of the total 261\,GHz-continuum flux measured toward \cs), the
main-beam brightness temperature ranges from $\sim$8\,K (at 
the outer 3$\sigma$ layers) to $\sim$470\,K at the center in our
10\,mas-resolution maps.  For an optically thin dusty disk, the
observed range in \tmb\ is a lower limit to the dust temperature
distribution across the disk. For an average dust temperature in the
disk of $\sim$350\,K (consistent with our initial rough estimate
presented in \san) and after subtracting the contribution by the
point-like continuum emission flux (by \qx\ and the compact hot dust
shell), the dust mass in the disk is \md$\sim$1.5\ex{-5}-1.5\ex{-4}\,\msun\ for a grain radius of $r_{\rm g}$=100-1000\,$\mu$m.

A rough estimate of the average H$_2$ number density in the extended
dust-disk of $\gtrapprox$10$^8$-10$^9$\,\cm3 (assuming
optically thin 100-1000\,$\mu$m-sized dust) and
$\gtrapprox$10$^{10}$\,\cm3 (assuming optically thick dust) is obtained
considering the gas masses derived above and the dimensions of the
disk, with an outer radius of $\sim$40\,au and thickness of
$\lsim$15\,au (since it is not spatially resolved in the vertical
direction).

Finally, we note that in any case, the dust (and gas) mass computed
above in the optically thin scenario, for both the hot compact and warm disk-like dust components
at the core of \ohs, have large uncertainties mainly
due to the largely unknown properties (absorption coefficient) of the
dust, particularly for big grains, and poorly constrained gas-to-dust mass ratios.


\subsubsection{An \alo\ compact shell around \qx?}
\label{alo}



The high dust temperature and small radius of the dust structure around \qx\
inferred above are most consistent with \alo\
dust, which is the solid with the largest condensation temperature that is
expected to form abundantly in the extended atmospheres (within
2-3\rs) of O-rich AGB stars based on chemical equilibrium models
\citep[e.g.][and references therein]{hof19,agun20}.
Based on thermodynamic equilibrium (TE) chemical models, \alo\ is
indeed the first condensate (often referred to as {\sl seed} particle)
emerging in the atmospheres of O-rich AGB stars. The presence of hot
($\sim$1400\,K) \alo\ dust at the inner boundaries ($\sim$2\rs) of the
dust shells around Mira (O-rich AGB) stars is not only theoretically
predicted but it has been empirically confirmed by spatially and
spectrally resolved mid-infrared interferometric observations of
characteristic dust features \cite[e.g.][]{zhao12,karo13,kho15}.
Observations presented by these authors are consistent with
\alo\ grains condensing in a thin gravitationally bound dust layer
close to the stellar surface, co-located with the extended atmosphere
and SiO maser emission.

The presence of a significant amount of amorphous \alo\ in \ohs\ has
been previously proposed by \cite{mal04} to explain both the
long-wavelength wing of the 10$\mu$m silicate feature and to account
for the enhanced opacity between the 10- and 18$\mu$m silicate bands.
We also note that the radius of the hot-dust component inferred from
our continuum maps (\rd$\sim$7.5\,au) is comparable to that of
the torus-like SiO masing region around \qx\ ($r$$\sim$6\,au,
\S\,\ref{intro}), as expected in case of \alo\ grains.

\section{Molecular line emission maps: dissecting the core and the compact SiO/SiS-outflow}
\label{res-mol}

We have observed a large number of molecular transitions as part of
this project. In this paper, we focus on a few species that have been
found to selectively trace a number of small-scale nebular components
at the core of \ohs, down to linear sizes of $\sim$20-30\,au
($\sim$10-15\,\rs) from the central mass-losing star. In this
section, we present the results from the observations of these
species, namely, NaCl, KCl, \water, SiO and SiS. 
KCl is a new detection in this object and also
represents the first detection of this molecule in an O-rich AGB CSE.
A list of the transitions reported in this work is given in table
\ref{tab:mols}.
\begin{table}[ht]
\small
\caption{Molecular transitions reported in this work} \label{tab:mols}
\begin{tabular}{l l r c}
\hline
\hline
Transition & Rest Frequency & \eu & A$_{\rm ul}$\\
                &  (MHz) & (K) & (s$^-1$) \\  
\hline
NaCl\,($v$=0, $J$=18-17) & 234251.915 & 106.9 & 5.890$\times 10^{-3}$ \\ 
NaCl\,($v$=0, $J$=20-19) & 260223.117 & 131.2 & 8.108$\times 10^{-3}$ \\ 
NaCl\,($v$=1, $J$=17-16) & 219614.934 & 614.5 & 4.917$\times 10^{-3}$ \\ 
NaCl\,($v$=1, $J$=19-18) & 245400.990 & 637.5 & 6.881$\times 10^{-3}$ \\ 
NaCl\,($v$=1, $J$=20-19) & 258287.755 & 649.9 & 8.034$\times 10^{-3}$ \\ 
NaCl\,($v$=2, $J$=17-16) & 217980.229 & 1128.4 & 4.872$\times 10^{-3}$ \\ 
NaCl\,($v$=3, $J$=19-18) & 241758.920 & 1659.9 & 6.755$\times 10^{-3}$ \\  
\naclts\,($v$=0, $J$=19-18) & 247239.646 & 118.7 & 6.940$\times 10^{-3}$  \\
KCl\,($v$=0, $J$=30-29) & 230320.560  & 171.5 & 7.377$\times 10^{-3}$  \\ 
KCl\,($v$=0, $J$=32-31) &   245623.530 & 194.7 & 8.956$\times 10^{-3}$  \\
KCl\,($v$=1, $J$=32-31) &   244114.520 & 592.7 & 8.895$\times 10^{-3}$ \\
\waterv\ & 232686.700 & 1133.2 & 4.630$\times 10^{-6}$ \\ 
SiS\,($v$=0, $J$=12-11) & 217817.473 & 68.0 & 1.738$\times 10^{-4}$ \\
SiO\,($v$=1, $J$=6-5) &  258707.324  & 1812.7 & 9.042$\times 10^{-4}$ \\

\hline

\hline                  
\end{tabular}
\tablefoot{Spectroscopic information of the transitions from the {\sl Madrid  Excitation Code} \citep[MADEX][]{cer12}}
\end{table}


\subsection{A salty and watery (brine) rotating disk inside \cs.}
\label{res-nacl}
   \begin{figure*}[t]
     \centering
     \includegraphics[width=0.95\hsize]{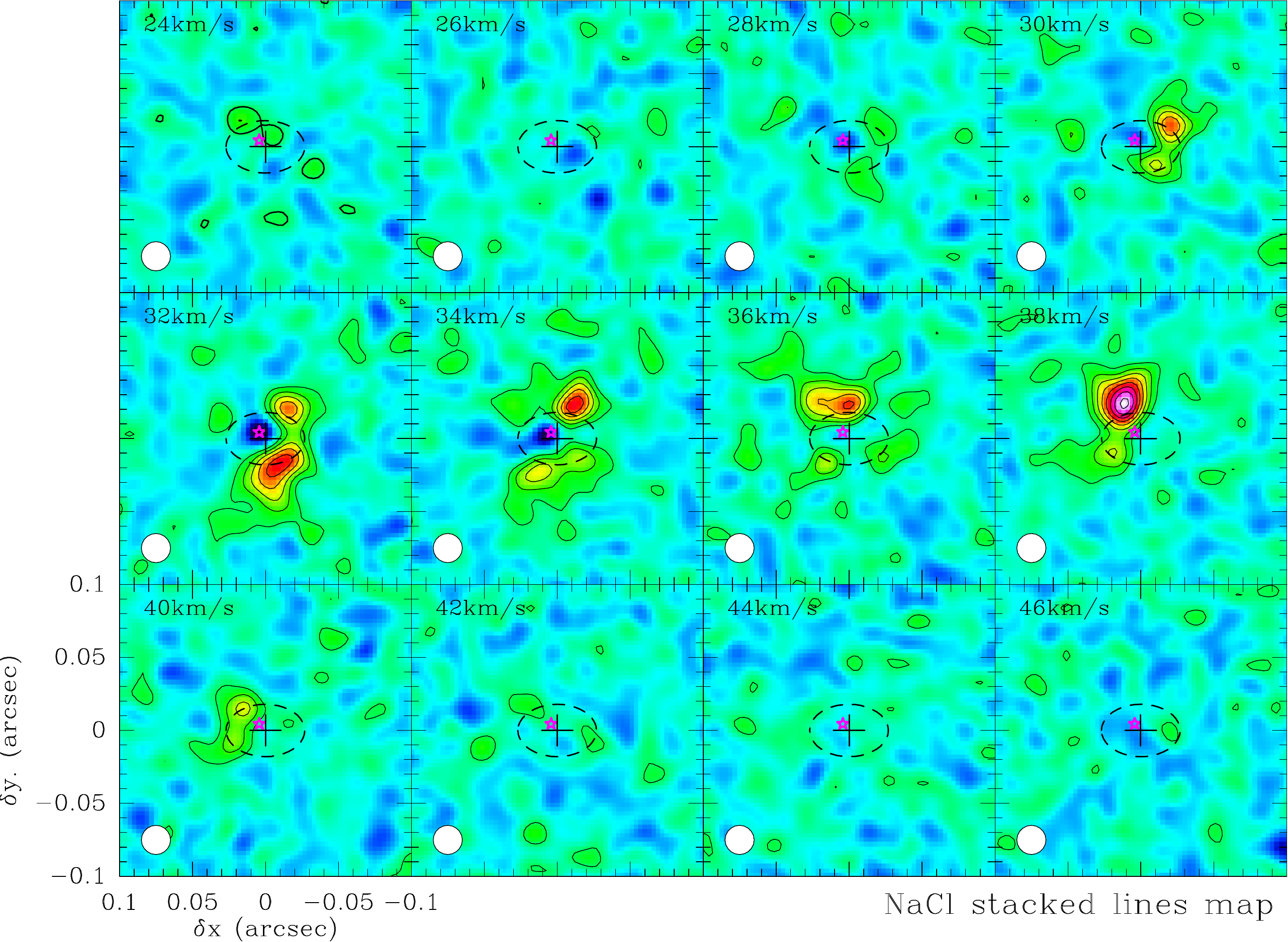}
     \includegraphics[width=0.33\hsize]{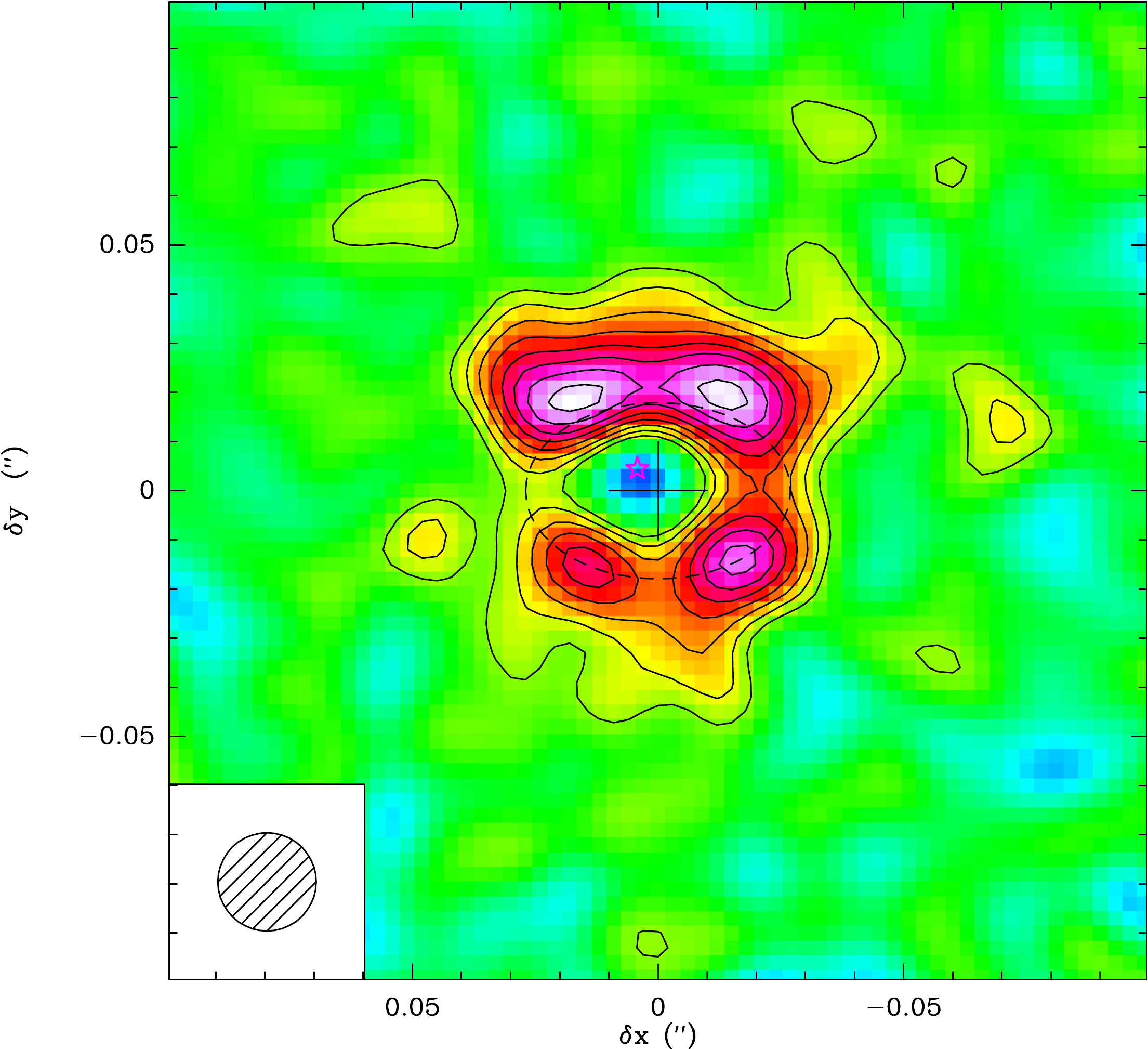}
     \includegraphics[width=0.33\hsize]{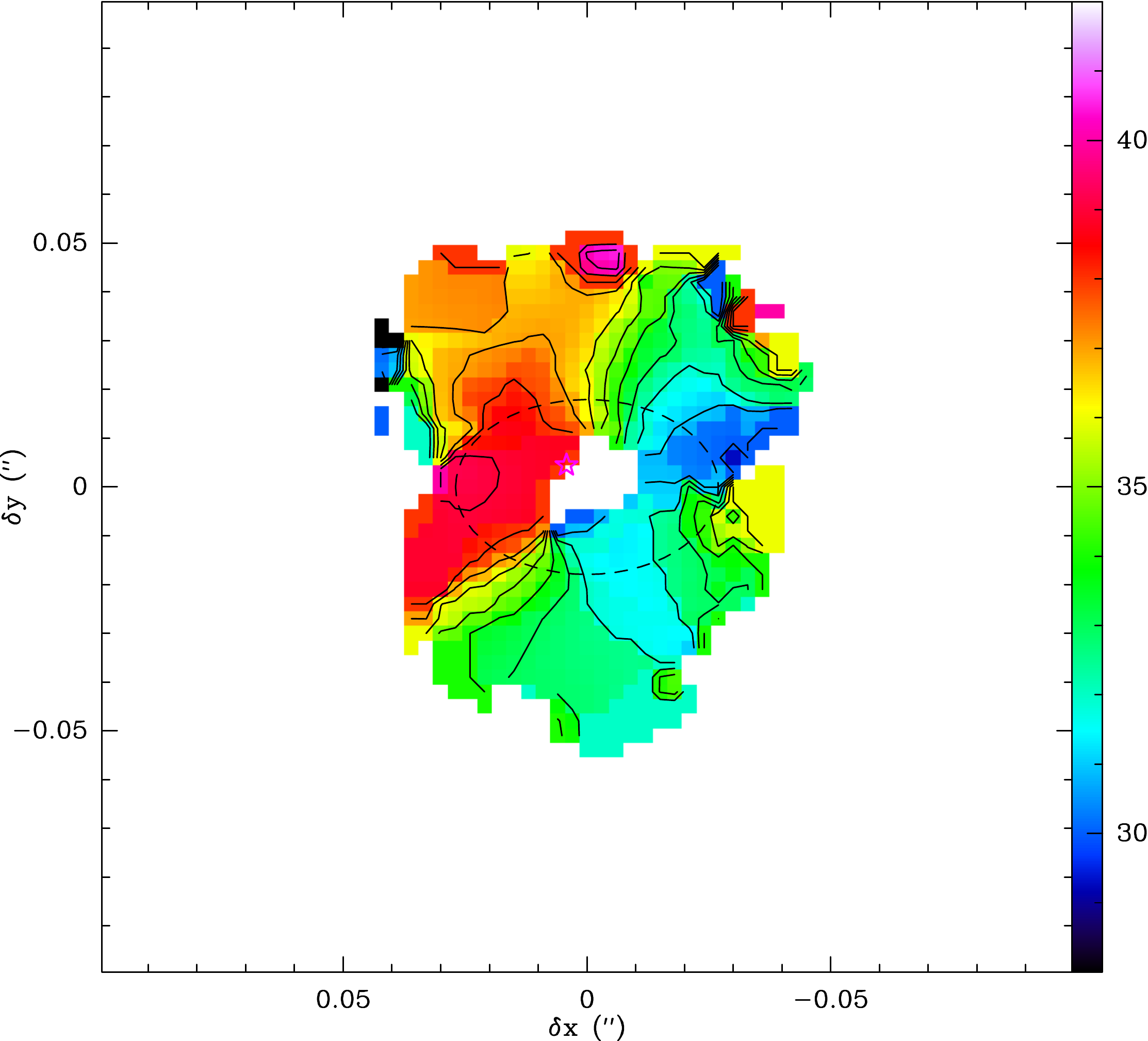}
     \includegraphics[width=0.30\hsize]{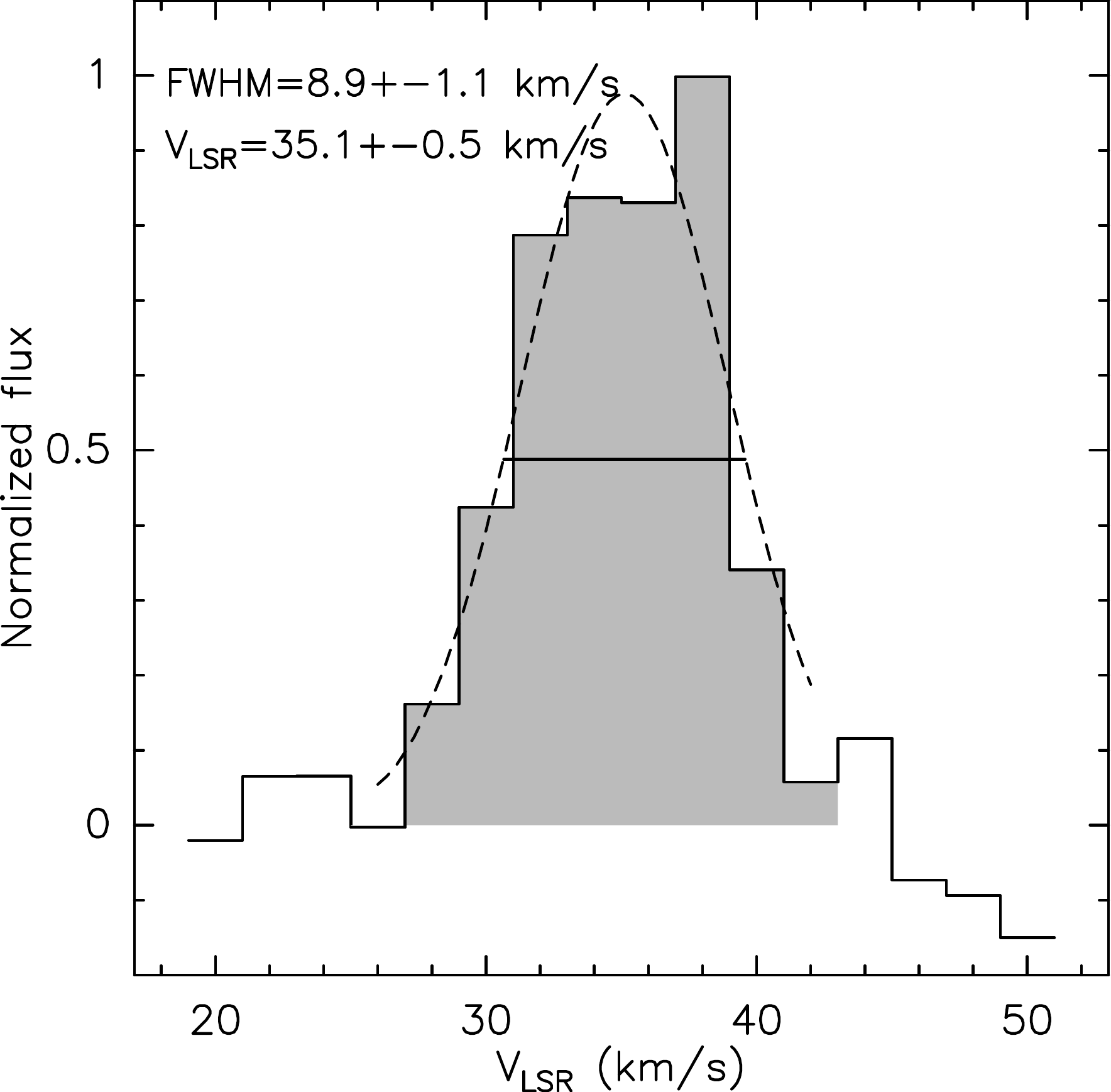}     
     \caption{ALMA data of NaCl after stacking together the
            individual NaCl transitions detected in this work --
            Table\,\ref{tab:mols} and Fig.\,\ref{f-nacl-app}. {\bf
              Top)}Velocity channel maps rotated by 25\degr\ clockwise
            so the symmetry axis of the disk is vertical; contours
            are 2$\sigma$, 4$\sigma$,... by 2$\sigma$
            ($\sigma$=0.225\,mJy/beam).  The clean beam
            (HPBW=0\farc02$\times$0\farc02) is plotted at the
            bottom-left corner of each panel.  The center of the dust
            disk (dashed ellipse) inferred from the continuum emission
            maps is marked with a cross (R.A.=07\h42\m16\fs91500 and
            Dec.=$-$14\degr42\arcmin50\farc0716, J20000) and has been
            adopted as the origin of positional offsets in these and
            all subsequent figures illustrating image data. The
            position of the central mass-losing star \qx, slightly
            offset from the map/disk center, is marked with a starlike
            symbol. {\bf Bottom-left)} zeroth moment map over the
            \vlsr=[24:46]\,\kms\ velocity range; contours are
            2$\sigma$, 3$\sigma$,... by 1$\sigma$
            ($\sigma$=1.7mJy/beam\kms). {\bf Bottom-center)} first
            moment map; contours go from \vlsr=28 to 43 by
            1\,\kms. The wedge indicates the
            color-\vlsr\ relationship. {\bf Bottom-right)} Integrated
            1d-spectrum normalized to the peak of the stacked line. A
            Gaussian fit to the line profile (dashed line) is shown
            together with the values derived for the line centroid and
            full width at half maximum.}
         \label{f-nacl}
   \end{figure*}
%

   \begin{figure*}[t]
     \centering
     \includegraphics[width=0.3\hsize]{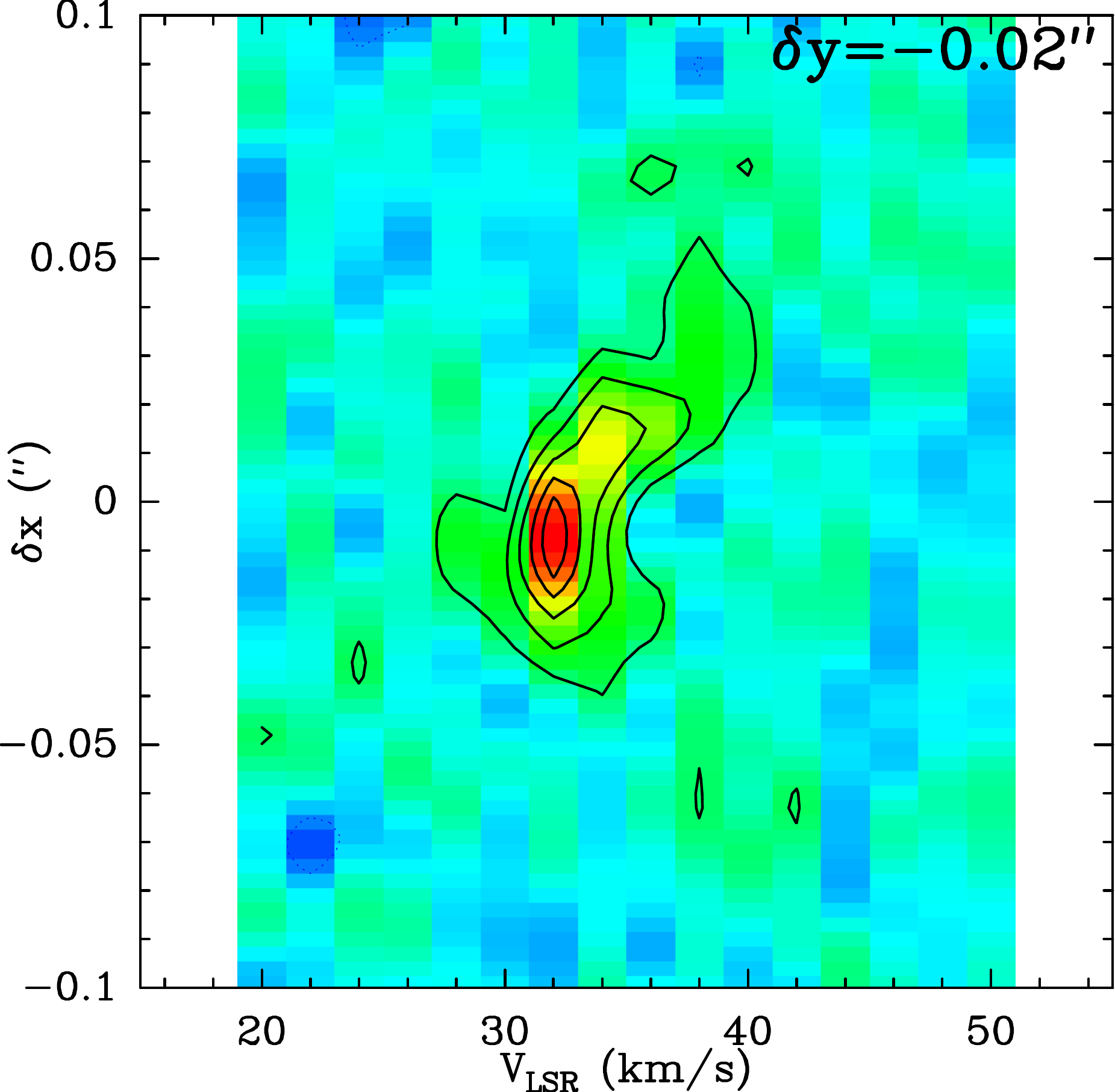}
     \includegraphics[width=0.3\hsize]{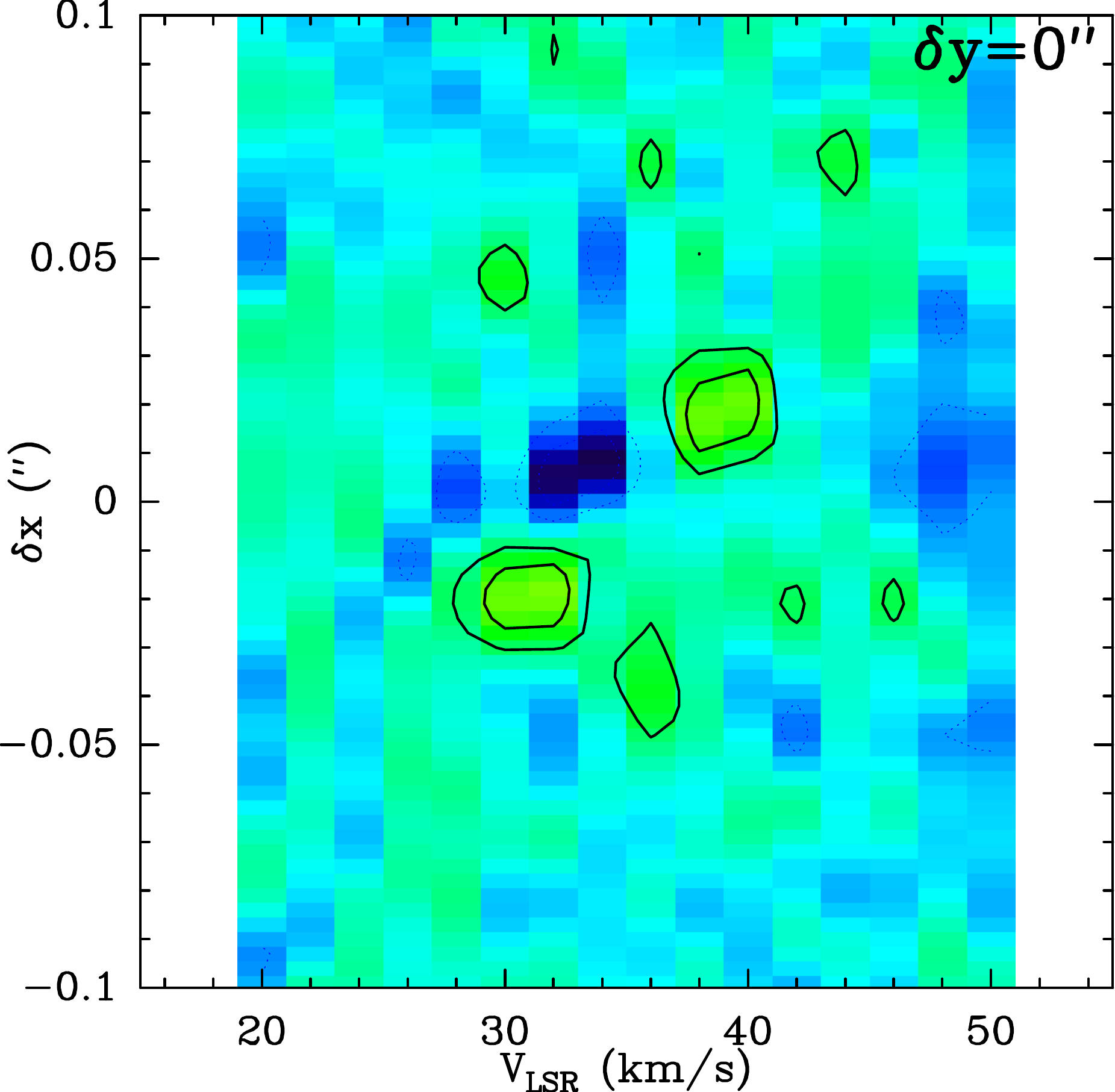}
     \includegraphics[width=0.3\hsize]{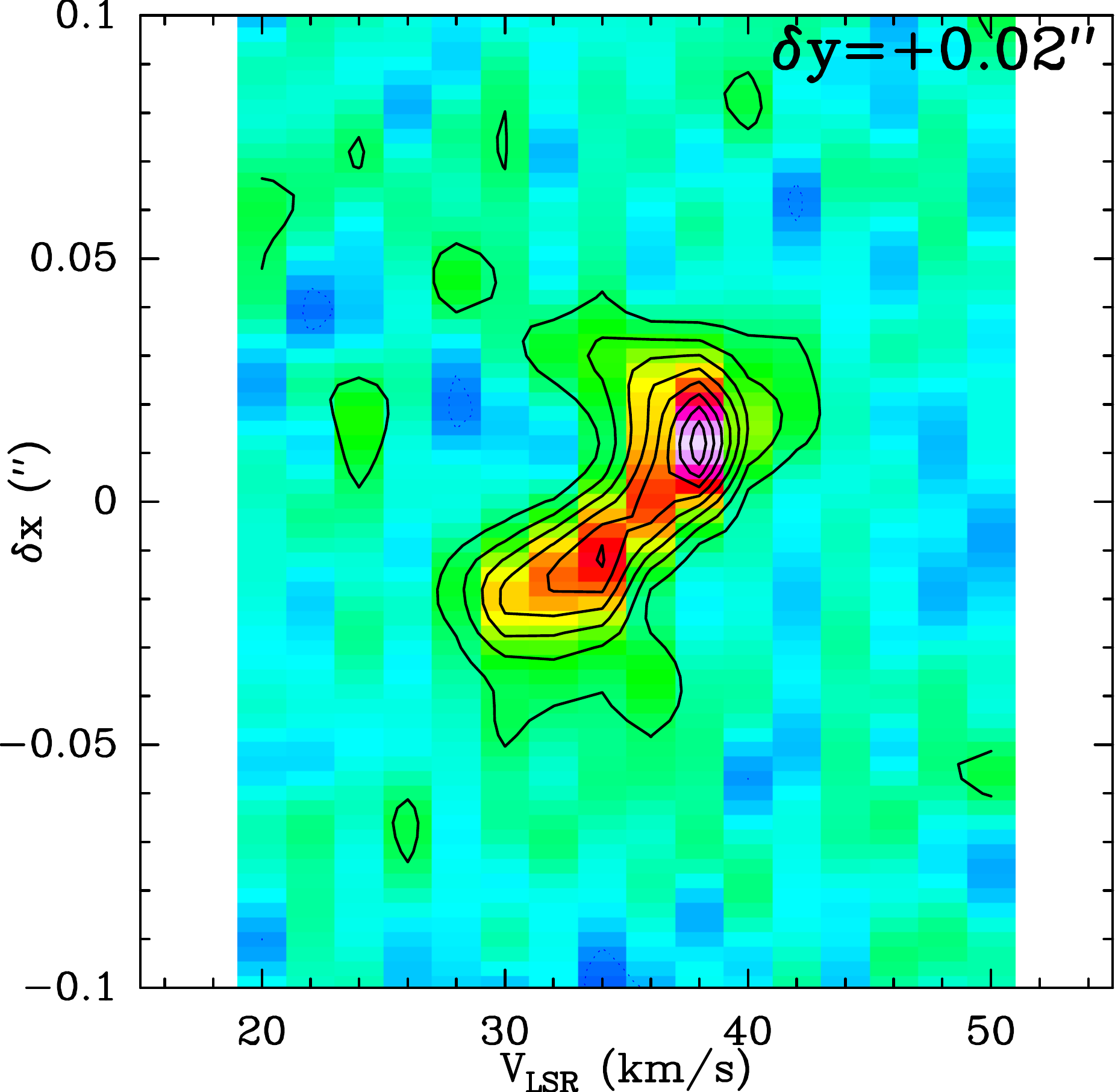} \\
     \includegraphics[width=0.3\hsize]{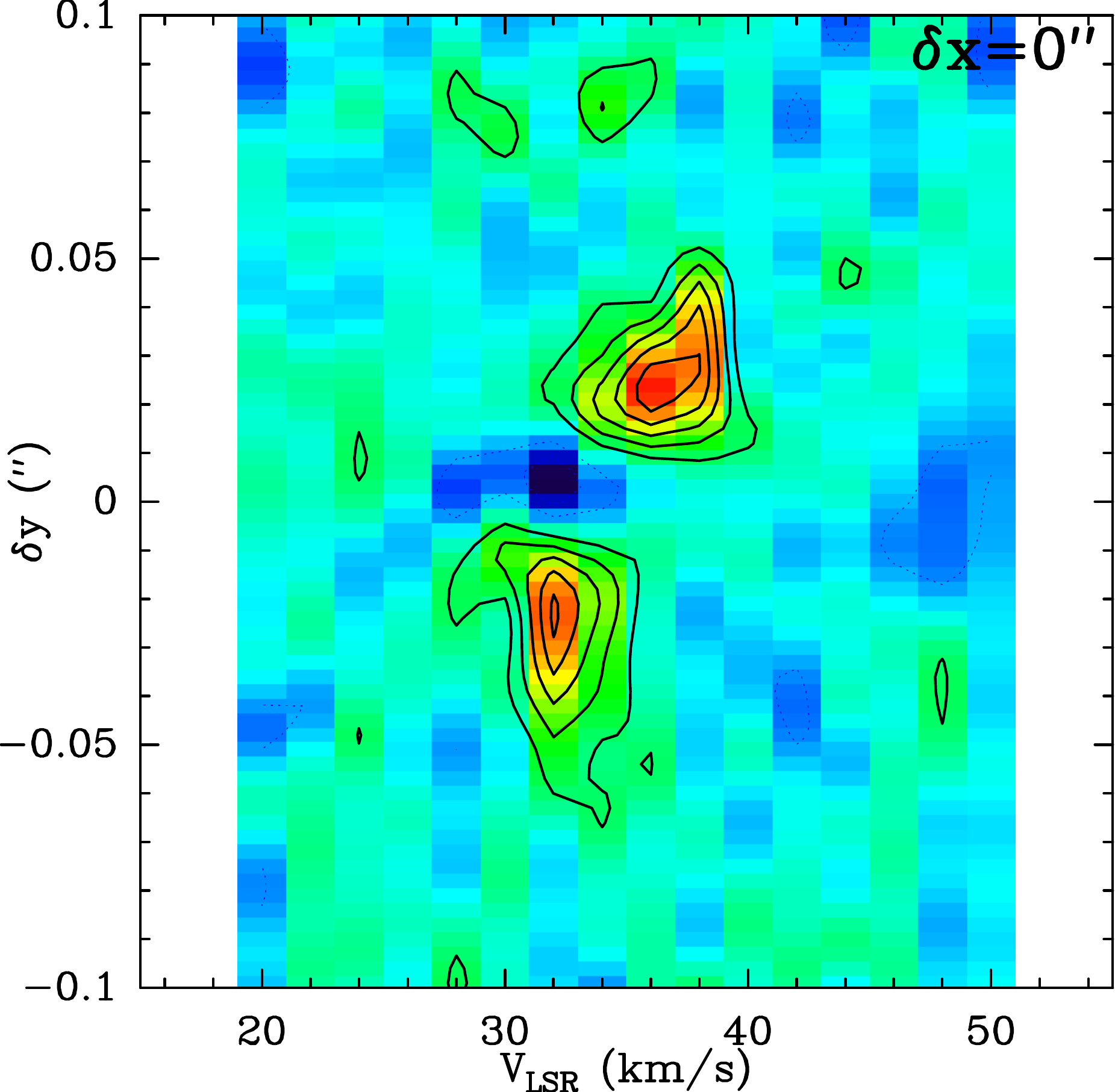}
     \caption{Position velocity (PV) cuts of the NaCl line-stacked cubes along  the direction of the equator (PA=115\degr) through different $\delta y$ offsets ($\delta y$=$-$0\farc02, 0\arcsec, and +0\farc02 --- top panels) and the nebula
     axis (PA=25\degr) through the nebula center ($\delta x$=0\arcsec -- bottom) . Contour levels as in Fig.\,\ref{f-nacl}.}
         \label{f-pvnacl}
   \end{figure*}
   \begin{figure*}[htpb]
     \centering
     \includegraphics[width=0.95\hsize]{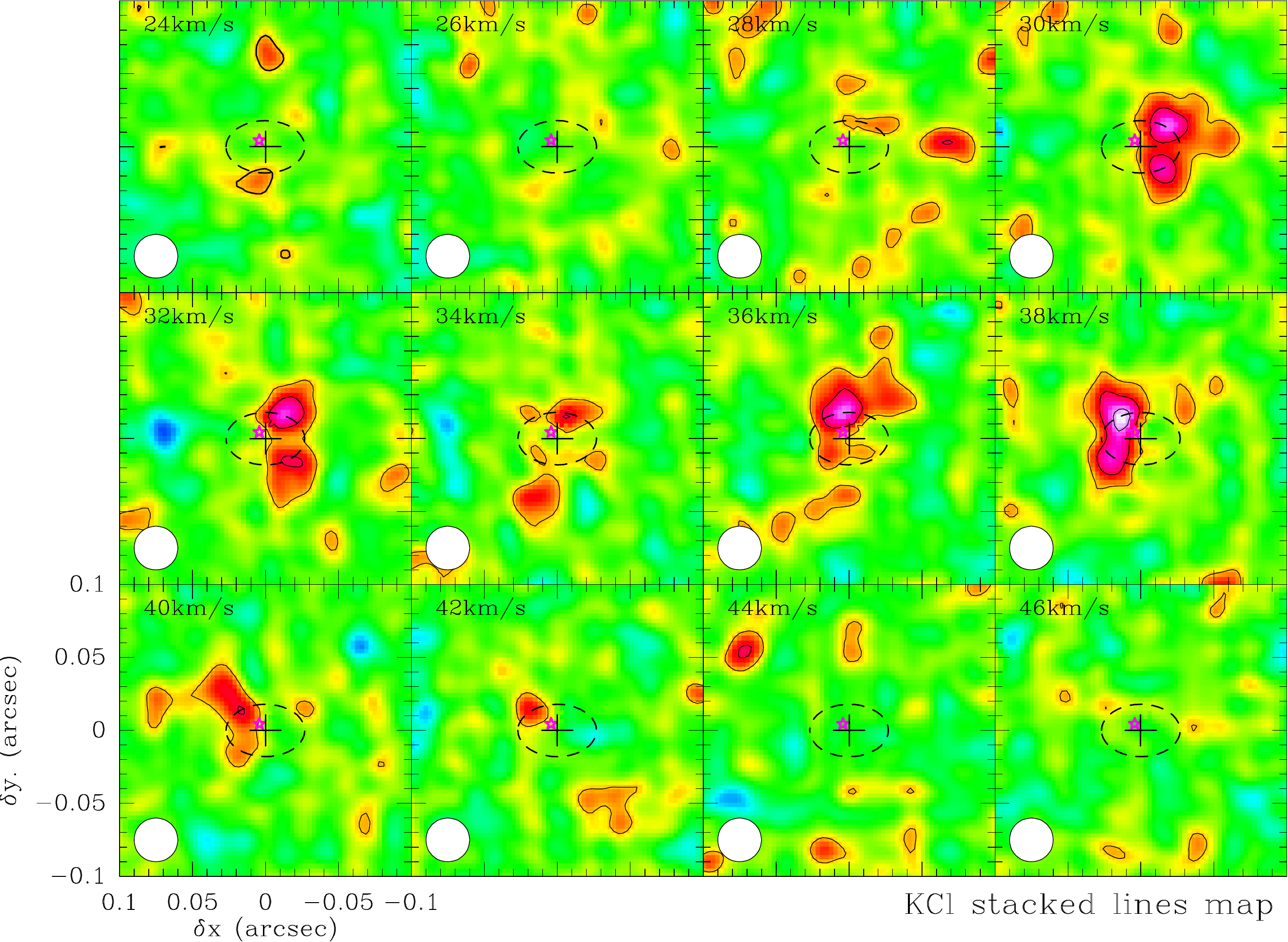}
     \includegraphics[width=0.33\hsize]{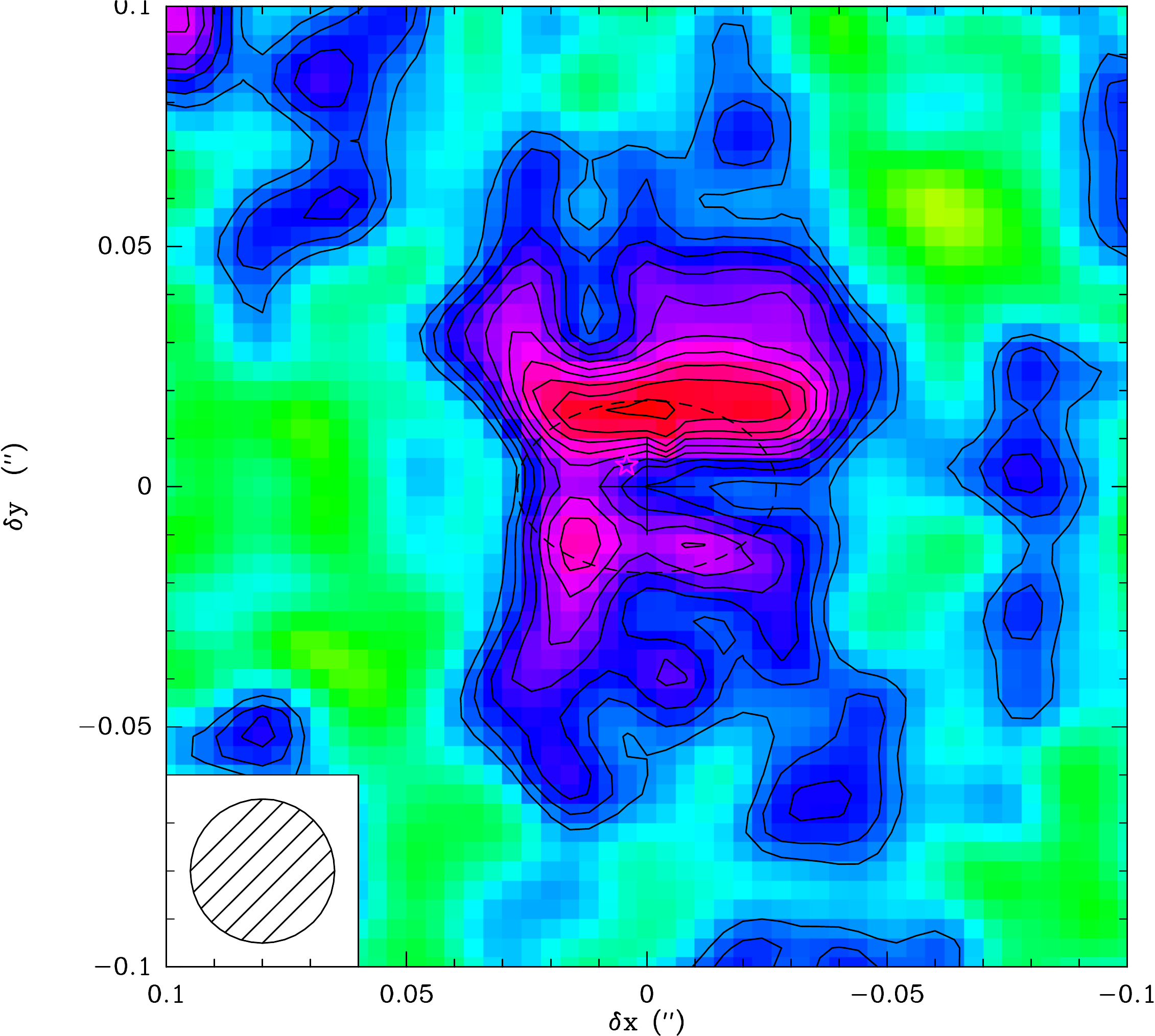}
     \includegraphics[width=0.34\hsize]{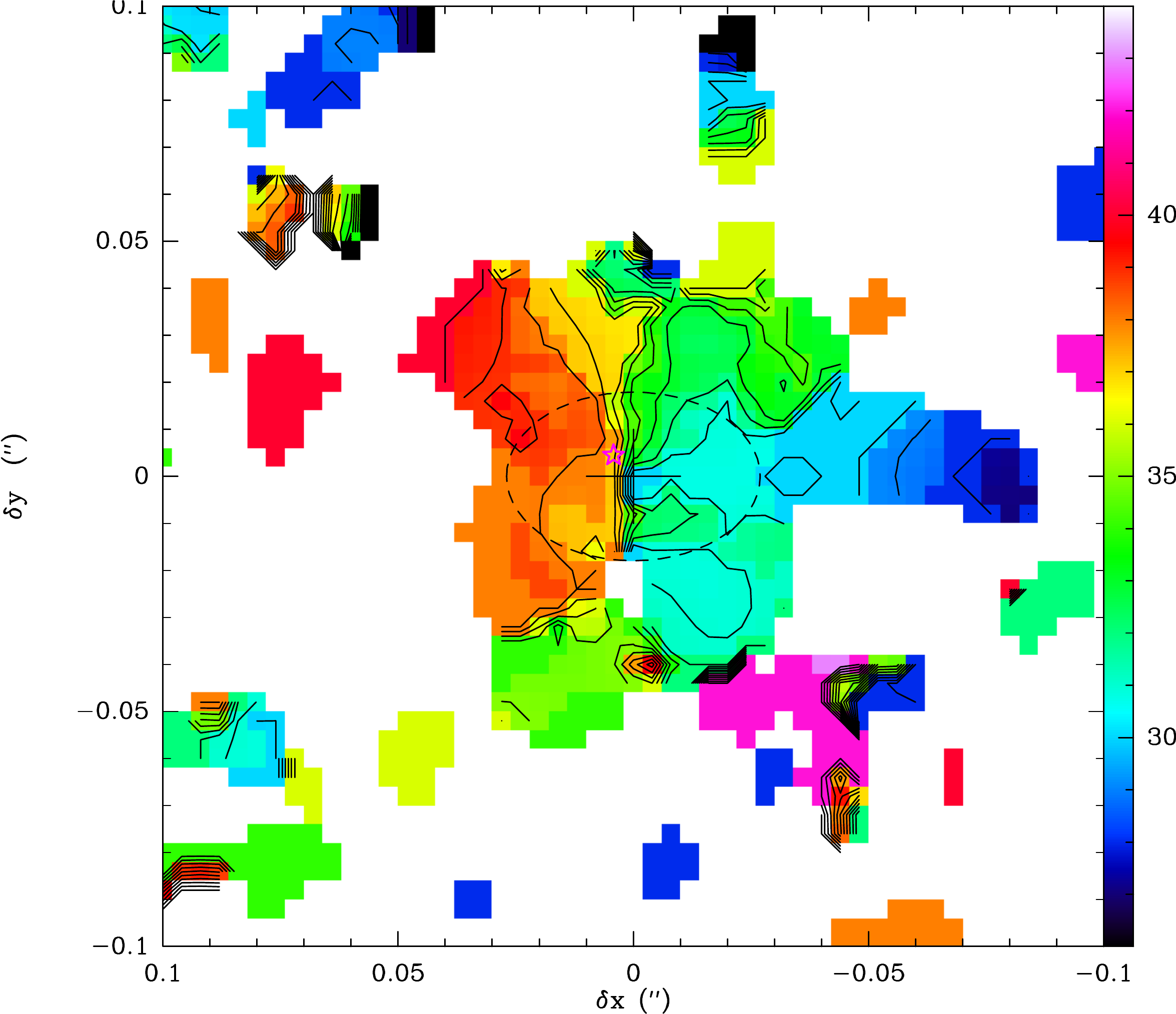}
     \includegraphics[width=0.30\hsize]{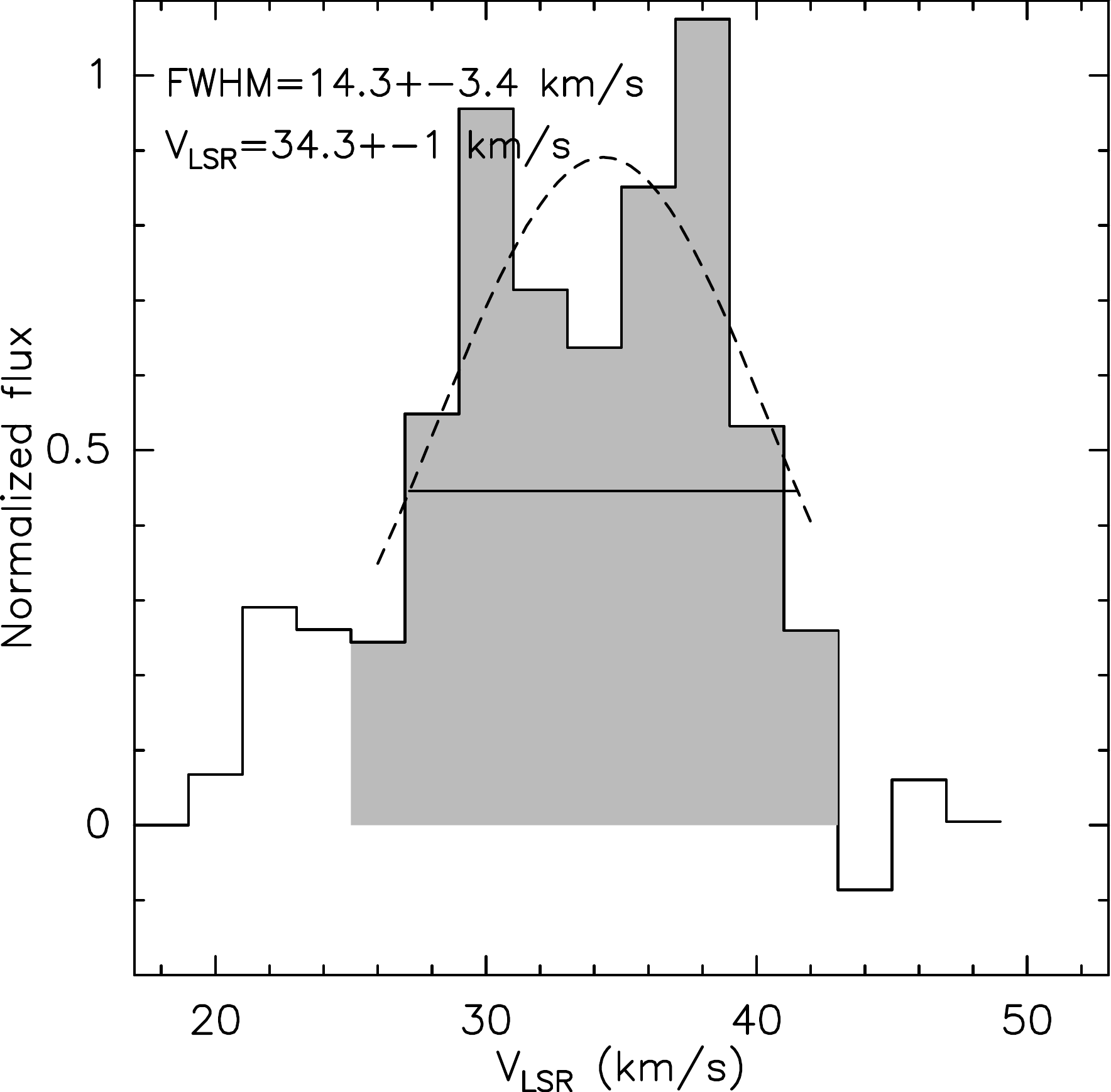}     
     \caption{As in Fig.\,\ref{f-nacl} but for KCl (line-stacked maps: individual transitions are shown in Table\,\ref{tab:mols} and Fig.\,\ref{f-kcl-app}).
       In this case, maps have been restored using natural weight and a clean beam with
       HPBW=0\farc03$\times$0\farc03.}
       \label{f-kcl}

   \end{figure*}
%
   \begin{figure*}[htpb]
     \centering
     \includegraphics[width=0.95\hsize]{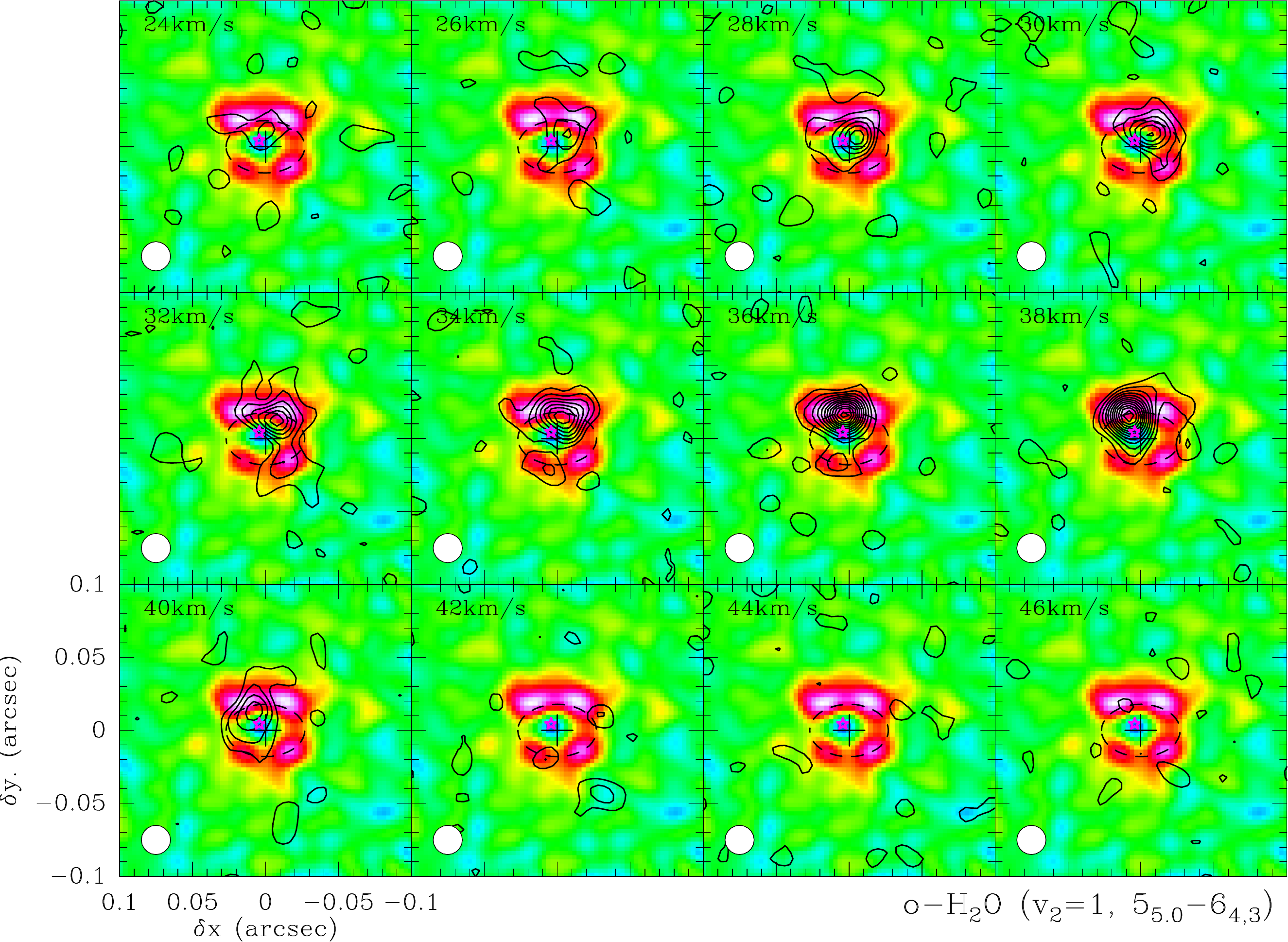}
     \includegraphics[width=0.33\hsize]{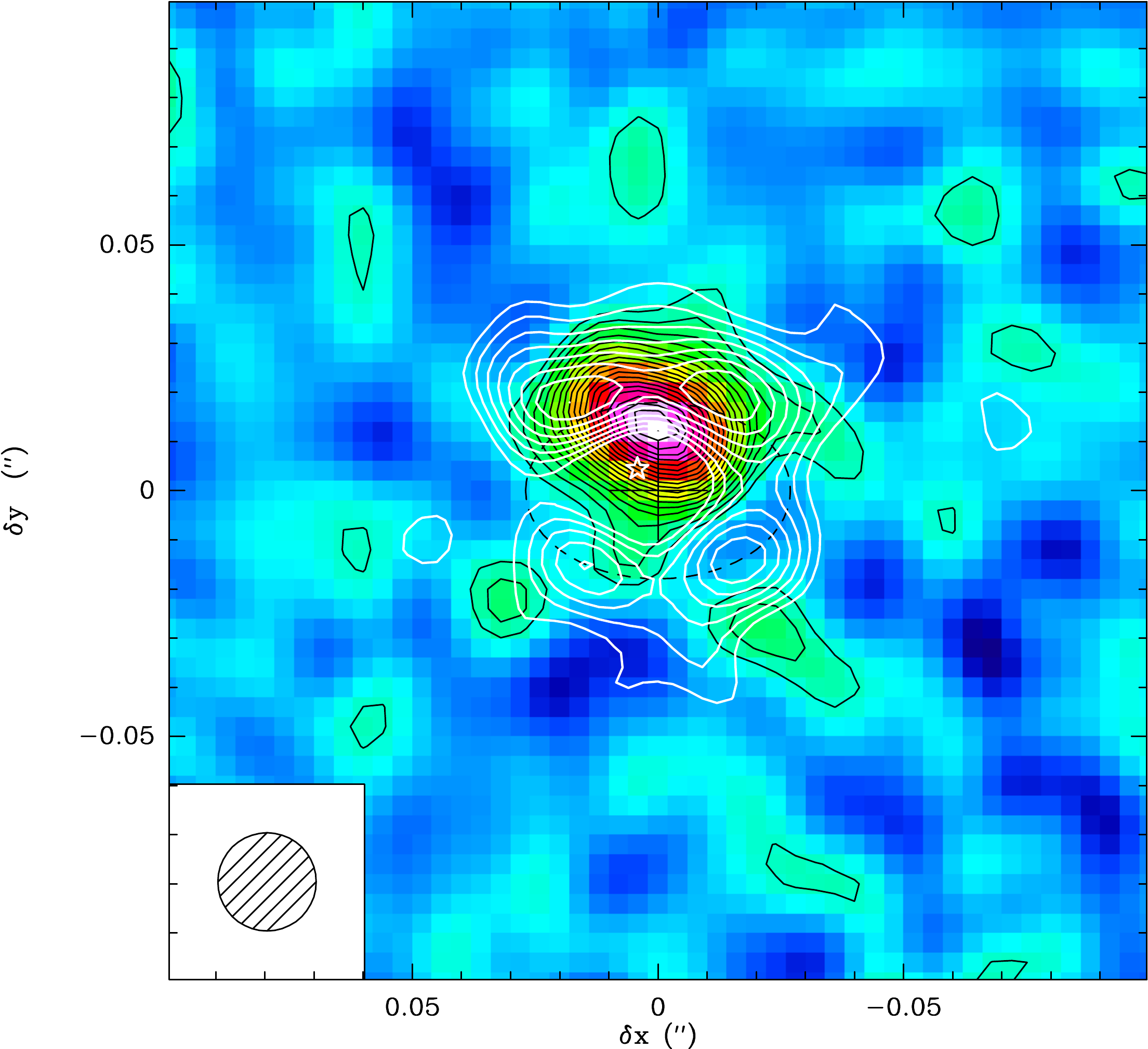}
      \includegraphics[width=0.33\hsize]{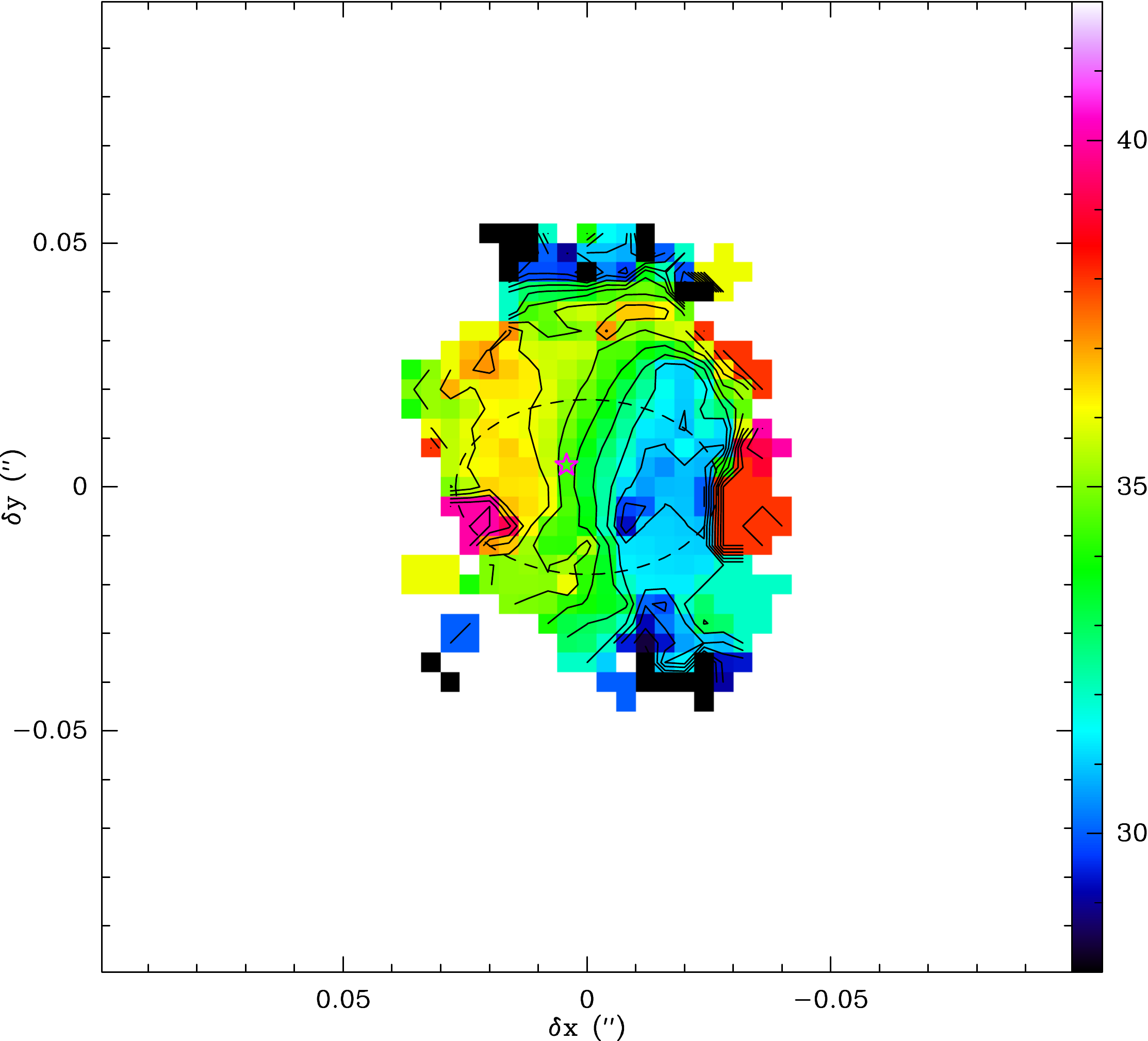}
           \includegraphics[width=0.31\hsize]{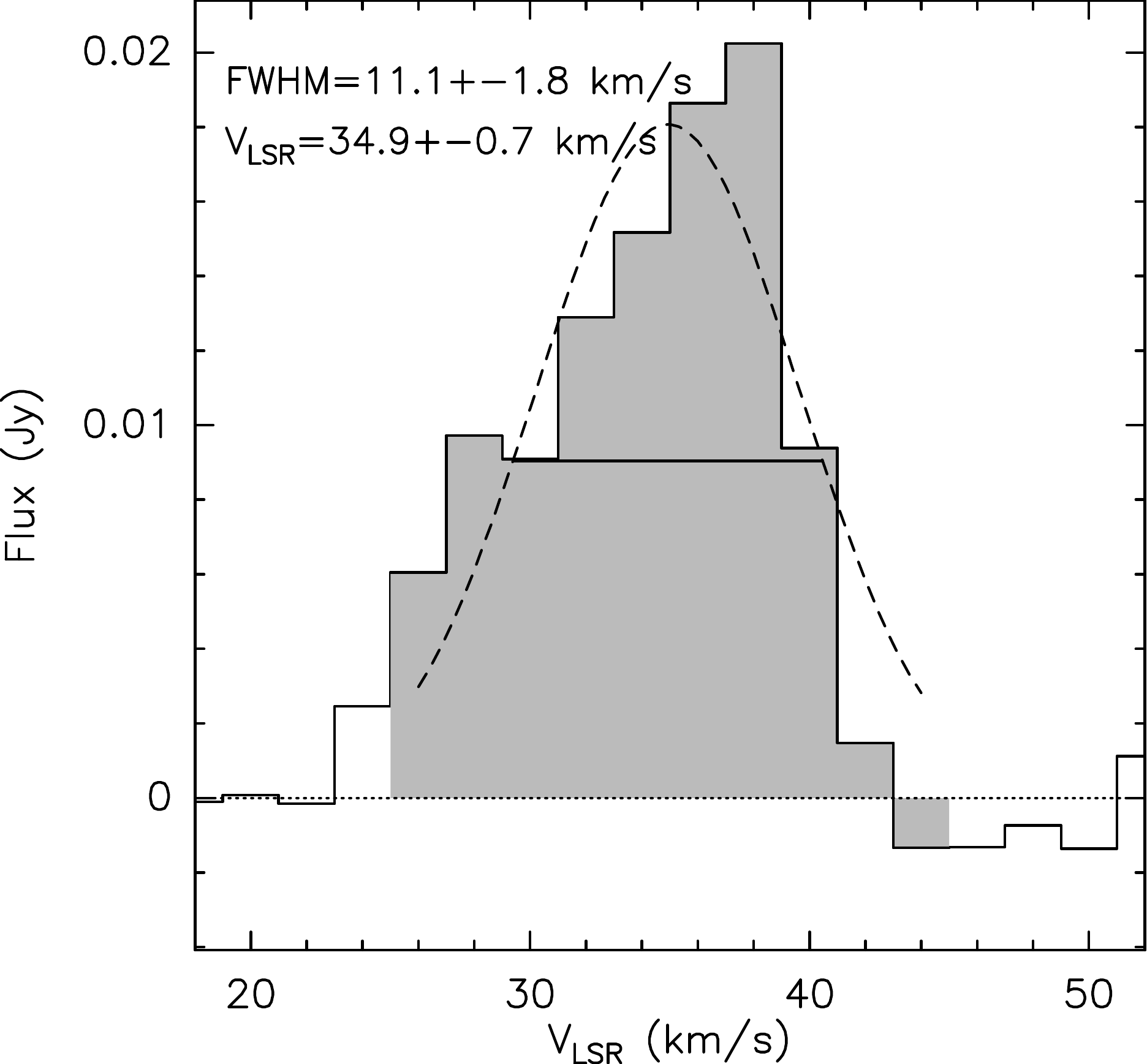}     
     \caption{As in Fig.\,\ref{f-nacl} but for the transition  \waterv.
       Level contours are 2$\sigma$, 4$\sigma$,... by 2$\sigma$ with
       $\sigma$=0.5\,mJy/beam for the velocity-channel maps and (top)
       $\sigma$=8\,mJy/beam\kms\ in the first moment map
       (bottom-left). The NaCl line-stacked integrated intensity map
       are overplotted (white contours) on the \water\ first moment
       map.}
      \label{f-h2o}
   \end{figure*}
%

   The first detection of \naclts\ in \ohs, with emission arising
   entirely from \cs, was reported by \san.  In this work, we
   present $\sim$20\,mas-resolution emission maps of NaCl, which
   represent the first detection of the main isotopologue of sodium
   chloride ({\it salt}) in this object, and that enable us to
   spatially and spectrally resolve the close surroundings of \qx.

We have detected emission from seven NaCl and one \naclts\ rotational
transitions in different $\upsilonup$=0, 1, 2, and, tentatively 3,
vibrational levels (Table\,\ref{tab:mols}). Velocity-channel maps of
the \naclvd\ transition, with the highest S/N amongst the observed
NaCl lines, are shown in Fig.\,\ref{f-nacl51}.  Integrated intensity
(zeroth moment) maps and 1d spectra (integrated over the emitting
area) of all the lines detected are presented in
Fig.\,\ref{f-nacl-app}.  Although the emission is generally weak,
specially for vibrationally excited lines, the surface brightness
distribution of all transitions consistently appears as a hollow,
squared region of dimensions $\sim$0\farc08$\times$0\farc08 (at the
rms level) suggestive of a cylindrical shell-like structure
surrounding or {\it coating} the dust disk traced by the continuum
(illustrated by the dashed ellipse in the figures).  All transitions
show a clumpy/non-uniform surface brightness distribution with an
overall depression
near the equatorial plane, 
a dip at the center,
and with the northern side being in general brighter than the
southern.

The NaCl emission spreads over a modest full velocity range of
$\sim$12\,\kms. The centroid of the NaCl lines points to a systemic
velocity near 35\,\kms\ (LSR), as already deduced by \san\ 
based on a number of molecular lines with emission arising from
\cs\ (including an additional \naclts\ transition) and from the
compact SiO-outflow that emerges from \cs. The maps of all the
individual NaCl transitions observed consistently show blue- (red-)
shifted emission from the west (east) side of the compact, squarish emitting region (see e.g.\ Fig.\,\ref{f-nacl51}).

The data of the individual NaCl transitions have been combined to
produce a unique NaCl line-stacked cube with improved S/N -- see
Fig.\,\ref{f-nacl}.  Line stacking has been done in the $uv$ plane
(using the GILDAS task {\tt uv\_merge}); subsequent image
reconstruction and cleaning has been done using standard
GILDAS/MAPPING tasks as for the rest of the transitions reported (see
\S\,\ref{obs}).  The velocity-channel maps of the NaCl stacked lines
corroborate the main spatial and kinematic characteristics of the
emitting region guessed from the individual transitions. Our NaCl
stacked-line maps clearly point to equatorial rotation 
in the central, inner regions of \ohs: the
emission from the east (west) side of the cylindrical NaCl-emitting
volume is systematically red- (blue-) shifted, which is indicative of
receding (approaching) gas. This velocity gradient along the
equatorial direction, i.e.\ orthogonal to that resultant from the
dominant expansive kinematics of the bipolar lobes of \ohs, is a clear
signature of equatorial rotation -- this is can be also appreciated in
the first moment maps shown in Fig.\,\ref{f-nacl} (bottom-center).
The west-to-east sense of the rotation derived from these maps
coincides with that of the inner ($r$$\sim$3\rs) SiO-maser torus
\citep{san02}.

The radius and the height (above or below the equatorial plane) of the hollow, shell-like
cylindrical structure where the NaCl-emission arises are similar, 
which is the reason for the squarish appearance of the NaCl surface brightness distribution. The
walls of the cylinder are unresolved and, thus,
the wall-thickness is expected to be a small fraction of the radius.  The centroid of both
the velocity field and the overall surface brightness distribution of
NaCl roughly coincides with the center of the dust disk probed by the
mm-continuum, strongly suggesting that the rotating
structure traced by NaCl is also circumbinary. This is also the most
likely scenario based on previous observations, since all rotating
structures spatially resolved to date around binary post-AGB stars,
a.k.a.\ dpAGBs, are circumbinary (\S\,\ref{intro}). The relative
distribution of the emission from NaCl and the dust disk deduced from our ALMA maps
suggest
indeed that the salt emission arises in the surface layers of the
disk, both then being part of the same equatorial structure rotating around
the binary system at the center of \ohs.

The velocity-channel maps, show weak absorption below the continuum
level near the nebula center, at velocities \vlsr$\sim$32\,\kms; the
absorption is appreciated in some of the individual NaCl transitions,
particularly the $\upsilonup$=0 lines. This indicates that the line
excitation temperature is smaller than that of the background source,
the dust continuum, at that position (Fig.\,\ref{f-nacl51}). As
discussed in \S\,\ref{s-rd}, the presence of absorption enables
constraining the rotational temperature and opacity of the NaCl lines observed.

In Fig.\,\ref{f-pvnacl}, we show position-velocity (PV) cuts along the
direction of the nebula equator (top) and along the axis (bottom) of
the NaCl-stacked maps.  The peak-to-peak velocity separation measured
in the equatorial PV cut through the center ($\delta y$=0\arcsec) is
$\sim$8\,\kms, implying a projected rotation velocity of about
\vrot$\sim$4\,\kms. The relative separation between the red- and
blue-peak emitting areas along the equator is $\delta x$$\sim$0\farc04
(60\,au).  The projected rotation velocity deduced at offsets $\delta
y$$\pm$0\farc02 above and below the equatorial plane, i.e.\,
coincident with the two brightest north and south regions, is slightly smaller than (but comparable to) that measured at $\delta y$=0\arcsec.

In addition to rotating, the gas in the NaCl-emitting structure is
expanding. This is best seen in the PV cuts along the nebula axis
through the center (Fig.\,\ref{f-pvnacl}-bottom panel) where the
emission from the north and south surfaces of the NaCl cylinder is
observed to be red- and blue-shifted, respectively. This axial
velocity gradient is consistent with an inclined cylindrical structure
in equatorial expansion with its north side pointing toward the
observer, similarly to the orientation of the small- and large-scale bipolar outflows of the nebula.  The projected
expansion velocity deduced from the peak-to-peak velocity separation
in the axial PV cut is very low, $\sim$3\,\kms. This value is
a factor $\sim$5-10 smaller than the wind terminal
velocity expected for a high mass-loss
(\mloss$\approx$10$^{-4}$-10$^{-5}$\,\my) AGB star like \qx, specially
considering that the NaCl emission traces the
surface layers of the continuum disk, i.e., is present beyond the 
regions where dust has already been formed massively and, thus, where
the stellar wind should have been strongly accelerated approaching its terminal velocity. 
We discuss this further in Sect.\,\ref{dis-nacl}.




We have found two other molecular species that, apart from NaCl, 
selectively trace the rotating (and expanding) equatorial structure at
the core of \ohs, namely, potassium chloride (KCl, i.e.\,another salt)
and water (\water). The detected transitions are listed in
Table\,\ref{tab:mols}.

The emission from KCl is extremely weak, specially that of the $\upsilonup$=1
line (Fig.\,\ref{f-kcl-app}). As for NaCl, the $uv$ data of the individual
KCl lines have been combined to obtain maps with higher S/N. The KCl
line-stacked cubes (Fig.\,\ref{f-kcl}) confirm a very similar 
spatial distribution and kinematics for the two salts, revealing themselves as optimal probes of
the long-sought rotating disks in pPN candidates with massive bipolar outflows like \ohs.

Our ALMA maps of the \waterv\ transition (Fig.\,\ref{f-h2o}) 
reveal as well the presence of water in the surface layers of the rotating
continuum disk traced by the salts, with the water emission being
slightly more compact, i.e.\, probing regions slightly closer (in
radius and height) to the center.  The brightness contrast between the
north and south surfaces of the water-emitting volume is much higher
than for the salts, with the emission from the south being barely
above our detection limit. The water emission
spreads over a full velocity range slightly larger than the salts, with faint water
emission detected in channels near \vlsr=24-26\,\kms\ close to the center, where the weaker NaCl and
KCl transitions are not detected.
This possibly denotes slightly larger rotation velocities at smaller distances to the center as expected for Keplerian or sub-Keplerian rotation.   

\subsection{The compact bipolar SiO/SiS-outflow}
\label{res-sissio}

From these observations, we find that the compact ($\sim$1\arcsec$\times$4\arcsec)
bipolar outflow discovered in \san\ is traced by several
rotational transitions in the $\upsilonup$=0 and $\upsilonup$=1
vibrational states of SiO and SiS (including some isotopologues). We refer to this component as the
SiO/SiS-outflow (hereafter, \sso).
In this work, we concentrate exclusively on two
transitions: the \sisdo\ line (\S\,\ref{res-sisv0}), which best delineates the morphology and kinematics of the dense walls
of the bipolar \sso, and the vibrationally excited \siov\ line (\S\,\ref{res-siov1}), which
traces the innermost layers of the \sso\ (i.e.\ closer to the center)
where the gas is exposed to a more intense infrared stellar radiation field (probably largely responsible for SiO $v$=1 level pumping) and, also, is
presumably denser and hotter. The line emission maps of
additional Si-bearing molecules, particularly SiO\,$\upsilonup$=0, with
more intense and opaque lines than SiS and that mainly trace the lobes
interior, will be the focus of a dedicated publication in the future.


\subsection{SiS\,$\upsilonup$=0}
\label{res-sisv0}

\begin{figure*}[htbp]
     \centering
     \includegraphics[width=0.95\hsize]{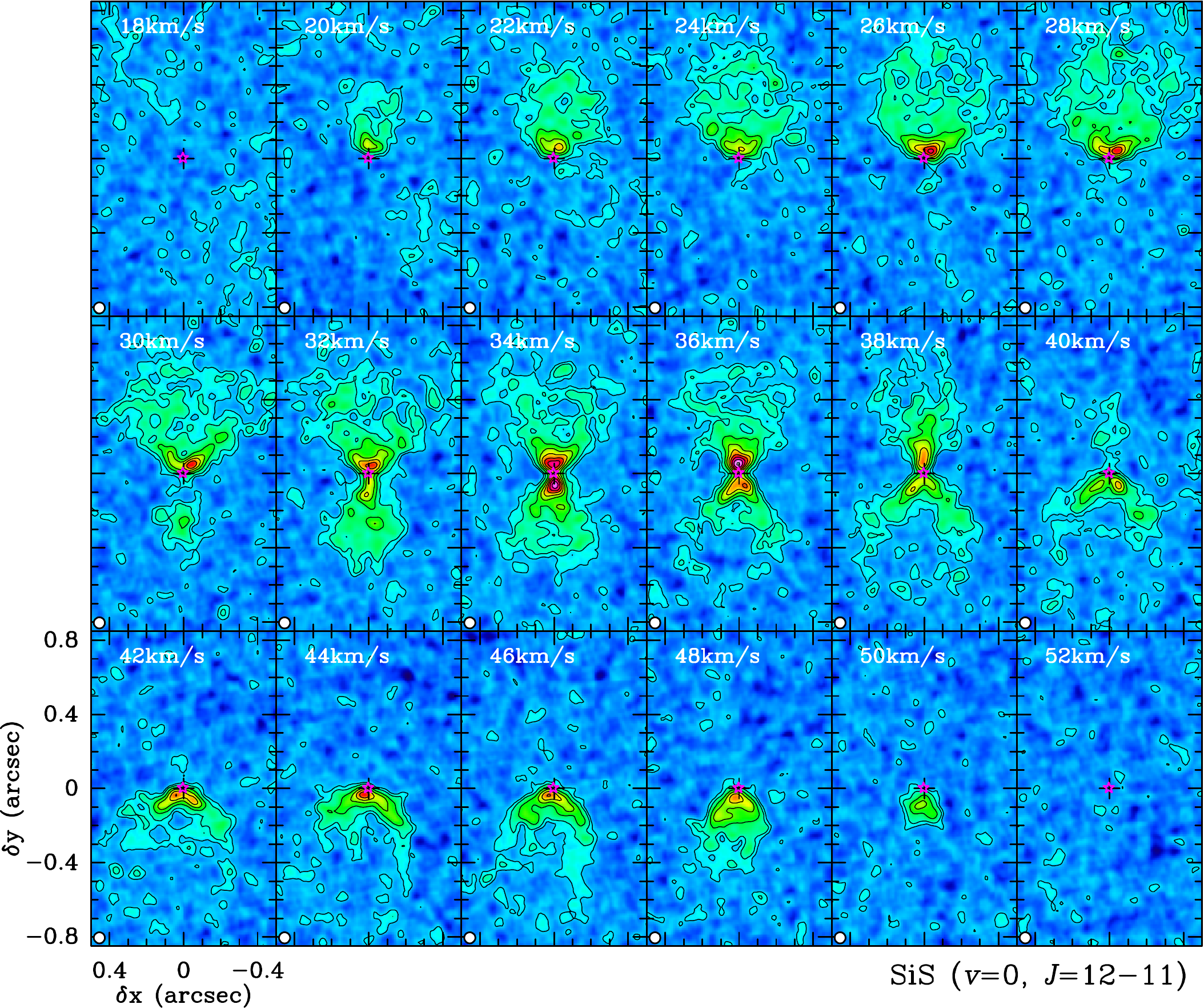} \\ \vspace{2.5mm}
     \includegraphics*[bb= 0 0 345 555, width=0.235\hsize]{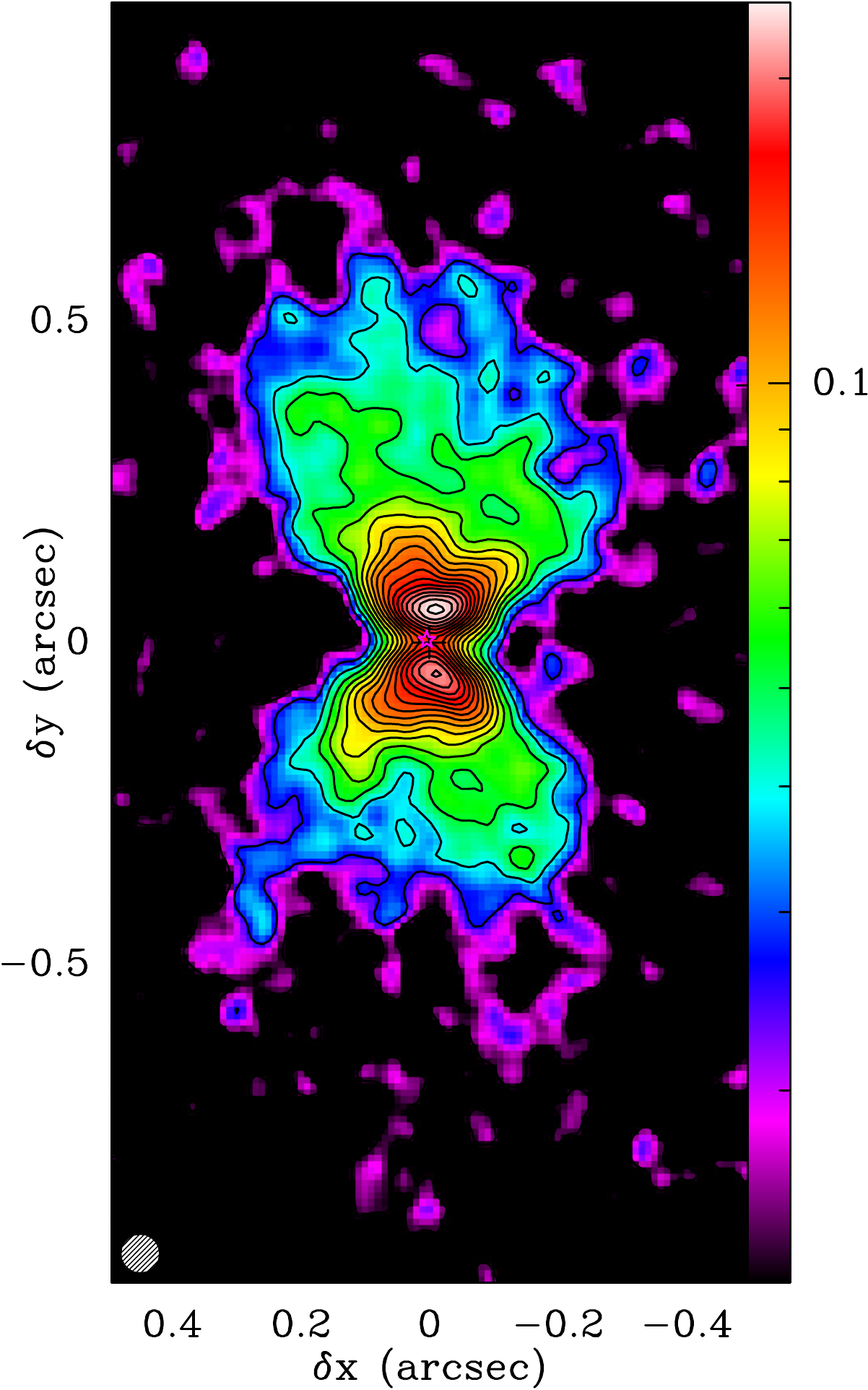}
     \includegraphics*[bb= 0 0 345 555, width=0.235\hsize]{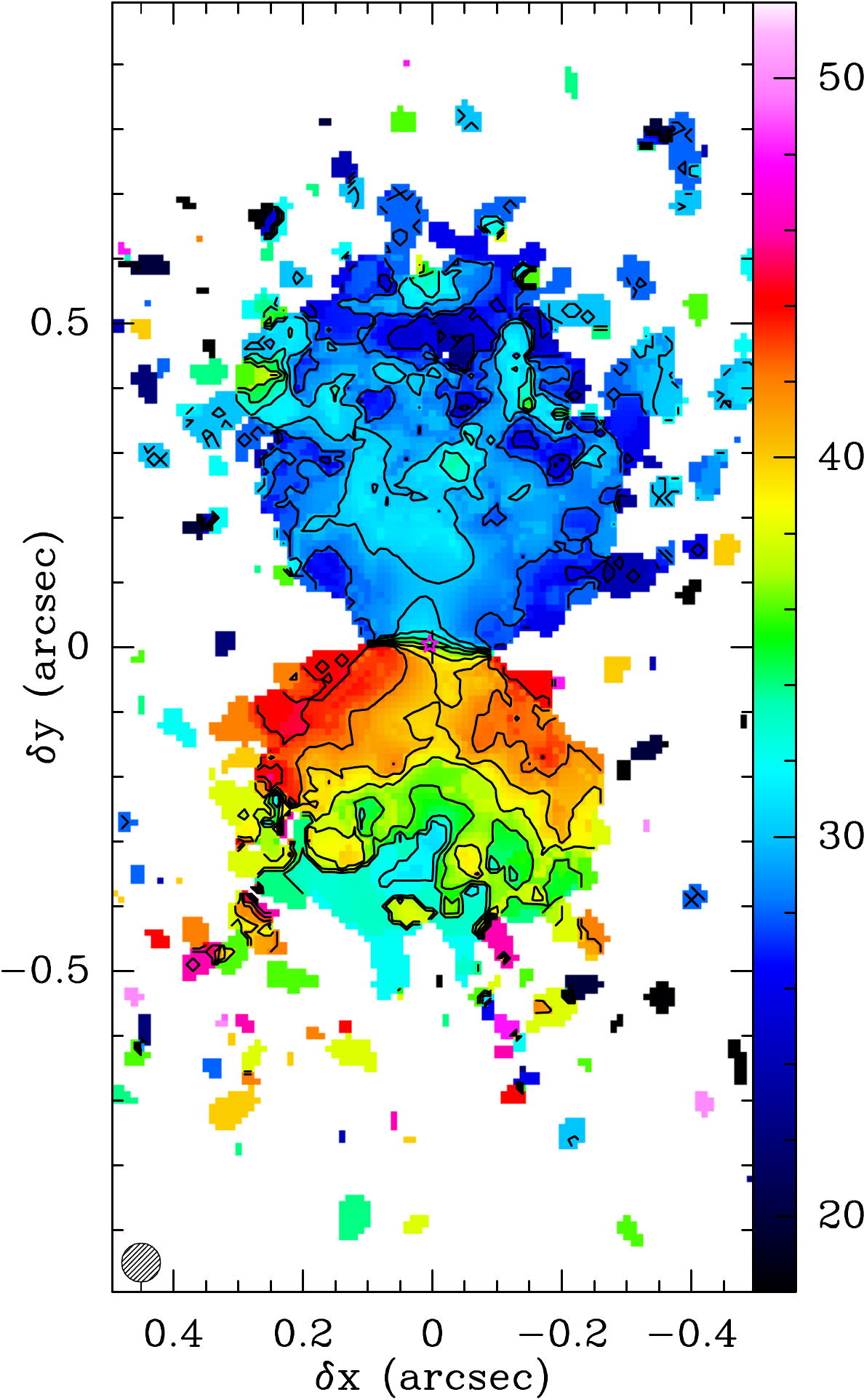}
     \includegraphics[width=0.38\hsize]{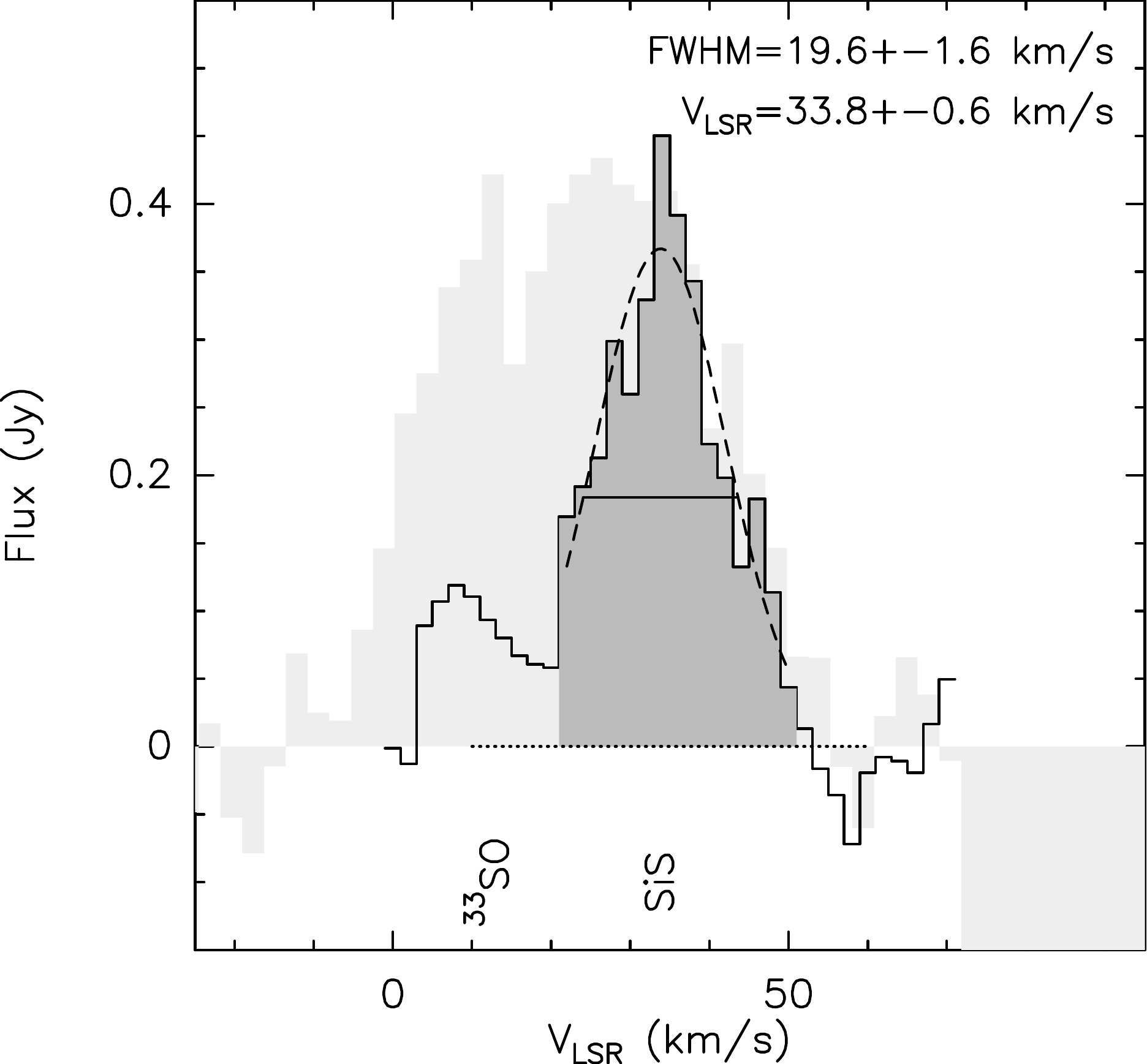} 
     \caption{SiS\,$\upsilonup$=0\,($J$=12-11) ALMA data. Natural
       weighting and tapering have been used to restore the emission
       maps with a half-power clean beam width of
       HPBW=0\farc06$\times$0\farc06. {\bf Top)} Velocity-channel maps
       (contours: 2$\sigma$, 4$\sigma$,... by 4$\sigma$;
       $\sigma$=0.57\,mJy/beam). {\bf Bottom-left)} Integrated
       intensity map over the velocity range \vlsr=[18:52]\,\kms. {\bf
         Bottom-center)} First moment map; contours go from \vlsr=18
       to 52 by 2\,\kms. {\bf Bottom-right)} Integrated 1d-spectrum
       obtained with ALMA (solid line) and with the \iram\ telescope
       (light grey area, Velilla-Prieto et al.\,in prep). The ALMA
       spectrum has been obtained integrating the line surface
       brightness over an area comparable to the beam of the
       \iram\ telescope at this frequency (HPBW$\sim$11\arcsec).  In
       contrast to SiS, the adjacent $^{33}$SO emission is strongly
       filtered out by the ALMA interferometer.  A Gaussian fit to the
       SiS profile (over the dark grey area) is represented by the
       dashed line; the FWHM and \vlsr\ of the centroid from the fit
       are indicated in the top-right corner of the box.}
         \label{f-sisv0}
   \end{figure*}

\begin{figure*}[htbp]
     \centering
     \includegraphics[width=0.99\hsize]{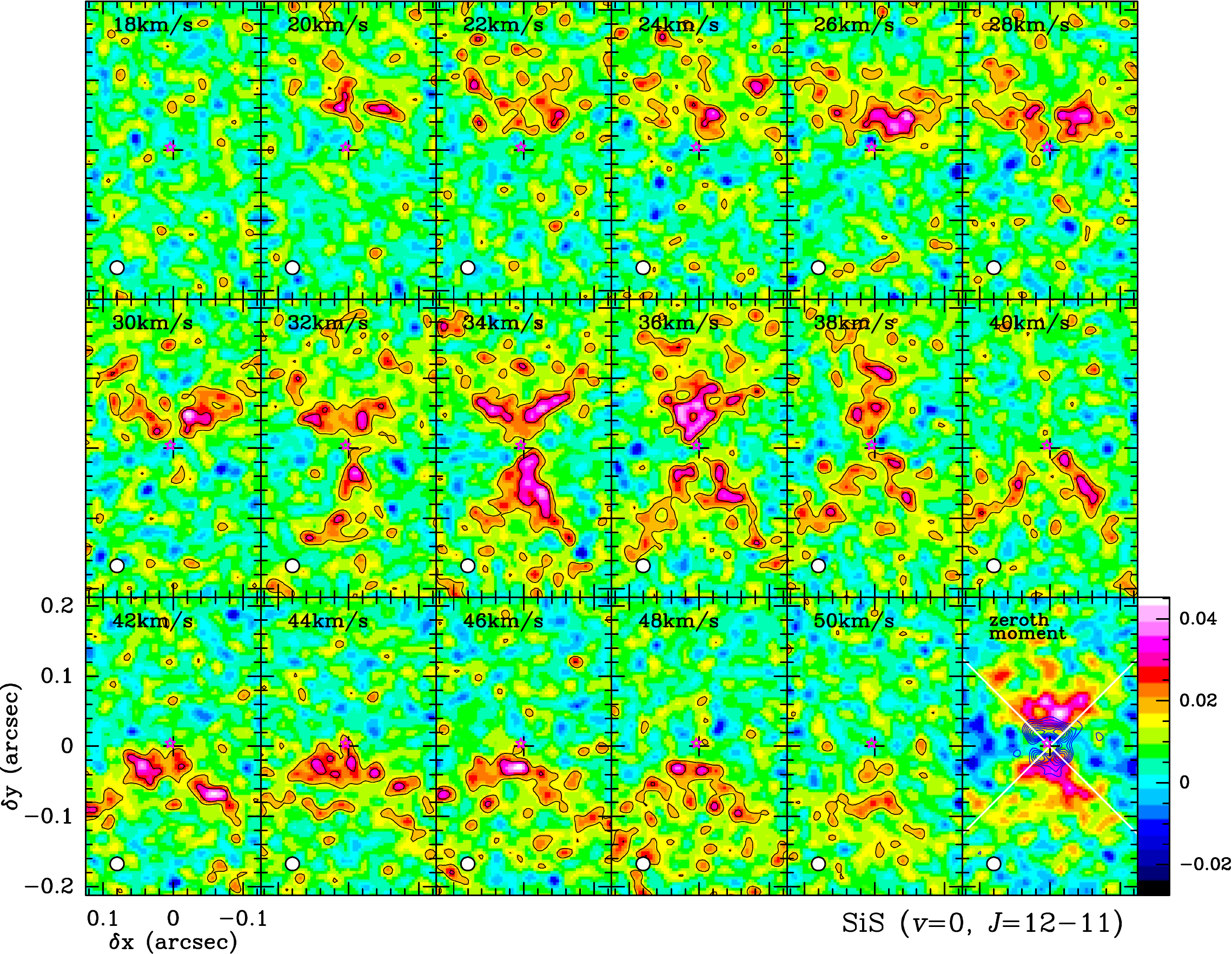}
      \caption{{\bf Top)} \sisdo\ velocity-channel maps with a beam
        with HPBW=20\,mas$\times$20\,mas; note the smaller field of
        view compared with that of the 60\,mas-resolution maps show in
        Fig.\,\ref{f-sisv0}. Contours are 2$\sigma$, 4$\sigma$,... by
        4$\sigma$ ($\sigma$=0.5\,mJy/beam). In the last panel, the
        zeroth moment map of SiS (integrated over
        \vlsr=[18:52]\,\kms) is shown (color scale and wedge) together
        with that of the NaCl line-stacked data (contours). Units of
        the wedge are Jy\,beam$^{-1}$\,\kms.
        Two orthogonal segments are used to outline the opening angle of the \sso\ at its base ($\sim$90\degr).}
     \label{f-sisv0-20mas}
   \end{figure*}

\begin{figure*}[t]
     \centering

 \includegraphics[width=0.95\hsize]{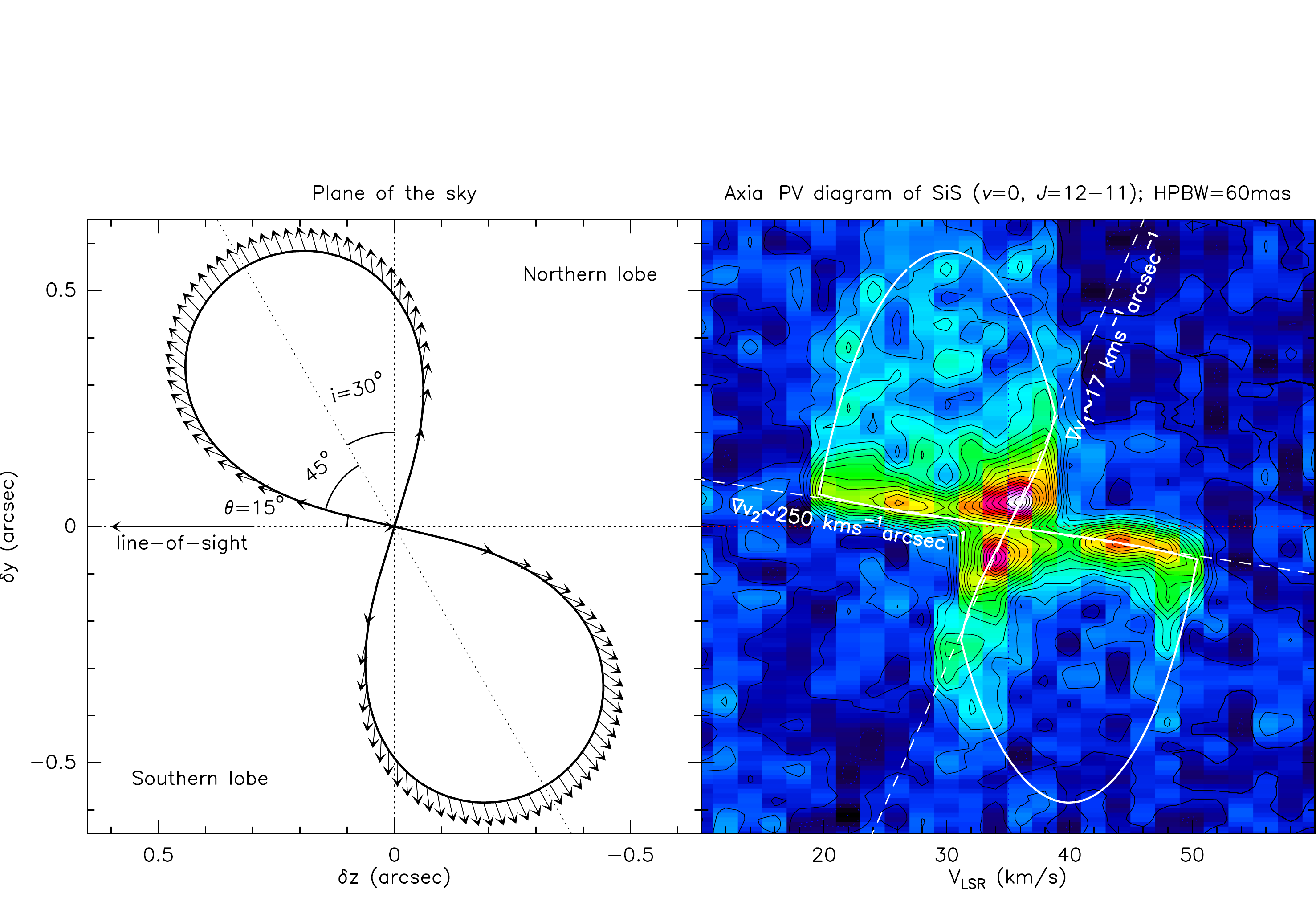}
 \caption{{\bf Left)} Schematic geometry and velocity field (arrows) of the compact \sso\ of \ohs. This plot represents a slice of the \sso\ through a plane
   perpendicular to the plane of the sky oriented along PA=25\degr. In this figure, 0\farc1 corresponds to 150\,au and the largest arrows correspond to \vexp=16\,\kms. 
{\bf Right)} Position-velocity (PV) diagram of the \sisdo\ data with 60\,mas-resolution along the direction of the nebula axis (PA$\sim$25\degr)
through the nebula center ($\delta x$=0\arcsec) and synthetic axial-pv diagram (white line) resulting from the model in the left. The values of the projected velocity gradients,
$\nabla$v$_1$ and $\nabla$v$_2$, measured along the rear (front) side of the Northern (Southern) lobe and the  rear (front) side of the Southern (Northern) lobe, respectively,
are indicated (dashed lines).}
         \label{f-pvsis}
   \end{figure*}

The \sisdo\ ALMA data are shown in Figs.\,\ref{f-sisv0}-\ref{f-pvsis}. 
The SiS maps have been restored using
natural weighting and tapering (with a tapering distance of 5700\,m),
resulting in a beam with HPBW=60\,mas, which offers a good compromise
between sensitivity to smooth medium-size ($\sim$0\farc1-0\farc2)
structures and angular resolution (Fig.\,\ref{f-sisv0}). In order to have a sharper view
of the inner regions of the \sso, the \sisdo\ maps have been also
restored using robust weighting, resulting in a HPBW=20\,mas
(Fig.\,\ref{f-sisv0-20mas}).

%
%

By comparing the SiS\,$\upsilonup$=0, $J$=12-11 line profile obtained
from our 60\,mas-resolution maps with single-dish measurements
(Fig.\,\ref{f-sisv0}, bottom-right), we infer rather small interferometric flux
losses, consistent with the SiS emission being mainly (or almost
entirely) circumscribed to the compact \sso.
Note that, in contrast, the emission from $^{33}$SO that partially overlaps with
that of SiS is significantly filtered out by the interferometer,
consistent with most of the $^{33}$SO emission arising predominantly
from the large-scale molecular outflow. This is also the case of many other
species with intense single-dish emission, including CO, HCO$^+$, HCN, CS, OCS,
SO$_2$, etc, observed in this project (but not discussed here).

Our \sisdo\ ALMA maps define rather precisely the shape of the \sso,
adding detailed information on its morphology due to the improved
angular resolution over the discovery ALMA SiO emission maps
(\san). The compact \sso\ is formed by two opposing lobes with a
nearly conical geometry at their base and a rounded morphology at
their ends, resulting in a characteristic flame-like or $\infty$-like
shape. The orientation of the \sso's symmetry axis
in the plane of the sky is similar to
that of the NaCl equatorial structure. The size of the lobes as seen in the
\sisdo\ 60\,mas-resolution maps is $\sim$0\farc6$\times$0\farc8 (at a
3$\sigma$ level; Fig.\,\ref{f-sisv0}). This is a lower limit to the
full extent of the \sso' lobes, which appear to be somewhat larger
along the axis in the SiO maps (Fig.\,4 of \san), due to partial flux
losses of the SiS emission from these structures at medium angular
scales (particularly at their outer regions) and due to the
sensitivity limit of our maps. The shape of the lobes, which appear
limb-brightened in our SiS maps indicating a dense-walled structure,
is consistent with a wide-angle wind with an opening angle of
$\sim$90\degr\ at its base.

The pinched-waist appearance of the integrated intensity maps of \sisdo\ (Fig.\,\ref{f-sisv0}) partially results from 
a lack of emission in the equatorial plane. This is best seen in the
20\,mas-resolution maps that show a $\sim$0\farc2-thick waist where no
emission from SiS is detected (Fig.\,\ref{f-sisv0-20mas}). We compare
the distribution of the \sisdo\ integrated intensity emission
(zeroth moment map) with that of NaCl in the last panel of this
figure.  As it can be clearly seen, the \sso\ emerges immediately
beyond the surface layers of the rotating equatorial disk at the
center traced by NaCl.  The \sso\
runs almost perfectly through the diagonals of the squarish surface
brightness distribution of NaCl, oriented at
$\pm$45\degr\ from the nebula/disk axis.  The relative distribution of
the NaCl and SiS emission strongly suggests that both molecules probe
adjacent layers of the same nebular structure at different scales or
that the two underlying structures, the rotating disk and the \sso,
are very closely physically related.


The SiS emission is spread over the velocity range \vlsr$\sim$18-50\,\kms,
similar to the full velocity width of the SiO\,($\upsilonup$=0,
$J$=7-6) emission mapped with 0\farc2 resolution by \san. This confirms
that the moderate flux losses in our 60\,mas-resolution \sisdo\ 
maps do not result in any kinematic components of the \sso\ to be
missed but only to a partial filtering of the emission from smooth medium-size
structures mainly within the central/outer regions of the flame-shaped
lobes. The kinematics of the \sso\ is predominantly expansive
as denoted by the overall blue- (red-) shifted emission from the north
(south) lobes, respectively (Fig.\,\ref{f-sisv0}, bottom-center) given
the inclination of the nebula, with the north lobe pointing toward
the observer. 

The outflow kinematics can be best constrained by further exploring
the data using position-velocity (PV) cuts (Fig.\,\ref{f-pvsis}).
To guide the eye and to help the reader more quickly visualize the
kinematic information contained in the axial-pv diagram described
in the next paragraphs, we include in Fig.\,\ref{f-pvsis} a very simplistic spatio-kinematic model of
the \sso\ (left panel) and the axial-pv diagram resulting from the
model (overplotted on the data, right panel).

The PV diagram of \sisdo\ along the major axis of the \sso\ 
shows an easily recognisable X-shape in the central regions (within
$\delta$y$\sim$$\pm$0\farc2-0\farc25), where the emission is brightest.
The X-shape is consistent with the conical geometry of the \sso,
inferred directly from the velocity-channel maps, and with an
expansion velocity that increases (linearly or almost linearly) with
the distance to the center (\vexp$\approxprop$$r$), i.e.\,with a
constant velocity gradient along the walls of the SS-outflow at its
base.
This type of kinematics is common in the outflows of pPNe and
yPNe and, indeed, is also observed in the medium-to-large scale
nebular components of \ohs, namely, the large-scale CO outflow
(including the high-velocity lobes and the large equatorial waist) and
the so-called mini-hourglass that surrounds the
\sso\ (\san). 


The slope (or projected velocity gradients) of the 
two crossing straight lines that form the X-shaped SiS emission distribution in the axial-PV diagram (Fig.\,\ref{f-pvsis}), can be
used to derive the inclination $i$ of the \sso's axis using elementary maths and 
assuming a conical geometry with a semi-opening
angle of $\sim$45\degr\ at the base (directly deduced from the
velocity-channel maps). 
On the one hand, the smallest projected velocity gradient between the two
\sisdo\ emission peaks ($\nabla$v$_1$$\sim$17\,\kms\,arcsec$^{-1}$)
results from the gas outflowing along the rear (front) side of the
North (South) lobe, which is inclined by an angle $\theta$$\sim$45\degr-$i$
with respect to the plane of the sky.  On the other hand, the largest
projected velocity gradient observed,
$\nabla$v$_2$$\sim$250\,\kms\,arcsec$^{-1}$, results from the gas
outflowing along the front (rear) side of the North (South) lobe,
inclined by the same angle, $\theta$=45\degr$-$$i$, but from the line-of-sight.
For an axially symmetric conical flow and assuming that the deprojected velocity
gradient is the same in both lobes, i.e.\ that $\nabla$v$_2$$\times$tan($\theta$)=$\nabla$v$_1$/tan($\theta$), 
we derive $\theta$$\sim$15\degr, implying that
$i$$\sim$45$-$15$\sim$30\degr\ (see Fig.\,\ref{f-pvsis}). This value of $i$ is in very good
agreement with the inclination of the symmetry axis of the different
nebular components identified so far in \ohs\  from previous studies
(\S\,\ref{intro}) and from this work, including the dust continuum
disk (\S\,\ref{res-cont}) and the equatorial rotating structure traced
by the salts and water (\S\,\ref{res-nacl}). The deprojected velocity
gradient along the \sso\ in regions close to its base is then
$\nabla$v$\sim$250$\times$tan(15\degr)$\sim$17/tan(15\degr)$\sim$65\,\kms\,arcsec$^{-1}$.

In the intermediate-to-outer regions of the \sso\ lobes, where these acquire
a rounded appearance, the distribution of the SiS emission in the
axial-pv diagram indicates that the
\vexp$\approxprop$$r$ law is not sustained any longer, since otherwise
much larger projected velocities would be observed at the lobe tips
given their dimensions.  As shown in Fig.\,\ref{f-pvsis}, the axial-pv
diagram is indeed consistent with rounded lobes radially expanding at
a constant (terminal) speed of \vexp$\sim$15-17\,\kms\ at radial
distances from the center larger than $\sim$0\farc2-0\farc25
($\sim$350\,au).  Note that the limited angular resolution of the
  observations does not enable us to spatially resolve the compact
  region over which the velocity law changes from a gradual increase
  to a terminal constant value or to precisely determine the exact
  velocity law, not being possible to rule out, for example, a (quasi)
  $\beta$-wind velocity profile (\S\,\ref{dis-nacl}).
 

Rotation at the base of the \sso\ is tentatively identified in
our \sisdo\ maps: note the slightly curved shape of the iso-velocity
contours at $\delta$y=0\arcsec\ in the first moment maps
(Fig.\,\ref{f-sisv0}, bottom-center). In these inner equatorial regions,
however, there is a lack of strong SiS emission that, together with
the limited angular resolution, precludes a reliable description of
the velocity field from these maps. As we show in
\S\,\ref{res-siov1}, rotation at the base of the \sso\ is confirmed 
in our \siov\ ALMA data and, thus, we defer to the next section for further developing on this. 



\subsection{SiO\,$\upsilonup$=1}
\label{res-siov1}
   \begin{figure*}[htp]
     \centering
     \includegraphics[width=0.95\hsize]{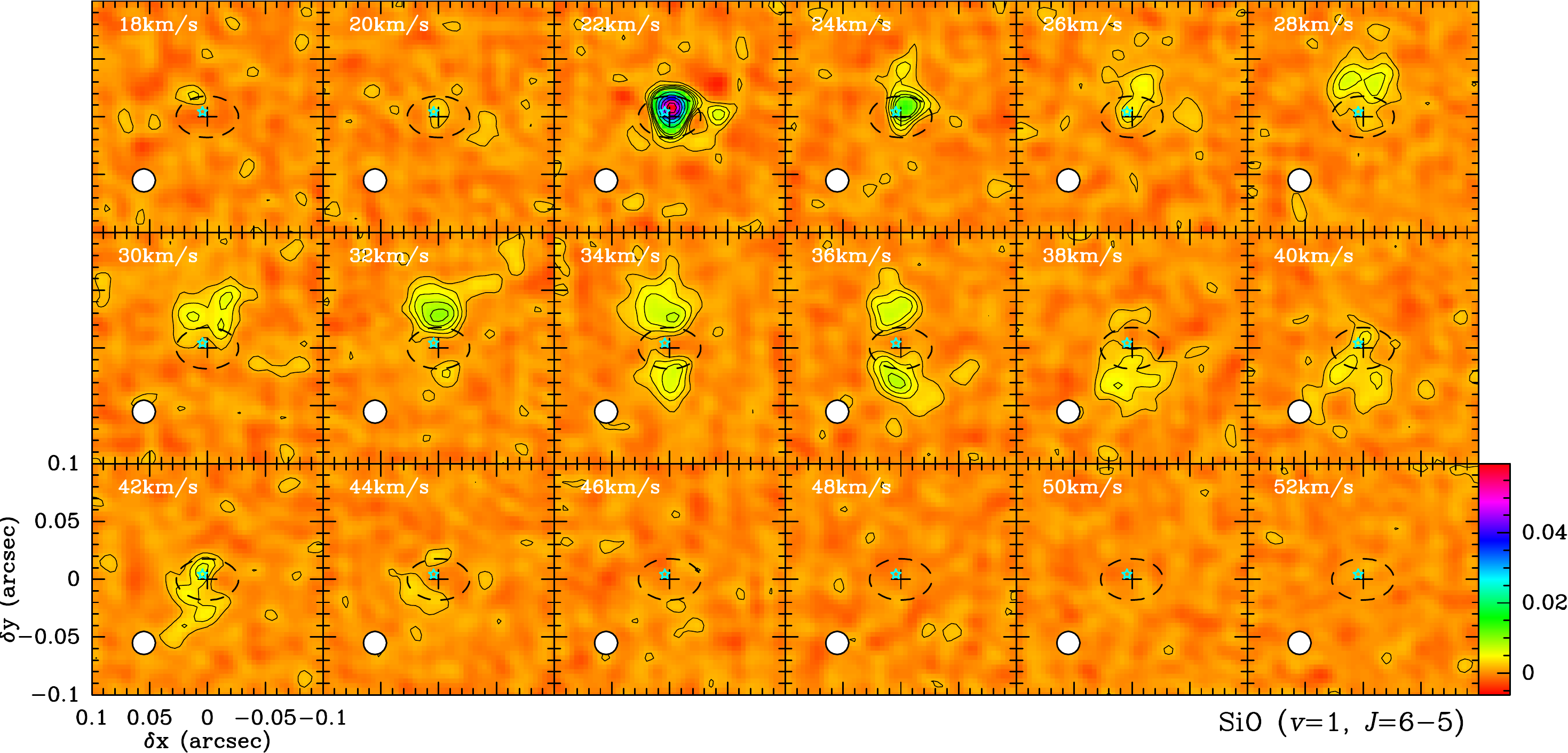}
     \includegraphics[width=0.3\hsize]{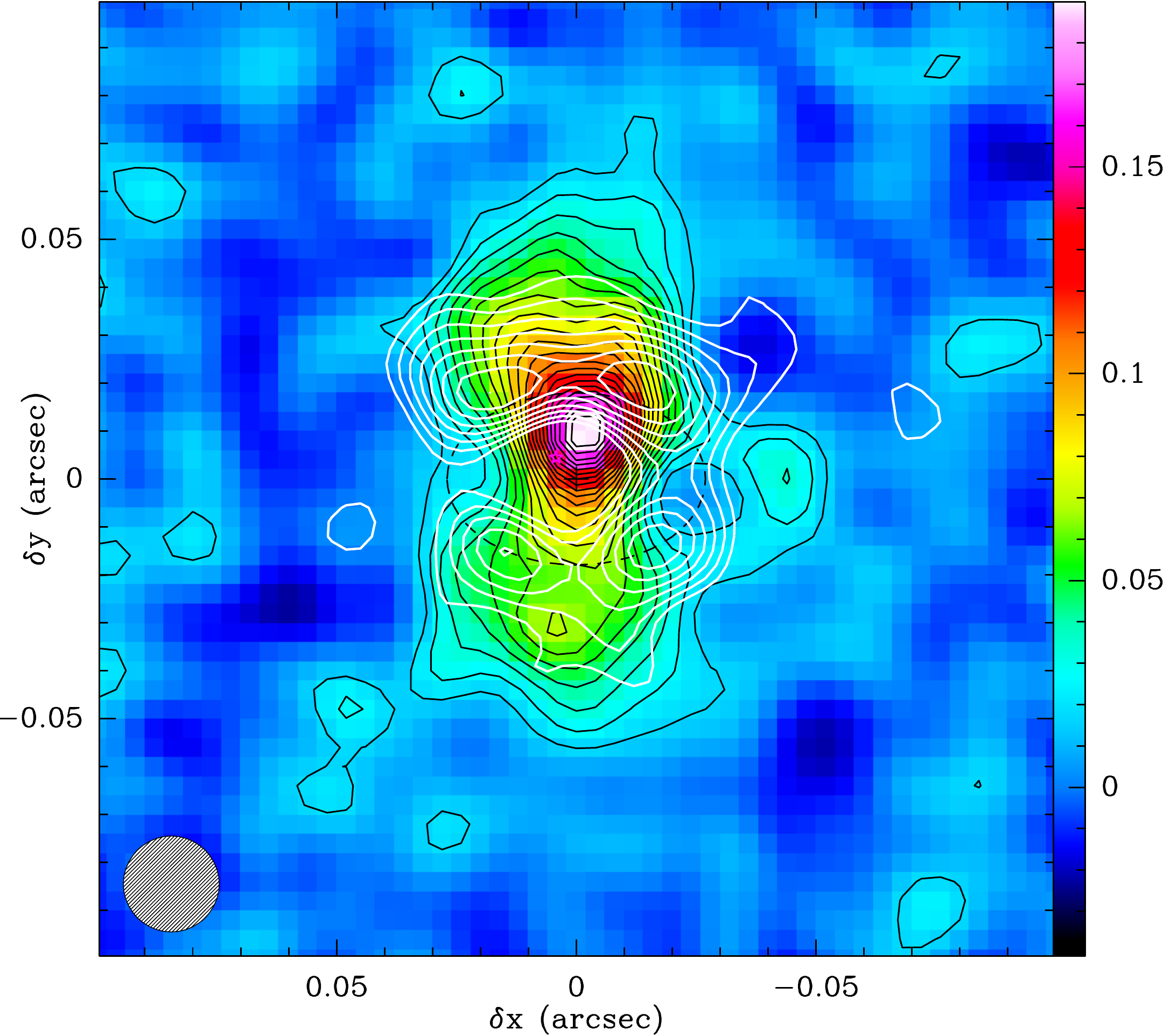}
     \includegraphics[width=0.3\hsize]{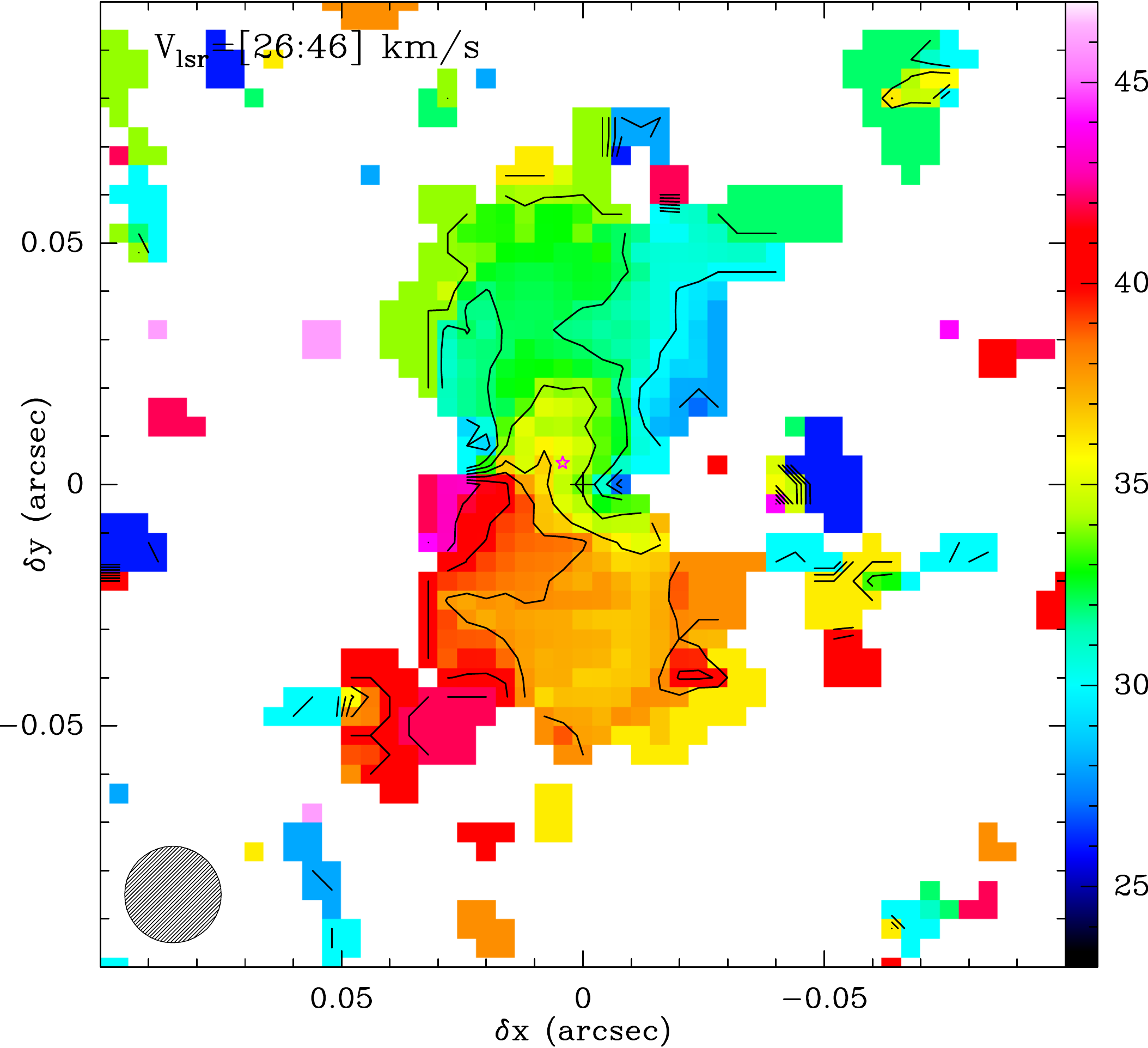}
     \includegraphics[width=0.28\hsize]{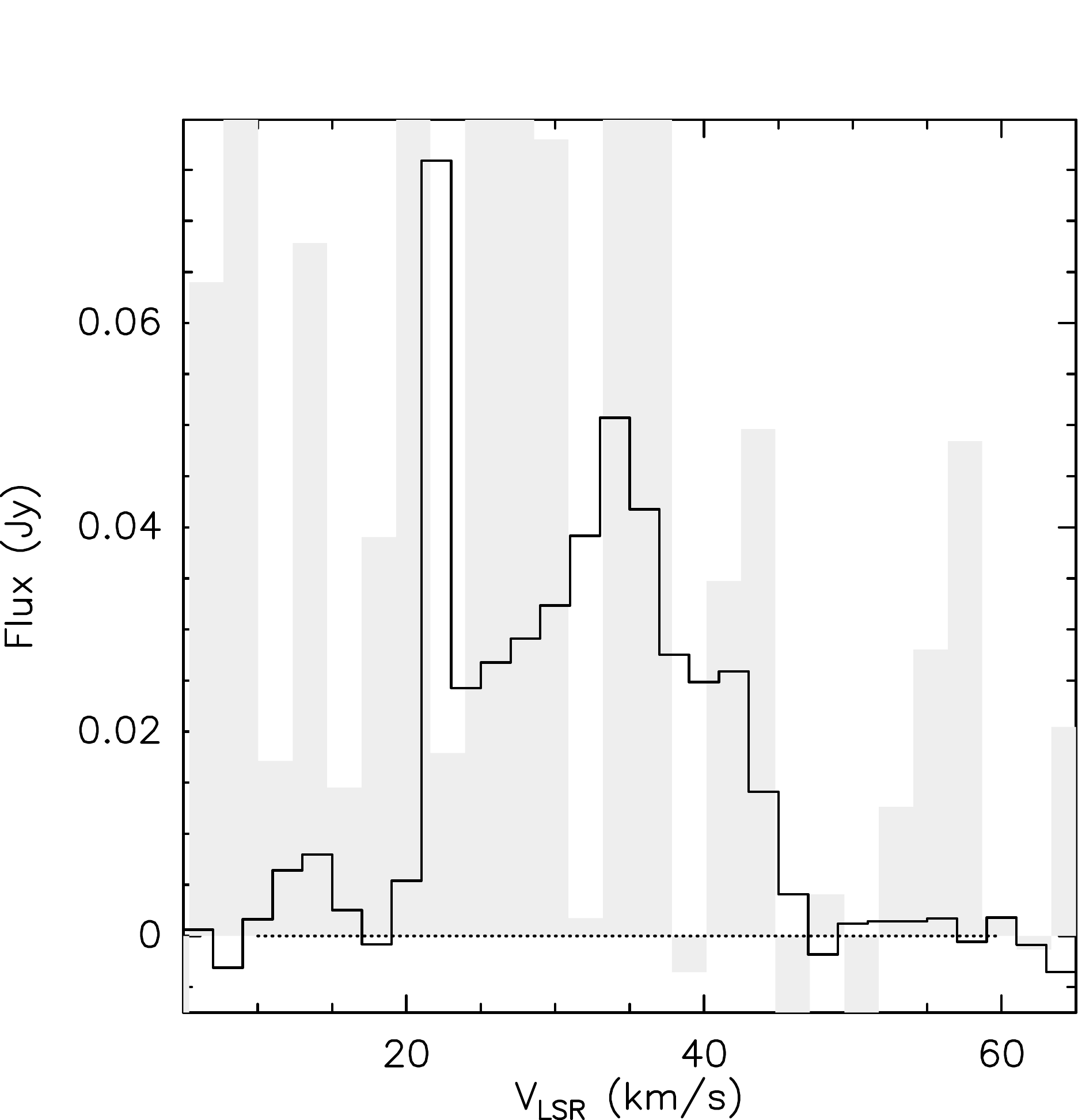}
\caption{\siov\ ALMA data. {\bf Top)} Velocity-channel maps; contours
  are: 2$\sigma$, 4$\sigma$ to 10$\sigma$ by 2$\sigma$ and from
  20$\sigma$ to 60$\sigma$ by 10$\sigma$
  ($\sigma$=0.9\,mJy\,beam$^{-1}$). {\bf
    Bottom-left)} Integrated intensity (zeroth moment) map over the
  velocity range \vlsr=[18:46]\,\kms;
  contours are 2$\sigma$, 3$\sigma$,... by 1$\sigma$
  ($\sigma$=8\,mJy\,\kms\,beam$^{-1}$).   The NaCl line-stacked integrated
  intensity map is on top (white contours) for comparison.  {\bf Bottom-center)} Velocity
  (first moment) map over the \vlsr=[26:46]\,\kms\ range,
  i.e.\,excluding channels \vlsr$\le$24\,\kms\ (with amplified
  emission) for clarity.  {\bf Bottom-right)} ALMA integrated
  1d-spectrum over the emitting area (black solid line)  compared with
  the \iram\ single-dish spectrum at this frequency where the
  \siov\ line is undetected (light grey). }
         \label{f-siov1}
   \end{figure*}
%
   \begin{figure*}[t]
     \centering
\includegraphics[width=0.3\hsize]{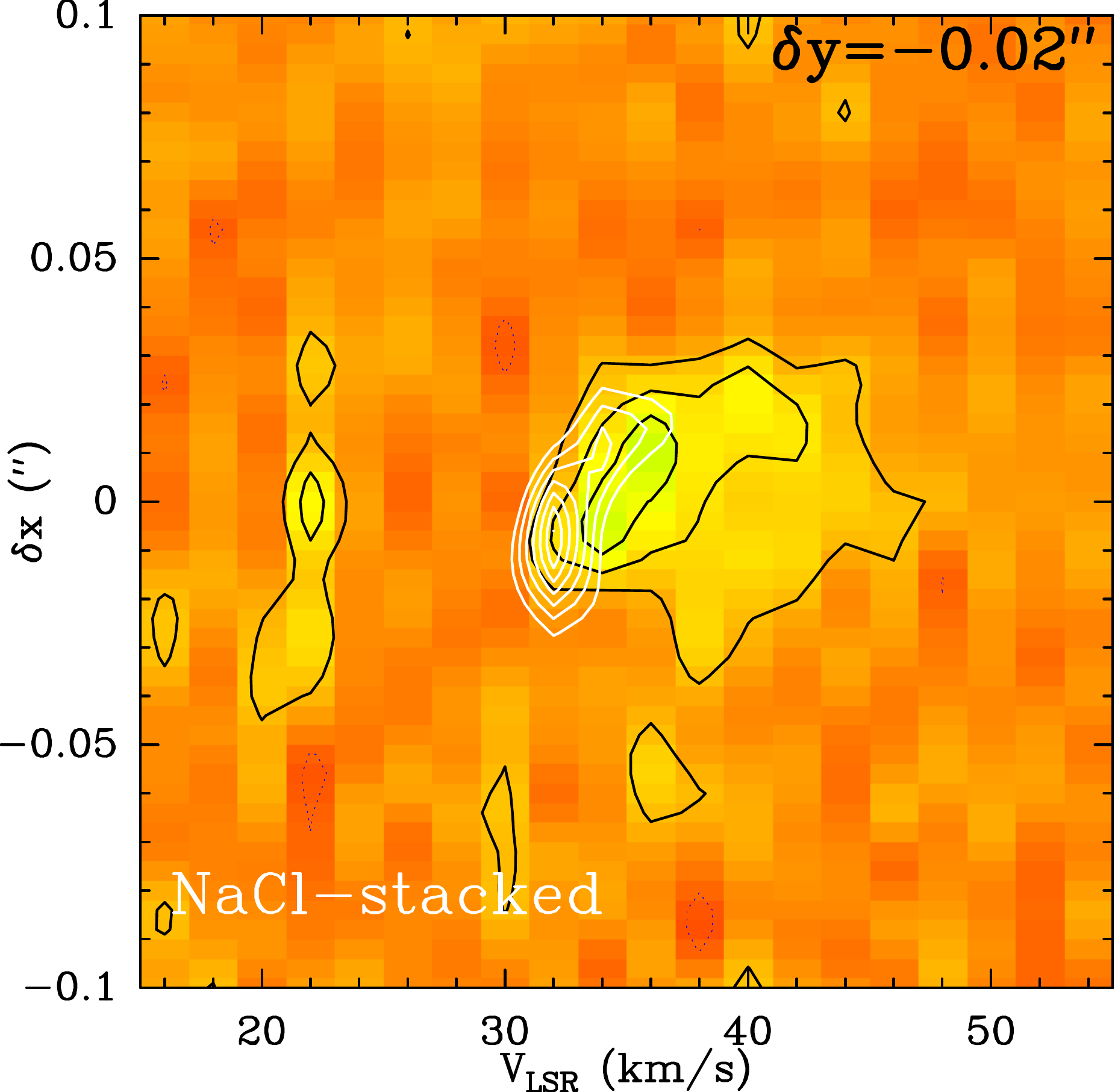}
\includegraphics[width=0.3\hsize]{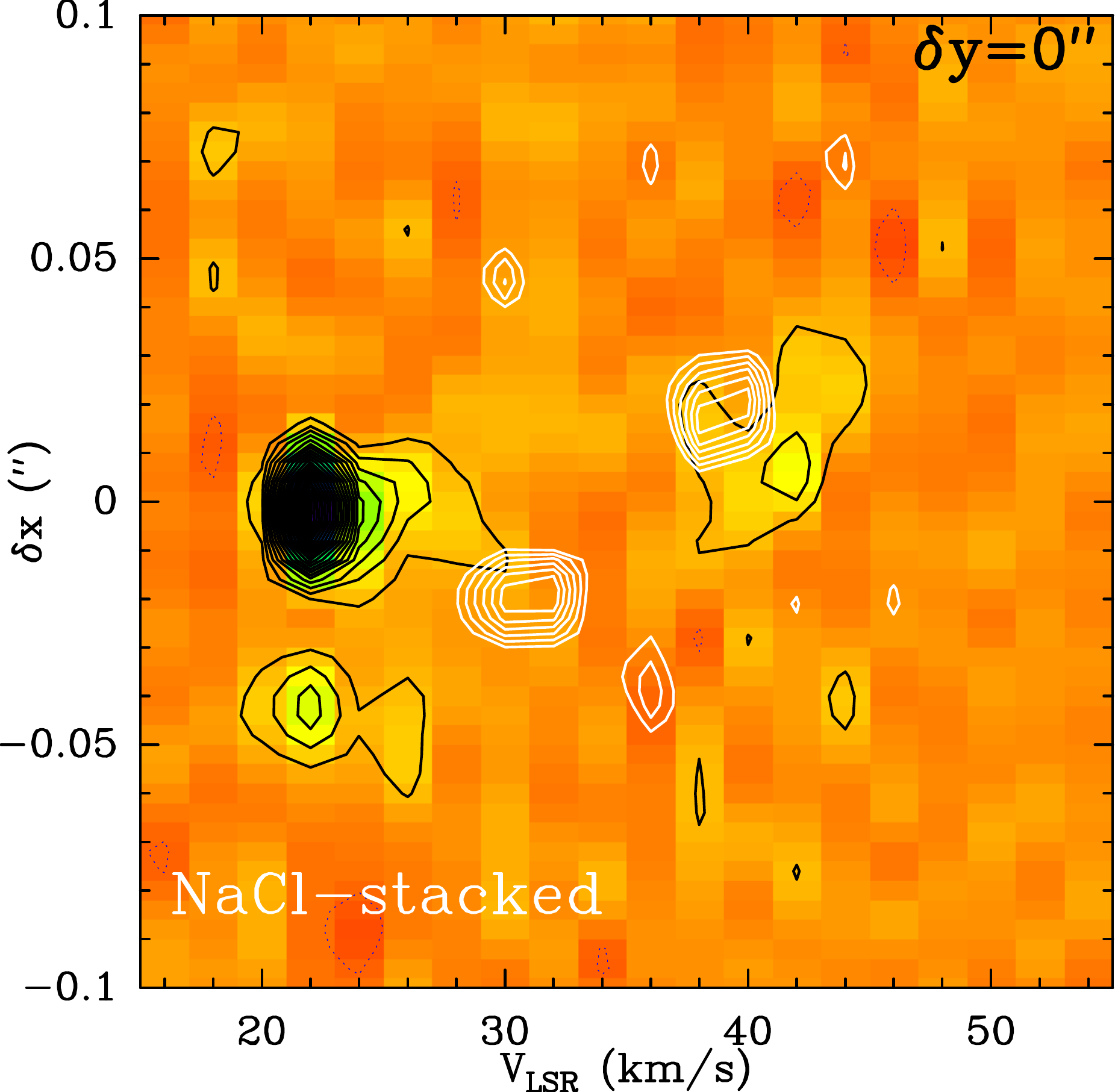}
\includegraphics[width=0.3\hsize]{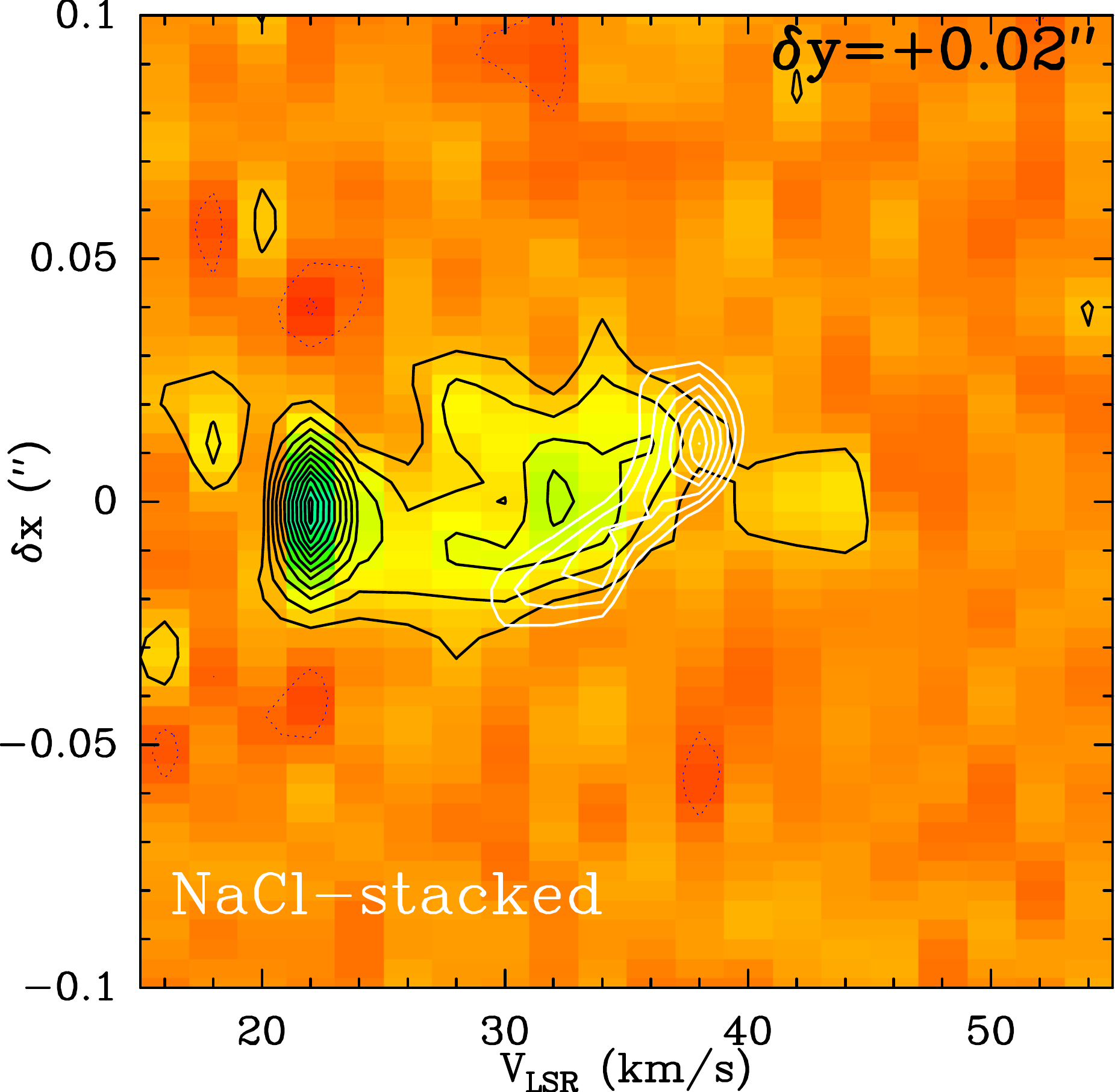} 
\includegraphics[width=0.3\hsize]{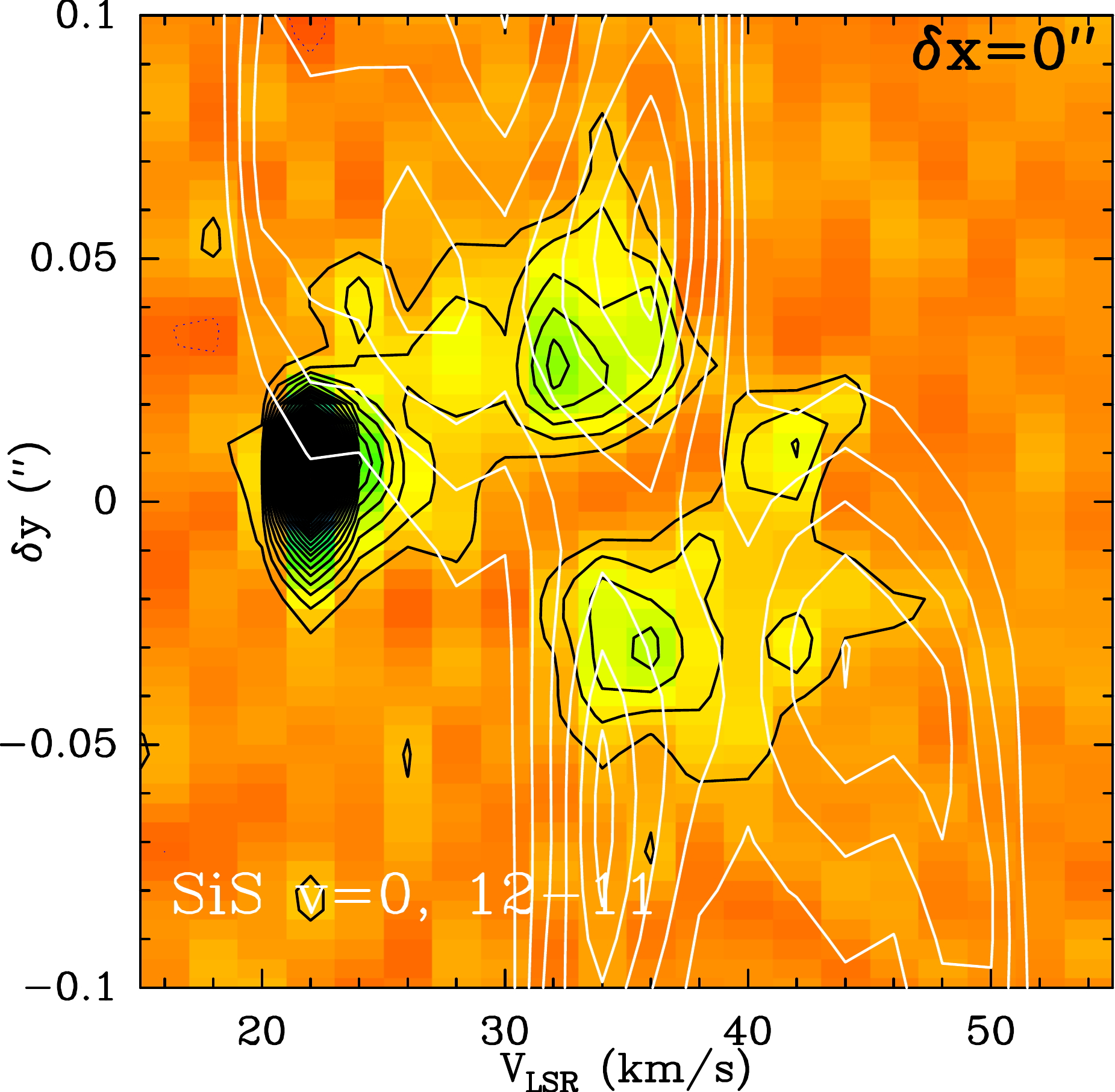}   
\caption{As in Fig.\,\ref{f-pvnacl} but for \siov; NaCl line-stacked
  pv diagrams along the equator are overplotted (white contours) in
  the top panels. In the bottom panel, we overplot the \sisdo\ axial
  pv-diagram instead to help visualizing the locus of the
  \siov\ emission at the base of the \sso\ arising from the lobes'
  interior as well as the comparable expansive kinematics of these two transitions.}
           \label{f-pvsiov1}
   \end{figure*}
%

   The SiO $J$=6-5 emission in the first vibrationally
   excited state $\upsilonup$=1 (\eu$\sim$1800\,K, Fig.\,\ref{f-siov1}) 
   arises from a central elongated region, of dimensions
   $\sim$0\farc06$\times$0\farc12 (at a $\sim$3$\sigma$ level),
   with the long axis oriented along PA$\sim$25\degr, consistent with a bipolar wind
   running inside the hollow rotating disk traced by the salts and
   water. The distribution of the \siov\ emission overlaps (spatially and spectrally) with that
   of the SiS $\upsilonup$=0 transitions at the base of the \sso,
   indicating that the $\upsilonup$=1 emission is selectively probing
   high-excitation regions of the \sso\ closer to the center.   
   
   The kinematics of the SiO\,$\upsilonup$=1 emitting region is indeed consistent
   with an overall expansion (Fig.\,\ref{f-siov1}, bottom-center) closely following 
   the velocity gradient of the \sso\ at its base inferred from the
   SiS $\upsilonup$=0 ALMA maps. This is best seen in the axial-pv
   diagram (Fig.\,\ref{f-pvsiov1}-bottom), where the \siov\ emission is
   clearly constrained to the $\pm$0\farc06 central regions of the \sso\ with
   velocities spreading from \vlsr$\sim$20 to 44\,\kms. 
   The comparison with the axial-pv of the \sisdo\ transition, which best traces the walls of the \sso, suggests that the \siov\ emission mainly arises
   from the interior of the lobes and very close to their base.
   In particular, the position of the two relative maxima of the SiO\,$\upsilonup$=1 seen in the axial-pv diagram, at
   \vlsr=32 and 36\,\kms\ at $\pm$30\,mas ( $\pm$45\,au) from the center, 
   is consistent with brightening of the SiO\,$\upsilonup$=1 emission in regions close to the rear (front) side of the North (South) lobe due to a smaller velocity dispersion resulted from the bulk expansive
   motions being along (or near) the plane of the sky in these regions (also visible in the \sisdo\ axial-pv diagram, Fig.\,\ref{f-pvsis}).

   There is a very bright peak of \siov\ emission at
   \vlsr$\sim$22\,\kms\ arising from an unresolved clump near the
   center of the nebula (Fig.\,\ref{f-siov1}-top). The main-beam
   brightness temperature of this feature reaches T$_{\rm
     mb}$$\sim$2700\,K, to be compared with the much lower values of
   T$_{\rm mb}$$\sim$200\,K at other velocities.  The spatial and
   spectral location of this bright feature suggests that it is
   produced by a parcel of gas located between the central
   dust-continuum source and the observer, plausibly in the front side
   of the approaching North lobe. (Absorption below the continuum
   level, i.e. toward the center, at this velocity is observed in
   several other transitions -- not shown -- corroborating this
   interpretation).
   The profile of the bright \siov\ feature is extremely narrow,
   with a FWHM$\sim$2.2\,\kms. This, together with the remarkably high
   value of the main-beam brightness temperature observed, strongly
   suggest a certain degree of maser amplification of the background
   continuum.  (We have also considered that this feature could be due
   to the emission of another molecular species, but no probable
   candidates at this frequency are found.)
  The broad profile and moderate values of T$_{\rm mb}$ of the \siov\ line at the
  rest of the velocities indicates predominantly thermal emission.  

   In addition to expansion, the base of the \sso\ as traced by
   \siov\ emission is found to be rotating. This is clearly seen, for
   example, in the velocity (first moment) maps presented in
   Fig.\,\ref{f-siov1} (bottom-center): the East (West) part of
   the lobes is observed to be Doppler-shifted to redder (bluer)
   velocities than the average systemic velocity of each lobe. The velocity gradient
   across the lobes is most apparent in the
   South lobe; this is partially due to overlap with the prominent 
   blue-shifted maser spike of the \siov\ line, which arises from a
   very compact region in the front side of the North lobe that is
   primarily in expansion. For clarity, the first moment map
   show in Fig.\,\ref{f-siov1} has been obtained excluding the
   affected channels where maser amplification is produced (\vlsr$\leq$24\,\kms).
   
   Rotation can be also easily recognised in the \siov\ pv cuts along
   the direction of the equator, i.e.\ across the lobes (Fig.\,\ref{f-pvsiov1}, top panels). In this
   figure, we show pv cuts through the center ($\delta$y=0\arcsec) and
   through the base of the \sso\ lobes ($\delta$y=$\pm$0\arcsec02)
   together with those obtained from the NaCl-line stacked cubes for
   comparison. In the $\delta$y=$\pm$0\farc02 equatorial pv cuts, the
   distribution of the \siov\ emission from regions with the
   $\delta$x$>$0\arcsec\ ($\delta$x$<$0\arcsec), i.e.\ the east (west)
   side of the lobes, exhibits an overall shift toward the red
   (blue), consistent with rotation as observed also in the NaCl-line
   stacked pv diagrams.  Rotation is perhaps more clearly appreciated
   in the South lobe ($\delta$y=$-$0\farc02 pv diagram).
     
   The signature of rotation, a velocity gradient across the lobes,
   and expansion, a velocity gradient along the lobes, are
   simultaneously present in the \siov\ data, which partly makes it
   difficult to cleanly isolate/disentangle the two kinematic
   components. In the pv diagrams under discussion
   (Fig.\,\ref{f-pvsiov1}), the \siov\ emission from
   $\delta$y=+0\farc02 ($\delta$y=$-$0\farc02) is globally shifted to
   lower (higher) velocities compared to NaCl. This is reflecting
   moderately larger expansion velocities at the base of the
   \sso\ than at the surface layers of the rotating disk, which makes it somewhat more difficult to identify rotation in the
   \siov\ data than in the NaCl (KCl and water) data.

   The rotation velocity at the base of the
   \sso\ ($\delta$y=$\pm$0\farc02) deduced from \siov\ is similar to,
   perhaps slightly smaller than, that derived from NaCl; the limited
   S/N of the data precludes a very accurate estimate of the rotation
   velocity and its spatial distribution in these inner regions, which
   are also compact compared with the angular resolution of our ALMA
   maps. At the nebula equator ($\delta$y=0),  the \siov\ emission is
   very weak, except for the dominant bright \vlsr=22\,\kms\ emission peak from the
   center; for this reason identifying and quantifying rotation in the \siov\ data at this zero
   latitude regions is problematic and obviously less clear than in the maps of the salts and water.

\section{Analysis of the NaCl emission}
\label{res-anal}

\subsection{Physical conditions in the rotating disk}
\label{s-rd}
   \begin{figure}[h]
     \centering
     \includegraphics[width=0.95\hsize]{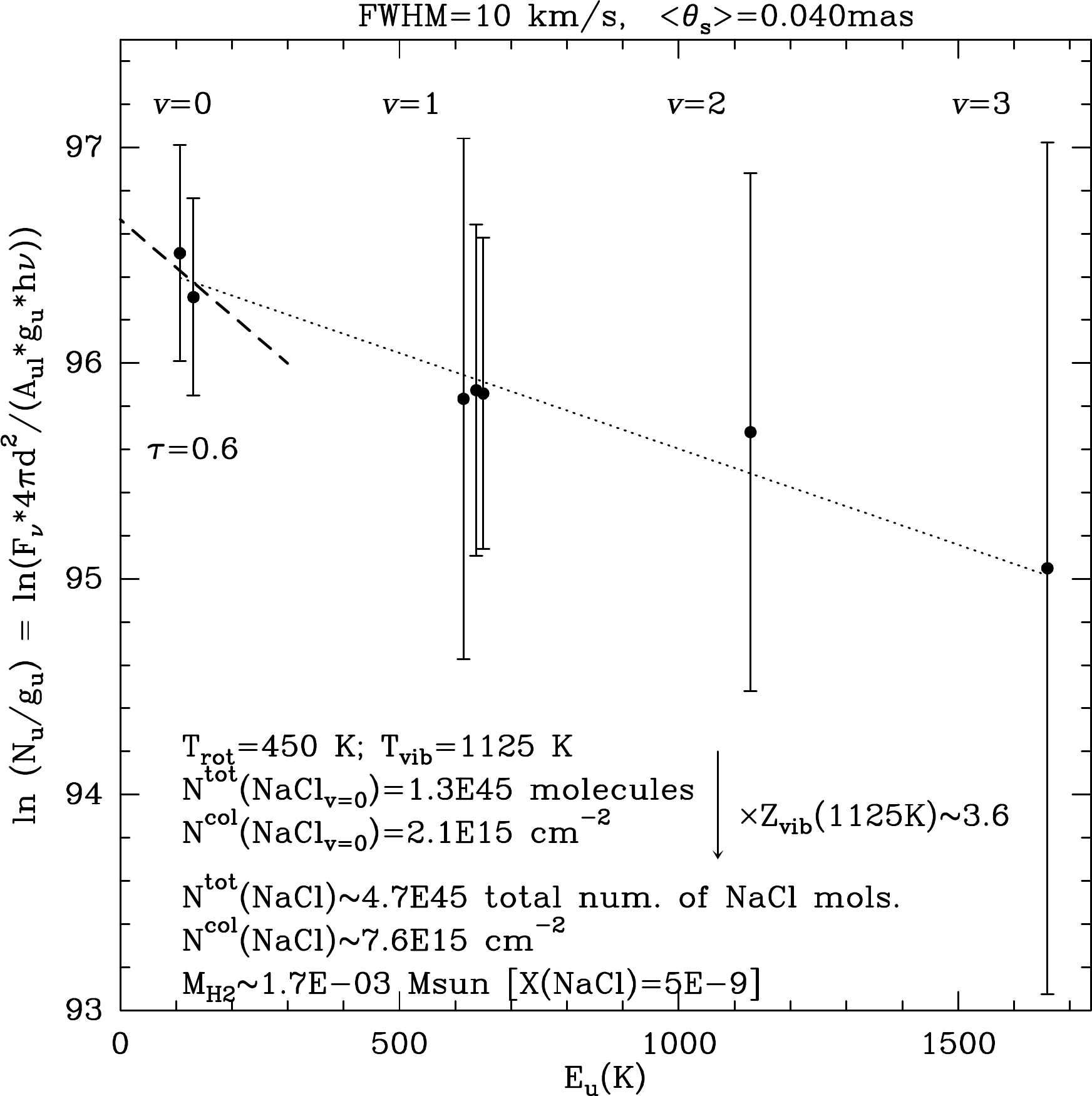}
     \caption{Population diagram of NaCl. A linear fit to the data using all the NaCl
       transitions detected across all vibrational levels from $\upsilonup$=0 to $\upsilonup$=3 indicates a vibrational
       temperature of \tvib$\sim$1125\,K (dotted line).  The slope of
       the dashed line, consistent with the observed ln(N$_{\rm u}/$g$_{\rm u}$) vs. \eu\ values for the only two NaCl $\upsilonup$=0
       transitions observed, corresponds to \trot$\sim$450\,K. The
       values for the total number of NaCl molecules and column
       density in the ground vibrational level are indicated, as well as the
       corresponding values after multiplying the former by the vibrational partition function
       to account for the NaCl distribution across multiple vibrational levels. }

       \label{f-rd}
   \end{figure}
%

We have performed a NaCl population diagram analysis in order to
constrain the excitation temperature and optical depth of the
observed transitions and, thus, to obtain a rough estimate of the
physical conditions in the surface layers of the rotating,
geometrically thick disk where the NaCl emission arises. The
population diagram is a well-known and widely used analysis technique
under the assumption of optically thin line emission and local thermal
equilibrium (LTE). This method has been described in detail and
discussed extensively by, e.g., \cite{gol99} and it has been
successfully used in the analysis of molecular line emission of many
evolved stars, including \ohs\ \citep{san15,vel15,san18}.  Here, we use
the approach presented in \cite{ram18} where the total
line  fluxes integrated over the emitting area are used to
ultimately derive the total number of molecules of NaCl (\ntot(NaCl))
and the excitation temperature in the emitting volume.  The spectroscopic parameters
of the rotational transitions used in this analysis, including the line
frequencies ($\nu$), Einstein coefficients ($A_{\rm ul}$), and
the rotational and vibrational partition functions (\zrot\ and \zvib) of NaCl are
from MADEX \citep{cer12}, which has been updated to include
new laboratory dipole moment values and collisional rates for this species 
\citep{cab16,quin16}.

The population (or Boltzmann) diagram for NaCl in \ohs\ is shown in
Fig.\,\ref{f-rd}.  The linear fit to the data including all
transitions from $\upsilonup$=0 to $\upsilonup$=3 indicates a vibrational temperature of
\tvib=1125$\pm$160\,K.
The rotational temperature (\trot), derived using rotational
transitions within the same vibrational state, cannot be properly
constrained from this diagram alone since a wide range of values,
spanning from $\sim$40\,K up to $\approx$1000\,K, are consistent with
the data; this is because of the large data errorbars, derived from
the limited S/N of the individual lines, the small number of
rotational transitions detected within the same vibrational level (two and three within $\upsilonup$=0 and $\upsilonup$=1, respectively), and
the narrow range of \eu\ spanned between them ($\Delta$\eu\la\,35\,K).

Fortunately, it is possible to obtain an estimate of \trot\ directly
from the velocity-channel maps of the \naclvd\ transition by taking
advantage of the weak line absorption at the 
nebula center, around \vlsr$\sim$32\,\kms\ (Fig.\,\ref{f-nacl51}).  At this position,
the brightness temperature of the line (after continuum subtraction)
in the Rayleigh-Jeans limit\footnote{The {\it Rayleigh-Jeans} approximation is valid at these
frequencies for the temperatures expected in these inner regions, of $\approx$100\,K -- see below.} is given by

\begin{equation}
  \label{eq1}
  T_{l}^{\rm on}=(\trot-T_{\rm c})\times(1-e^{-\tau}), 
\end{equation}

where $T_{\rm c}$ is the brightness temperature of the continuum and $\tau$ is the line optical
depth. 
The fact that the line is observed in absorption against the background 
dust continuum means that \trot$<$$T_{\rm c}$. The line intensity outside the 
continuum source, where the dominant background emission source is 
the 2.7\,K cosmic microwave radiation, is given by 

\begin{equation}
\label{eq2}
  T_{l}^{\rm off}=(\trot-2.7)\times(1-e^{-\tau})\sim\trot\times(1-e^{-\tau}). 
\end{equation}

Combining Eq.\,\ref{eq1} and \ref{eq2} and taking into account the
values of the line emission and absorption measured in the \naclvd\ 
maps as well as that of the continuum peak in the
same spectral window\footnote{For a proper line-to-continuum
comparison, we use the continuum maps of the same spectral window
where the line was observed cleaned and restored exactly in the same
manner and, thus, imaged with the same beam.}  ($T_{l}^{\rm
  on}$$\sim$$-$130\,K, $T_{l}^{\rm off}$$\sim$150-200\,K, and $T_{\rm
  c}$$\sim$730\,K, e.g. Fig.\,\ref{f-nacl51}), we deduce \trot$\sim$400-450\,K
and $\tau$$\sim$0.5-0.6. (Similar values are inferred from the
\nacldo\ maps, not shown.)

As seen in Fig.\,\ref{f-rd}, a value of \trot$\sim$400-450\,K is
well within the range of temperatures consistent with the intra-$\upsilonup$=0 and $\upsilonup$=1 data
points in the population diagram.  More precisely, we find that
\trot$\sim$450\,K is the value that best explains simultaneously the
population diagram, the optical depth (around 0.5-0.6) and the mean line
brightness temperature of the
\naclvd\ transition outside the continuum source ($T_{\rm l}^{\rm off}$$\sim$150-200\,K).
A straight-line fit to the $\upsilonup$=0 data with a fixed slope of 1/\trot$\sim$1/450\,K  implies a total number of NaCl
molecules in the $\upsilonup$=0 level of \ntot(NaCl,$\upsilonup$=0)$\sim$1.3\ex{45}.
To compute the corresponding column density, we have used the
simplified equation: \ncol=\ntot/$\pi$$r_{\rm s}^{2}$, where $r_{\rm
  s}$$\sim$30\,au represents the characteristic radius of the region where the
NaCl emission is produced projected in the sky. 
The resulting column density and
\naclvd\ line opacity are \ncol(NaCl,$\upsilonup$=0)$\sim$2\ex{15}\,\cm2 and
$\tau$$\sim$0.6 (for a line FWHM$\sim$10\,\kms), respectively.

The total number of NaCl molecules, also including the population of
excited vibrational levels ($\upsilonup$>0), has been computed as
\ntot=\ntot(NaCl,$\upsilonup$=0)$\times$\zvib(1125K), where
\zvib(1125K)$\sim$3.6. The value obtained, implies a beam-averaged
NaCl column density of \ncol(NaCl)$\sim$7\ex{15}\,\cm2 and a total mass of 
\mtot$\sim$1.7\ex{-3}\,\msun\ in the emitting volume, adopting a
fractional NaCl-to-H$_2$ abundance of $X$(NaCl)$\sim$5\ex{-9} \citep{san18}.  A
canonical opacity correction ($C_\tau$=$\frac{\tau}{1-e^{-\tau}}$) can
also be applied (following, e.g., \cite{gol99,ram18}), which results in
a mass-correction factor of $\sim$30\%, thus implying
\ncol(NaCl)$\sim$8\ex{15}\,\cm2 and a total mass of 
\mtot$\sim$2\ex{-3}\,\msun. 
Considering the dimensions of the NaCl-emitting volume, we deduce 
an average H$_2$ number density of \dens$\sim$3\ex{9}\,\cm3 in the surface layers of the rotating disk, as expected for these inner regions.

As we have seen before, the relative intensities of the different NaCl
lines, including rotational transitions in the ground vibrational
state and from excited vibrational levels, suggest different values
for \trot\ and \tvib. This is not unexpected as similarly
\trot$<$\tvib\ values are found in other sources with NaCl detections, 
e.g.\, the red supergiant VY\,CMa \citep{alc13,kam13} and the Orion SrcI's disk \citep{gin19}.
The high values of \tvib\ here inferred suggest that the vibrationally excited levels of
NaCl could be predominantly populated by IR pumping as a result of
strong IR emission from dust grains in these central regions of the nebula.
(Note that, unlike the rotational levels inside a given vibrational
state, the vibrational ladders are radiatively connected by
$\sim$30\,$\mu$m radiation).

The value of the \trot\ derived from the NaCl population diagram in
\ohs\ is very similar to the typical gas kinetic temperatures observed
at distances of a few tens of au (few\ex{14}cm) in the envelopes of
other evolved stars -- see e.g.\, the compilation of temperature
radial distributions in \cite{ram18}. This probably indicates that, in
contrast to the vibrational levels, the rotational levels are
predominantly populated by collisions and, thus, are thermalized or
close to thermalization.  This is in good agreement with the critical
densities of the rotational transitions observed,
\nc$\sim$few\ex{7}\,\cm3, which are much lower than the mean density
in these regions of the rotating disk deduced from our population
diagram analysis (\dens$\approx$\dex{9}\,\cm3).  These results support
the assumption of local thermodynamic equilibrium (LTE) adopted in the
NaCl emission model presented in the next section.







\subsection{Spatio-kinematic model}
\label{s-model}
%

We have compared the ALMA NaCl maps with the predictions of a LTE
radiative transfer model for this species. We have used a code that
has been employed in numerous previous works by our team under the LTE
or non-LTE approximation \citep[e.g.\,most recently by][]{buj21}.
Modeling has been done in two steps. First, the model is exclusively
used to constrain the geometry and velocity field of the NaCl-emitting
volume by comparison with the NaCl line-stacked maps, which have the
largest S/N ratio and, thus, provide the best diagnostic of the
nebular spatio-kinematics. Once the geometry and kinematics of the
source have been established, we have used the model to reproduce the
surface brightness emission of one of the individual transitions
observed, in particular the \naclvd\ line, with the aim of further
constraining the physical conditions (density and temperature) of the
emitting region.


In order to compare the model predictions with the ALMA data in an
optimum way, we have first built NaCl model data cubes using the same
velocity resolution but a much better spatial resolution (and finer
pixel sampling) than the observed data. Later on, these synthetic data cubes have been Fourier-transformed
and sampled to mimic the original $uv$-data from the ALMA
observation. This task has been done using the GILDAS procedure
{\tt uv\_fmodel}. Then, the $uv$-data from the model are mapped and cleaned
exactly in the same way we did for the $uv$-data from the
observations. By doing so, we can directly compare the resulting clean
images from ALMA and from our modeling: both with exactly the
same spatial and spectral resolutions, similar MRSs and lost flux if
any (not expected in this case), etc.

In light of our ALMA data, we have approximated the geometry of the
NaCl-emitting volume as two co-axial tori displaced along the nebula
axis from the continuum disk midplane, emulating the two (north and
south) surface layers (above and below) the disk where the salts and
water are detected.  (This geometry can also be visualized as a unique
hollow cylindrical structure with a lack of NaCl emission in the equatorial
waist.). A value of $i$=30-35\degr\ has been adopted for the
line-of-sight inclination of the tori's equatorial plane, meaning that the tori are
roughly orthogonal to the bipolar nebula, as observed in the plane of
the sky. The model-data fitting process indeed rules out values of the inclinations departing by more than $\pm$5\degr\ from this range.  

The continuum emission source at the nebula center has been
approximated by an elliptical Gaussian of radius
26\,mas$\times$18\,mas and a characteristic brightness temperature of
$\sim$1200\,K, as inferred from a 2d Gaussian fit to the continuum maps at
260\,GHz, i.e., the spectral window in which \naclvd\ is observed.
The continuum has been included in the model with the only
purpose of reproducing the weak line absorption observed at \vlsr$\sim$32\,\kms, 
which is partially responsible for the emission dip at the center of the integrated intensity maps of the NaCl-stacked data.
For simplicity, the continuum source has been placed at the center of the
NaCl-rotating structure. For an optically thin dust continuum source that is relatively compact compared with the line-emitting volume, the
model predictions are expected to be adequate for our purposes.
A proper radiative transfer model of the dust continuum disk is out of
the scope of this paper and will be presented elsewhere.

\begin{figure*}[ht]
     \centering
           \includegraphics[width=0.4\hsize]{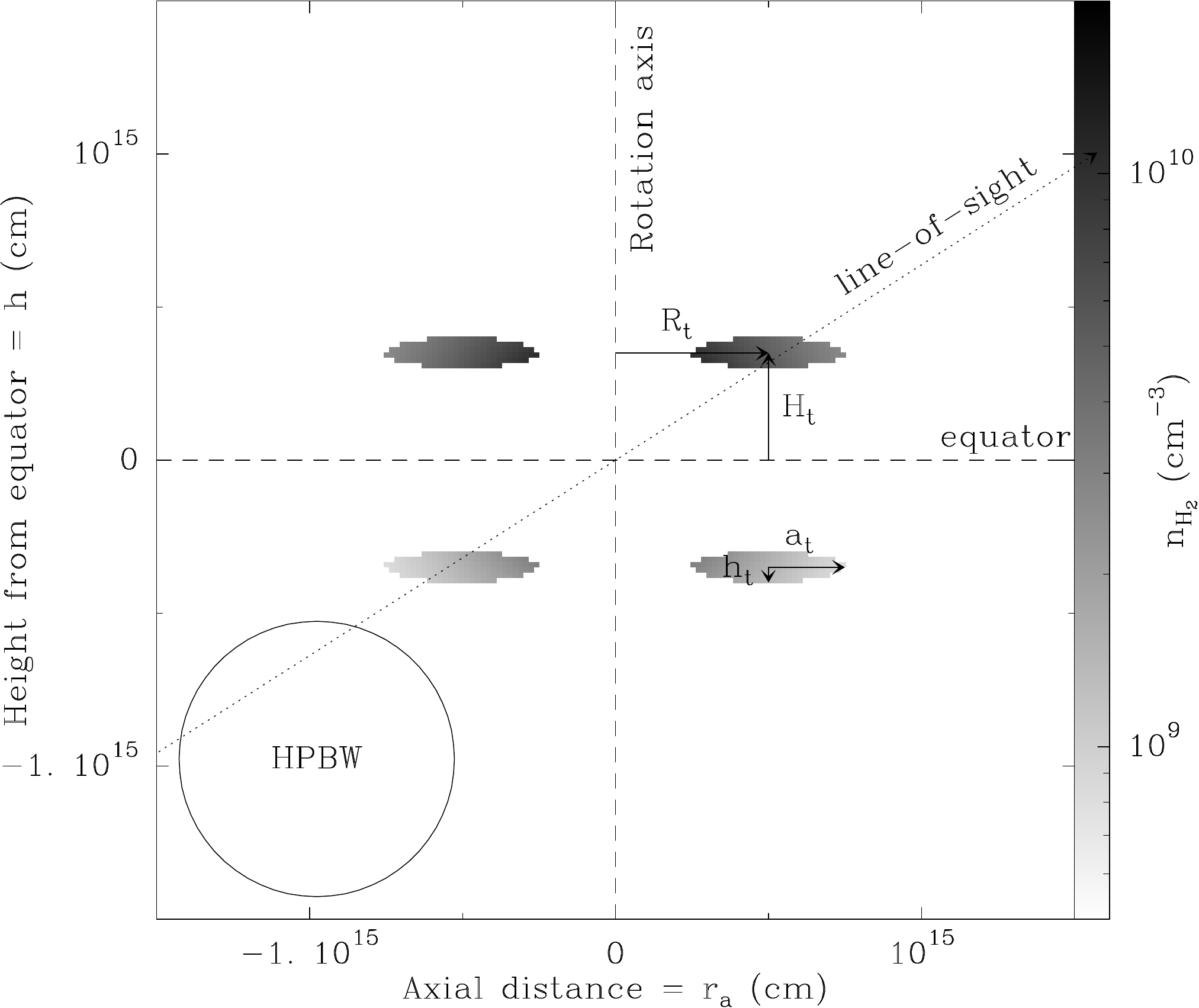} \\
     \includegraphics[width=0.33\hsize]{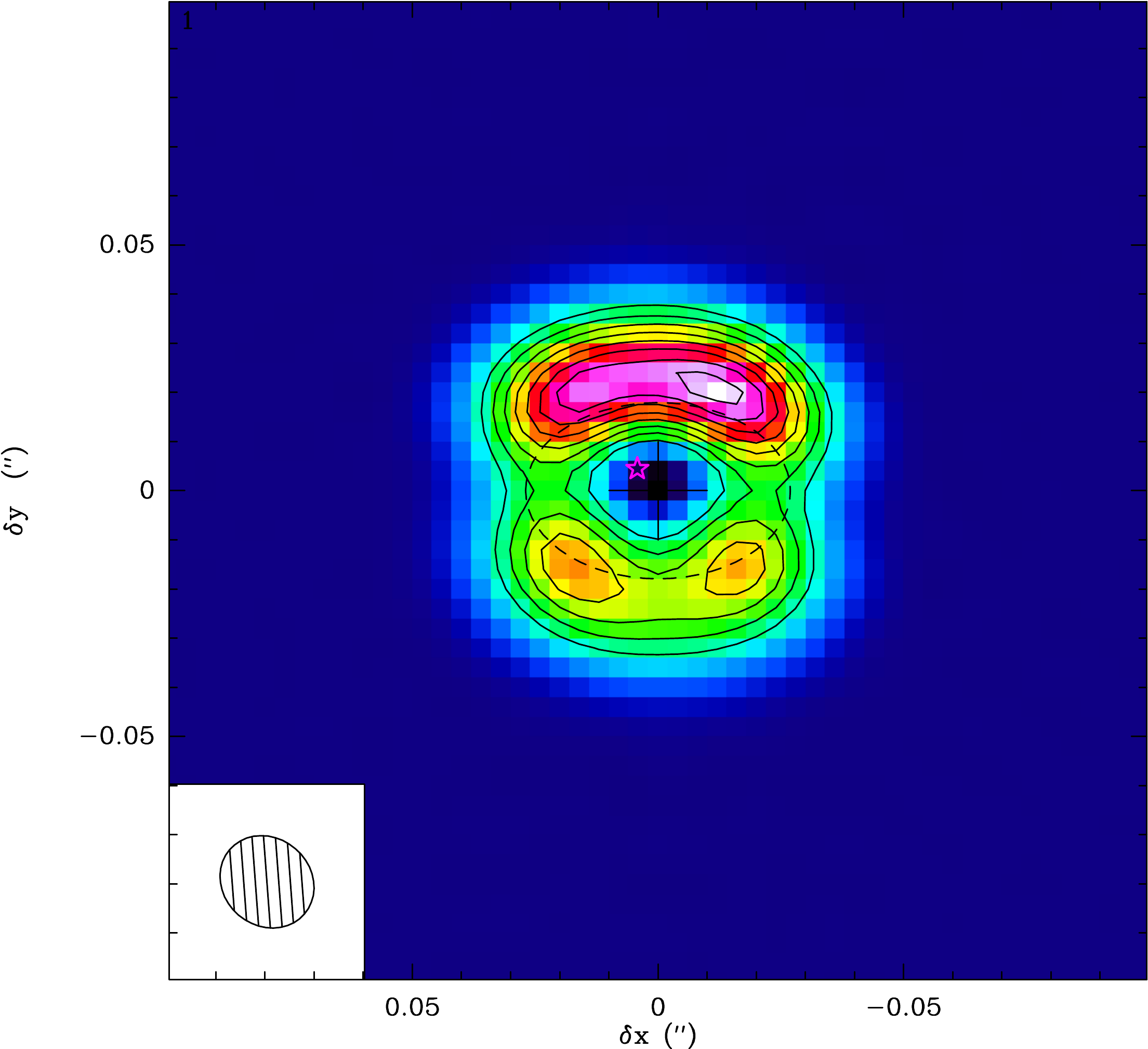}
     \includegraphics[width=0.33\hsize]{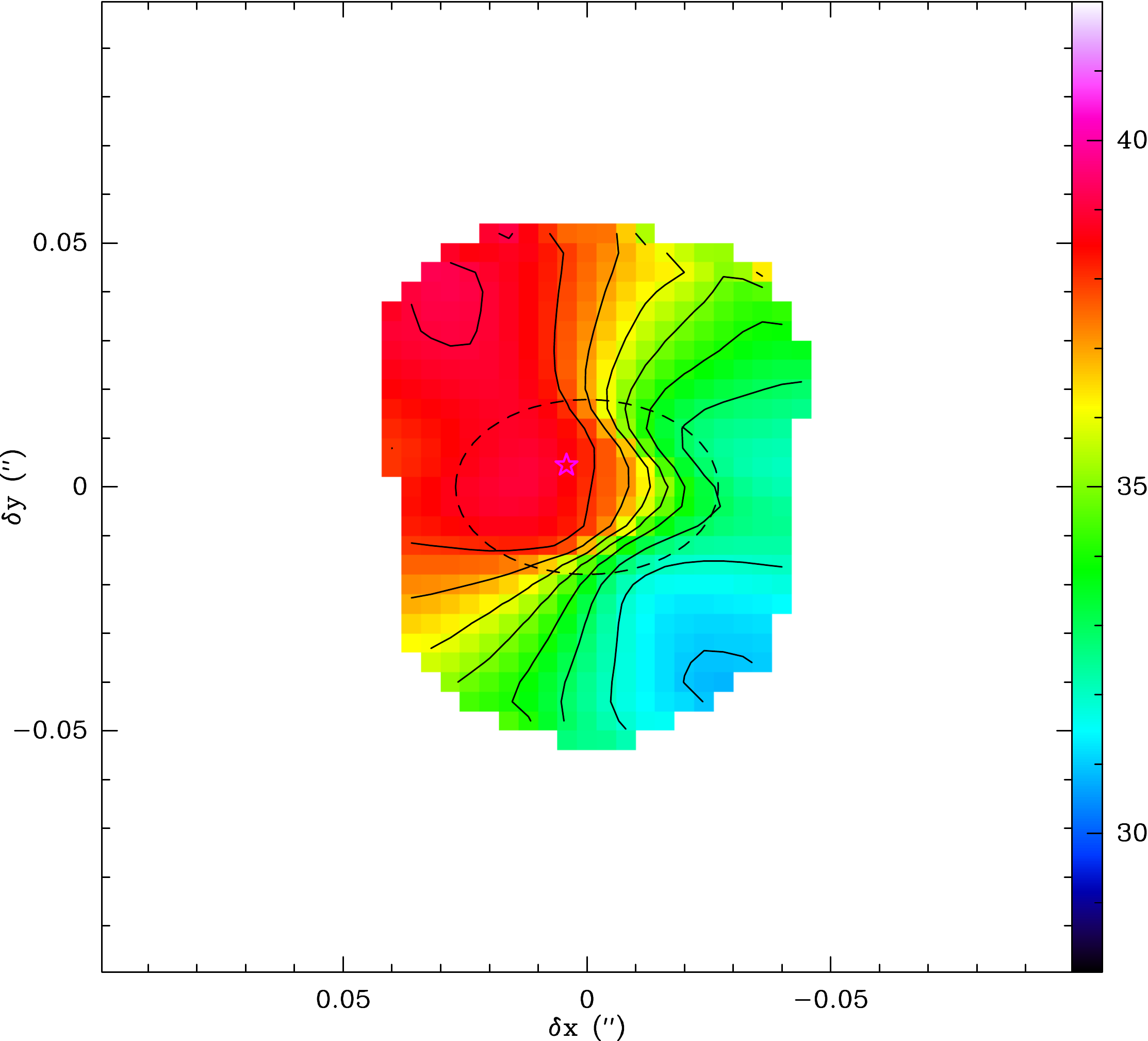}
     \includegraphics[width=0.30\hsize]{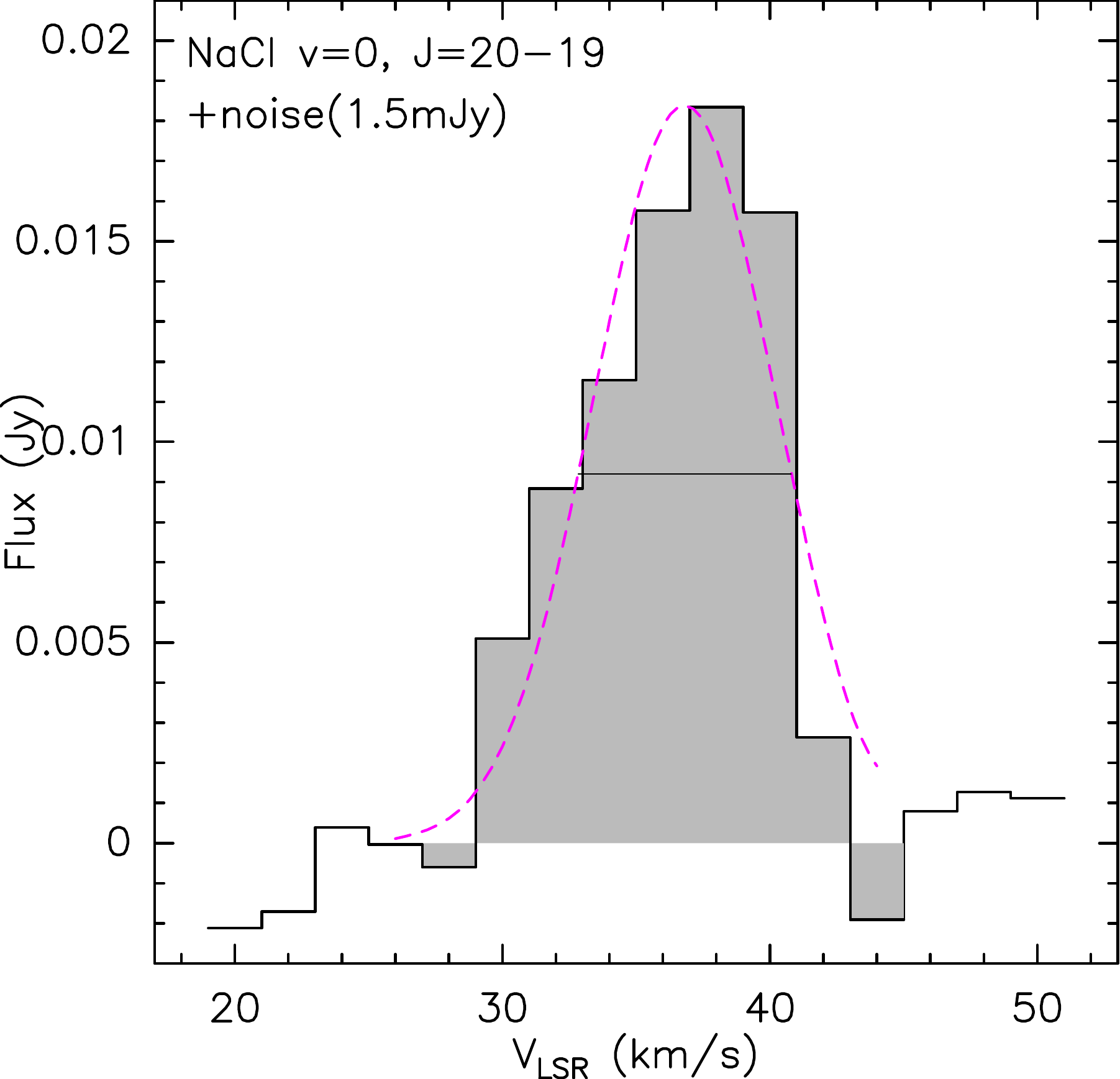} \\ 
     \caption{{\bf Top)} Schematic representation of the geometry and density distribution of 
       the NaCl-emitting region adopted in our model (\S\,\ref{s-model} and Table\,\ref{t-model}).
       The plot is a cut of this region by a plane that contains both the line of
       sight and the rotation axis of the nebula. The later is oriented in the sky along PA$\sim$25\degr. 
       The linear size of the
       20\,mas beam (at $d$=1500\,pc) of our ALMA NaCl maps is shown
       in the lower-left corner.  {\bf Bottom)} Synthetic NaCl ALMA
       data from our LTE model  plotted as in the bottom panels of
       Fig.\,\ref{f-nacl}. The synthetic cubes of the \naclvd\ transition are shown in Fig.\,\ref{f-nacl51}.}
         \label{f-naclmod}
   \end{figure*}

A sketch of the geometry adopted in the NaCl model, as well as some views of the synthetic data, are shown
in Fig.\,\ref{f-naclmod} (top and bottom, respectively). The
parameters of the best-fit model are given in Table\,\ref{t-model}.
%
\begin{table*}[h!]
\small
\caption{Model parameters used to reproduce the NaCl observations of \oh\ (\S\,\ref{s-model}). See Figs.\,\ref{f-naclmod} and \ref{f-nacl51}.}
\label{t-model}      
\centering                          
\begin{tabular}{l c}        %
  \hline\hline                      
Parameter & value \\ 
\hline
Distance ($d$) & 1500 pc \\ 
LSR Systemic velocity (\vsys) & $+$35\,\kms  \\ 
Inclination ($i$) & 32.5\degr\ \\ 
Radius of the tori ($R_{\rm t}$) & 5\ex{14}\,cm \\
Height of the tori ($H_{\rm t}$) & 3.5\ex{14}\,cm \\ 
Semi-Major axis of the tori's elliptical cross section  ($a_{\rm t}$) & 2.5\ex{14}\,cm \\ 
Semi-Minor axis of the tori's elliptical cross section ($h_{\rm t}$) &  5\ex{13}\,cm \\
Keplerian rotation (\vrot) & 4.0$\times$$\sqrt{40\,{\rm au}/r}$\,\kms\  \\
Equatorial expansion (\vexp) & 3.5$\times$($r_{\rm a}$/40 {\rm au})\,\kms\ \\
Turbulent velocity (\vtrb) & 2\,\kms\ \\
Molecular hydrogen density (\dens) & 5\ex{9}($\frac{r}{40 {\rm au}}$)$^{-2.0}$\,\cm3 \\
Gas kinetic temperature (\tkin) & 450($\frac{r}{40 {\rm au}}$)$^{-0.6}$ K \\
NaCl-to-H$_2$ abundance (X(NaCl)) & 5\ex{-9} \\ 
\hline 
\end{tabular} \\
\tablefoot{In the table, $r$ is the radial distance to the center and 
  $r_{\rm a}$ is the axial distance to the rotation axis. 
  The
  inclination of the rotation axis of the tori ($i$) is measured with
  respect to the plane of the sky.}
\end{table*}
%


%

The overall kinematics deduced from the ALMA maps is reasonably well described
with a composite velocity field that includes rotation in the
equatorial plane at \vrot$\sim$4\,\kms\ (at a mean radial distance of $\sim$40\,au) plus expansion, 
with equatorial expansion leading to slightly better
data-model agreement than a radial velocity distribution.
The average equatorial expansion velocity on the best-fit model is \vexp$\sim$3\,\kms. 
Since the cross section of the tori is small compared with the angular
resolution of our data, a very accurate determination of the radial dependence of
the velocity is not possible. In particular, we are not able to
discern between Keplerian or sub-Keplerian rotation, although the
presence of expansion in these regions makes the later more plausible
based on the results obtained for a number of dpAGB objects with
spatially resolved rotating and expanding circumbinary disks.
A turbulent velocity of \vtrb$\sim$2\,\kms\ has been used in our model to reproduce the width of the line absorption feature at \vlsr$\sim$32\,\kms,
with values of \vtrb$\sim$1 and 3\,\kms\ giving in general less satisfactory results. 

Adopting a uniform temperature of \trot$\sim$400-500\,K, as deduced in
\S\,\ref{s-rd}, the densities of the model that best reproduce the
flux and line profile of the \naclvd\ transition are of a few\ex{9}\,\cm3, also in
agreement with the average densities derived from our analysis
presented in \S\,\ref{s-rd}. For a (perhaps) more realistic model, 
radial power-laws of \dens($r$)$\propto$$r^{-2}$ and $T(r)$$\propto$$r^{-0.6}$ 
have been chosen for the best-fit model, although there are not significant differences in the synthetic maps compared with those obtained using uniform values of
the density and temperature within the NaCl-emitting volume. 
In our simple model, we have adopted
larger densities (by a factor of 3) for the North surface
disk layer than for the South one to reproduce the larger
surface brightness of the former observed in the data. Note, however,
that the presence of a density contrast between the two layers of the
disk cannot be accurately determined since the 
NaCl brightness asymmetry could also reflect different excitation
conditions and/or different fractional NaCl abundances between the two
layers above and below the dust disk. High S/N maps of several NaCl
transitions would be needed for a more accurate description of the
density and temperature conditions in these regions.


As we can see in Fig.\,\ref{f-naclmod}, our model reproduces
reasonably well the surface brightness distribution of NaCl, including
the overall shape and size of the integrated intensity maps (and the
velocity-channel maps) as well as the position of the four brightness
peaks and the narrow equatorial waist. The model also produces a weak
line absorption against the continuum, i.e.\, toward the center, near
\vlsr$\sim$32\,\kms, consistent with the observations.
The predicted velocity
distribution and the integrated line profile are in fair agreement
with the data as well, with the data probably suggesting slightly
larger velocities in regions close to the disk midplane than the
model. We note however that the S/N of the NaCl maps at these
low-latitude regions is particularly low and that the angular
resolution is in any case moderate as to precisely describe the
kinematics, which may involve velocity (radial or latitudinal)
gradients, in these compact regions.

\section{The central binary system}
\label{binary}
   \begin{figure}[ht]
     \centering
           \includegraphics[width=0.9\hsize]{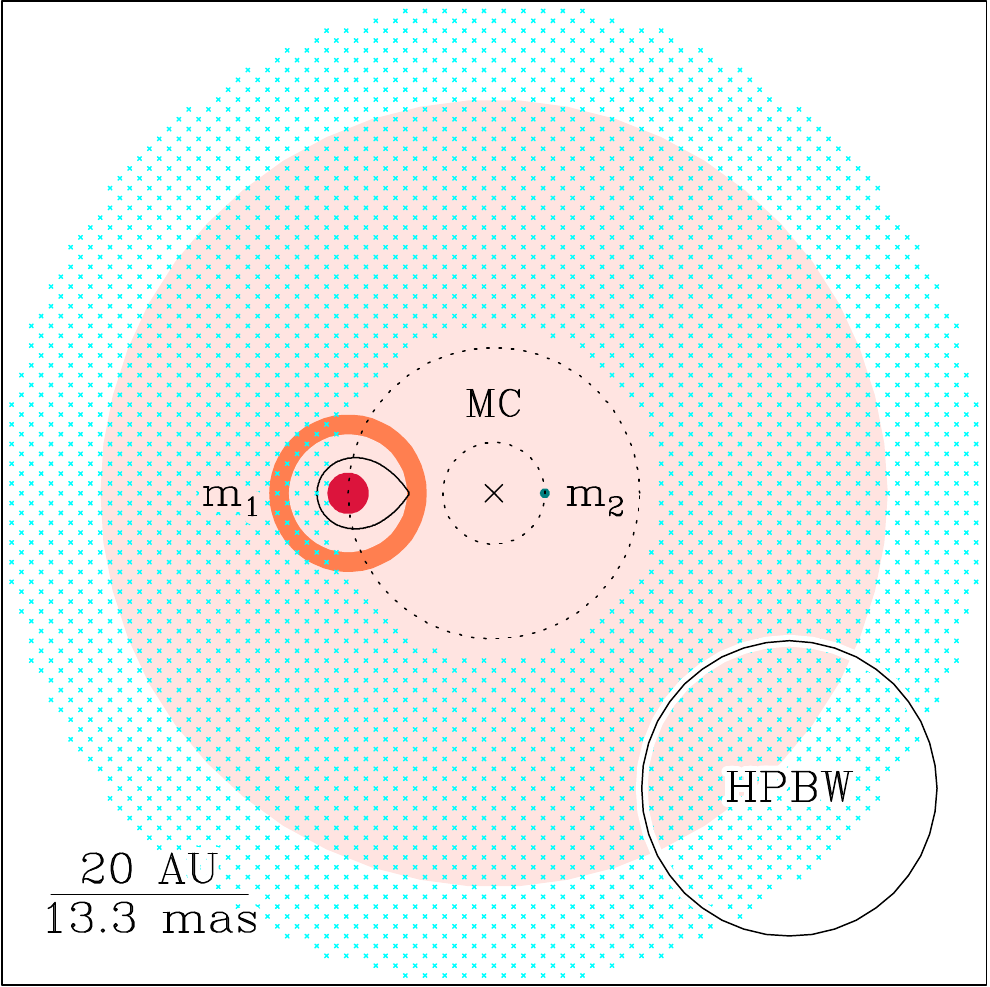} \\
     \caption{Sketch of the binary system and dust and salts
       components at the center of \ohs.  In this figure, the point of
       view is along the orbital axis of the binary system, i.e., the
       orbits are in the plane of the drawing.  The red crimson circle
       represent the primary, the Mira star \qx\ (m$_{\rm 1}$),
       showing its approximate size. The tale blue dot represents the
       location of the secondary (m$_{\rm 2}$), but the size is not to
       scale. The $\times$-symbol (MC) marks the location of the center of
       masses of the system. We have adopted a mass ratio $q$=m$_{\rm
         1}$/m$_{\rm 2}$ of 0.35. The two black dashed circumferences
       show the orbits of the two stars (we have assumed circular
       orbits for simplicity). The black ovoid shape around m$_{\rm
         1}$ shows the extent of its Roche lobe according to
       \cite{lea15}, assuming synchronous rotation, i.e.\,$p$=1 in
       their eq.\,2. The orange ring around m$_{\rm 1}$ shows the
       location of the SiO masers ($R_{\rm SiO}$$\sim$6\,AU) and the
       dust condensation zone (\rd$\sim$7.5\,AU) where hot dust is
       detected. The pale pink and cyan-crossed large circles show
       respectively the size of the (circumbinary) dust- and NaCl-disk
       detected. In the bottom-left corner, we include a scale in both
       physical (AU) and angular (mas) units for the adopted distance
       of 1500 pc. The size of the typical HPBW of these observations
       (20\,mas) is also shown in the bottom-right corner. }
         \label{f-binary}
   \end{figure}
%


To date, the orbital parameters of the binary system at the core of
\ohs, formed by the AGB star \qx\ and an A0 main-sequence companion
(\S\,\ref{intro}), have remained totally unconstrained. As shown in
\S\,\ref{res-cont}, our ALMA continuum maps spatially resolve a
disk-like structure at the center of \ohs\ and a point-like source
whose centers are not coincident, with offsets between the two along
the major and minor axis of the disk of about 4.5$\pm$0.3 and 4.9$\pm$0.2\,mas,
respectively, leading to an absolute offset of $\sim$6.6$\pm$0.3\,mas
($\sim$10$\pm$0.5\,au).  As already discussed, the point-like source marks the
position of the central AGB star \qx.

The relative offset between \qx\ and the center of the disk is
expected if the disk is circumbinary because in that case the disk's
centroid would coincide with the center of mass (CM) of the binary
system.  If this is the case, and assuming for simplicity a circular
binary orbit, the relative offset between the primary star \qx\ and
the CM (a$_{\rm 1}$) can be used to derive the orbital distance of the
secondary (a$_{\rm 2}$) as long as the stellar masses are known.
After deprojection of the distance between \qx\ and the disk plane,
considering a line-of-sight inclination for the disk plane of
$\sim$30\degr-35\degr, we deduce a value for a$_{\rm
  1}$=$\sqrt{(4.5^2+9.8^2)}$-$\sqrt{(4.5^2+8.5^2)}$=10.8-9.6\,mas$\sim$16.2-14.4\,au,
that is, a$_{\rm 1}$=15$\pm$1\,au.
Adopting a stellar mass for the companion and \qx\ of m$_{\rm
  2}$$\sim$2\,\msun\ and m$_{\rm 1}$$\sim$0.7\,\msun, respectively
(see Appendix \S\,\ref{mass-qx}), we deduce a value of a$_{\rm
  2}$$\sim$5\,au, which results in a total (a=a$_{\rm 1}$+a$_{\rm 2}$)
orbital separation of a$\sim$20\,au. A schematic representation
  of the central binary system and the dust and NaCl components found in this work
  is shown in Fig.\,\ref{f-binary}.

The orbital separation deduced for the system (a$\sim$20\,au) implies an orbital
period of $P_{\rm orb}$$\sim$55\,yr for the stellar masses adopted.
The long orbital period and orbital velocity of \qx, $V_{\rm
  1}$$\sim$8$\times$cos(30\degr-35\degr)$\sim$7\,\kms\ (after
deprojection) deduced are consistent with the variations of the
centroid of the SiO maser profile observed toward \ohs\ over the
course of more than two decades, with mean velocity offsets relative
to the systemic velocity ranging between $-$8 and $+$6\,\kms\ as
recently compiled by \cite{kim19} --- see their Fig.\,7d.

From the rotation velocity of the surface layers of the circumbinary
disk, \vrot$\sim$4\,\kms\ at a radial distance from the center of
$\sim$40\,au (Table\,\ref{t-model}), and adopting a purely Keplerian regime we deduce a value
for the central mass of $\sim$0.7\,\msun. This value of the mass is
clearly lower than the total mass of the binary system formed by the
mass-losing AGB star \qx\ and the A0 main-sequence companion
(m$_{\rm T}$=m$_{\rm 1}$+m$_{\rm 2}$$\sim$2.7\,\msun). As already mentioned in the previous section, the
expansive motions measured in the disk, with \vexp$\sim$3\,\kms,
suggest sub-Keplerian rotation in which case the value of the central
mass computed above has to be taken as a lower limit.

An upper limit to the central mass can be obtained from angular
momentum conservation considerations using the following expression:

$m_{\rm T}$$\sim$$m_{\rm Kepler}$$\times$$\frac{R_{\rm final}}{R_{\rm ini}}$, 

\noindent where $m_{\rm Kepler}$ is the central mass computed above in
a purely Keplerian regime, and $R_{\rm final}$ and $R_{\rm ini}$ are
the radial distances from the rotating gas to the central mass
observed at present ($\sim$40\,au) and when the disk started its
expansion, respectively. Assuming that the disk started its expansion
at the minimum radius that was possible, presumably $R_{\rm ini}$$\sim$a$_{\rm 1}$$\sim$15\,au, 
we derive an upper limit to the central mass of $\sim$2\,\msun. If we
take $R_{\rm ini}$ to be roughly half of the current orbital
separation ($\sim$10\,au), we derive an upper limit to the central
mass of $\sim$2.8\,\msun.  In spite of the vast uncertainties
associated to the simplified assumptions used and unknown value of
$R_{\rm ini}$, the crude values just obtained are close to the total mass
of the binary system inside \ohs.

A similar result is reached if we interpret that the gas we detect in
NaCl emission comes from a disk formed of gas ejected by \qx\ that is then initially expanding and corrotating with it (i.e.\,sharing
its orbital motion, $V_{\rm 1}$$\sim$8\,\kms\ at a$_{\rm
  1}$$\sim$15\,au). If we suppose that the velocity modulus decreases
with the distance to the disk axis, then we expect a rotation velocity of
8\,\kms$\times$$\frac{15\,\rm au}{30\,\rm au}$$\sim$\,4\,\kms\ in the NaCl-disk,
coincident to what is observed and, thus, in support of a central mass
of $\sim$2.7\,\msun.  However, in this case the forces cannot be
assumed to be exactly radial and the law of angular momentum
conservation does not obviously hold.  The sole purpose of these rough
calculations is to show that the upper bound on the mass of the central
system is probably not much larger than a few \msun. 



From our model of the NaCl emission, i.e.\, the gas density
distribution and velocity field, we deduce a value for the angular
momentum of the rotating disk of J$_{\rm
  disk}$$>$0.24\,\msun\,au\,\kms\,($>$7.2\ex{42} kg\,m/s). The lower
limit arises because the NaCl emission does not probe the disk in its
entirety but only its surface layers and within a certain range of
radii. Obtaining a reliable estimate of the total angular momentum of
the circumbinary disk is not possible, given the unknown density and
velocity structure of the disk interior, but it is unlikely to be
orders of magnitude larger than the lower limit given above taking
into account the disk dimensions (inferred from the continuum data)
relative to the NaCl-emitting volume (see Fig.\,\ref{f-naclmod}). For
comparison, we note that the angular momentum of the
circumbinary disk of the Red Rectangle (a prototype dpAGB object), 
which is relatively massive ($M_{\rm disk}$$\sim$1.3\ex{-2}\,\msun) and
significantly larger (R$_{\rm disk}$$\sim$1500\,au) than the disk in
\ohs, is $\sim$9\,\msun\,au\,\kms\ \citep{buj16}. Therefore we believe
the angular momentum of the circumbinary disk in \ohs\ can be safely
expected to be at most J$_{\rm
  disk}$$\sim$few$\times$1\,\msun\,au\,\kms.  This value is much
smaller than the angular momentum contained at present within the orbit of the
binary system formed by \qx\ and the A0\,V companion, J$_{\rm
  orb}$=$\mu$$\sqrt{Gm_{\rm
    T}a}$$\sim$115\msun\,au\,\kms\ (=3.4\ex{45}kg\,m$^2$/s), assuming
a circular orbit for simplicity.  This implies that a small decrease
in the orbital separation (of less than a few \%) in the past could
have accounted for the angular momentum of the circumbinary disk
(under the reasonable assumption that the binary imparted 
angular momentum to the circumbinary disk).






\section{Discussion}
\label{discussion}

\subsection{The locus and extremely slow expansion of the NaCl-emitting layers} 
\label{dis-nacl}

It is well known that in normal AGB stars NaCl forms by equilibrium
chemistry near the stellar photosphere and that this species, given its refractory character, 
disappears from the gas phase rapidly as it gets incorporated into 
dust grains \citep{mil07,mau10}.
In \ohs, we observe NaCl on the surface layers of the dust disk, that
is, beyond a region where dust has already formed massively and where,
in principle, NaCl should then be significantly gas depleted as a
consequence of condensation onto the grains.  At the base of the \sso,
the NaCl emission is found to be co-spatial with that of SiO
(\S\,\ref{res-siov1} and Fig.\,\ref{f-siov1}), which is a well known
shock tracer \citep{gin19}. This suggests that shocks are
probably efficiently extracting not only SiO but also NaCl from grains
and returning it to the gas phase in these regions.
We note that NaCl (or SiO) is not observed in the disk midplane, which could indicate
that this molecule is absent (significantly gas depleted) in these
dense and dusty, low-latitude regions that are most likely unaffected
by shocks or, alternatively, that the emission is extremely faint due
to excitation and radiative transfer effects (for example if the dust
and gas temperatures are similar).
In contrast to SiO, which is observed at high elevations throughout
the \sso, the salts emission is constrained to $\sim$20\,au above and
below the disk midplane.  This is expected given the large dipole
moment of NaCl and KCl, which results in these rare (low abundance)
species being extremely good/selective tracers of high-density regions
in contrast to the more abundant SiO, which is a better tracer of
low-density regions \citep[e.g.][]{quin16}.

We believe that, although it is not impossible, processes like
sublimation of NaCl produced by the stellar radiation of the companion
or thermal desorption from the grain surfaces \cite[e.g.\,][]{gin19}
have some difficulties in explaining satisfactorily the locus of the
salts in the outer layers of the disk (above $\sim$$\pm$20\,AU the
midplane) since both processes are expected to have notable effects
also in the inner regions of the disk, as the stellar radiation
propagates inside out heating up the dust throughout the disk to its
outer regions where the salts are observed.

Another surprising result from this work is the
extremely low expansion velocity (\vexp$\sim$3\,\kms) measured at the
surface layers of the rotating disk, at relatively large radial distances
from the center of $\sim$40\,au$\sim$6\ex{14}cm$\sim$20\,\rs.  At these
distant regions, clearly beyond the massive dust condensation (wind acceleration) zone, 
the wind velocity of a normal O-rich AGB star should be close to the
terminal velocity \cite[see e.g.\,][]{dec10}, that is, close to
$\sim$15-25\,\kms\ for a high-mass loss rate AGB star like \qx. The very
low expansion velocity observed around \qx\ implies an extremely slow
acceleration of the wind within $\sim$40\,au resulting indeed in the largest value of the
index $\beta$$\sim$8, for a classical $\beta$-wind velocity
profile $V(r)$\,$\propto(1-\rs/r)^\beta$, measured to date for
an AGB star \citep[][measured
$\beta$=5 for W\,Hydrae]{kho14}. 
Under the hypothesis of dust-driven wind, as for normal AGB stars, the
inefficient wind acceleration in the close environment of \qx\ could
be caused by the presence of very large grains or dust species that are
inefficient as wind drivers.



\subsection{Formation of an equatorial density enhanced (EDE) structure}
\label{dis-wrlof}


Mass-transfer from the AGB star \qx\ to the main-sequence companion is
a promising mechanism for the shaping of the dense equatorial
rotating disk/torus discovered in this paper.  As we show in this
section, \ohs\ brings together several favorable conditions for
effective mass-transfer from QX Pup to the companion (at a separation
of $a$$\sim$20\,au) in the so-called wind Roche lobe overflow (WRLOF)
mode \citep{moh07}. In this mode, the AGB wind material fills the
Roche lobe of the giant (primary) and is transferred to the compact
component (secondary), ultimately resulting the compression of the
AGB wind on the orbital plane and in the subsequent formation of an
equatorial density enhanced (EDE) structure that
remains gravitationally bound to the binary system.

The mass-transfer efficiency of WRLOF has been explored by several authors using
hydrodynamical simulations for a variety of parameters, including
binary mass ratios, orbital separations and initial wind velocities
\citep[e.g.][]{jah05,chen17,sal19,elm20}. 
These simulations indicate that
the strength of the interaction and, thus, the final morphology and
the pole-to-equator density contrast of the companion-perturbed
outflow,
mainly depend on the ratio of the wind velocity to the orbital
velocity ($\eta$=\vexp/\vorb), the primary-to-secondary mass ratio
($q$=M$_{\rm 1}$/M$_{\rm 2}$), and the dust condensation radius
filling factor \cite[$f$=\rd/r$_{\rm L}$, where r$_{\rm L}$ is the Roche
lobe of the primary given by][]{egg83}.  As shown in
\S\,\ref{res-cont}, the dust condensation radius of \ohs\ is found to be of about
\rd$\sim$7.5\,au, that is, comparable to the Roche lobe radius,
r$_{\rm L}$$\sim$6\,au, for an orbital separation of $a$$\sim$20\,au
(\S\,\ref{binary}).
This together with the very low values of $\eta$$\sim$3/8$\sim$0.4
and $q$$\sim$0.35, make \ohs's central system very prone to effective
mass transfer from \qx\ to the companion and the eventual compression
of the AGB wind on the orbital plane to form an EDE, as shown e.g.\,in
the recent hydrodynamical modeling work by \cite{elm20} (see their Fig.\,3).

In the context of WRLOF mass-transfer, it is not well known how long EDEs last
once the donor AGB star stops losing mass. The presence of SiO masers
at the core of \oh\ indicates that \qx\ is still undergoing mass-loss
at present and, therefore, it is not possible to constrain the age of
the rotating circumbinary disk from these type of
considerations. However, it is reasonable to assume that the formation
of the rotating disk has required at least a few orbits
\citep[e.g.\,][]{mae21}, implying that the equatorial disk
could be $\gsim$200\,yr old (since $P_{\rm orb}$$\sim$55\,yr,
\S\,\ref{binary}). On the other hand, given the rotation velocity and
radius of the NaCl-disk, a particle on its surface completes a
rotation in about 220\,yr, which can also be taken as a
lower limit to the age of the rotating disk since the later is observed
to be complete.

Finally, we do not observe a central cavity in the dust equatorial disk traced by
the continuum emission, which (if exists) would be difficult to
discern and characterize due to the presence of the relatively bright
point-like continuum source at the center (\S\,\ref{res-cont}), and the (limited) angular resolution of the data.
However, this is not inconsistent with the predictions by WRLOF binary
interaction/mass-transfer models, which in general
do not predict dense equatorial tori cleanly
detached from the central binary system but complex density
distributions also in regions interior to the binary orbit
for cases in that the mass loss is still ongoing \citep[e.g.][]{elm20,chen17}. Higher angular resolution observations
of the central disk of \ohs\ to be compared with WRLOF binary
interaction hydrodynamical simulations specifically accounting for the properties of
\ohs\ and its central binary are needed to further constrain the
formation history of the rotating disk discovered in this work.

\subsection{The \sso}
\label{dis-sso}

In principle, given its low-velocity and wide opening angle, the
\sso\ could simply result from the confinement of the on-going AGB
wind from \qx\ by the dense equatorial torus: the underlying stellar
wind escapes through the low-density polar regions but is impeded
along the dense equator of the companion-perturbed AGB-wind
environment (as in the so-called Generalized Interacting Stellar Winds
scenario of PN-shaping, \cite{bal87}).  There are, however, some
properties of the \sso\ that are not well understood in the simple
scenario of a confined dust-driven AGB wind, for example, the
radial velocity steady increase with the distance to the center (up to $\sim$16\,\kms\ at $r$$\sim$350\,au) and,
most importantly, the start of the wind acceleration beyond the region of massive dust formation (\S\,\ref{res-sissio}).
These properties suggest that we may be witnessing the active
acceleration and shaping process of the primary's stellar wind
(probably partially perturbed already by the companion) as it escapes
through the poles of the circumbinary disk.


If this is the case, our data indicate that such acceleration process
is acting on linear scales of up to $\sim$350\,au, which is the region
where the $\vexp$$\propto$$r$ kinematic pattern is observed. Beyond
this point, the \sso\ reaches its terminal velocity ($\sim$16\,\kms),
suggesting that the acceleration mechanism is not active anymore or
that the outward acceleration force is fully compensated by the
ambient ram pressure. Under the hypothesis of a stellar wind that is
currently being accelerated at a constant rate, 
the
kinematic age of the \sso\ can be computed as
2$\times$350\,au/16\,kms$\sim$200\,yr, or equivalently, using the
inverse of the deprojected velocity gradient observed at its base as
\tdyn$\sim$2$\times$1/$\nabla$v$\sim$2/65\,\kms\,arcsec$^{-1}$$\sim$220\,yr
(\S\,\ref{res-sisv0}).


Alternatively, if we assume that the $\vexp$$\propto$$r$ kinematic
pattern is the result of ballistic (self-similar) expansion after a
short acceleration (outburst-like) event that has already ended, the age of the
\sso\ would then be \tdyn$\sim$1/$\nabla$v$\sim$110\,yr, and the duration of the acceleration
burst event itself would then have been a small fraction of this, i.e.\, perhaps as short as a few years or a decade.

As already discussed in \san, there are several evidences showing that the \sso, in any case, 
is running into and carving out pre-existing circumstellar material, 
for example: the presence of shocks, demonstrated by the selective
SiO emission from this component, as well as the dense-walled
structure and rounded tips of its lobes, which naturally arise in a
two-wind interaction scenario but are otherwise difficult to justify in case of a purely {\sl pristine} (i.e.\,largely unaffected by wind interaction) stellar
wind \citep[see e.g.\, hydrodynamical simulations by][]{bal17}. 

Our data show that the base of the \sso\ and the surface layers of the
disk (traced by the salts and water) overlap spatially and, in
consequence, share a similar (rotation and expansion) kinematics. This
could indicate that the \sso\ is a disk wind, i.e.\,a wind launched
from the surface layers of the circumbinary disk, or, alternatively, that the confined
stellar wind is dragging along some of the material in the inner edge
of the disk as it escapes through the low-density poles.  We believe
that the second hypothesis is more probable since putative disk-winds
found to date in some dpAGBs objects are significantly more tenuous
than the circumbinary disks from which they emerge and are constrained
to low latitudes \citep[][and references therein]{gall21}, in contrast to the \sso\ in
\ohs.


In a future publication we will carry out a deeper investigation and
discussion of the origin of the \sso, which requires a comprehensive
analysis (by radiative transfer modeling) of multiple transitions of
SiO to precisely determine its physical properties, dynamics and
mass-loss rate.  These are crucial parameters to determine, e.g., if
the \sso\ is consistent with acceleration due to radiation pressure or if, on the contrary, its linear momentum is far too large to
be explained by this mechanism, as observed in the large-scale
molecular outflow of \ohs\ and many wpPNe (\S\,\ref{intro}). 


Regardless of the precise origin of the \sso, which remains to be
determined, it is clear from the absence of fast ejections
\cite[together with the lack of classical accretion indicators, such
  as H$\alpha$ emission from the nucleus,e.g.][]{san04} that high-rate
accretion and wind launching by a compact object is not happening at
present: the low-velocity of the \sso\ unequivocally indicates that
the wind is not launched by a compact object because wind ejection
speeds are typically of the same order of, but larger than, the escape
velocity of the ejector \citep[see e.g.\,][]{kwo07}. The situation was
clearly different $\sim$800\,yr ago, when the large-scale bipolar
nebula was shaped and accelerated up to velocities of
$\approx$400\,\kms, a process that necessarily required accretion on
(and jet-launching from) a compact main-sequence companion
\citep{san04}. The marked differences between the \sso\ and the fast
large-scale lobes indicate that the binaries interaction mode and wind
shaping process have changed over the course of the evolution of \ohs.
Perhaps the accretion disk around the companion has been exhausted in
recent times. Alternatively, as it was proposed by \cite{san04},
\ohs\ could be at present in a low-rate accretion (or `quiescent')
state in which the disk around the companion is steadily building up
its mass but there is no effective disk-to-companion accretion (and no
jet launching). Indeed, other astrophysical systems, like FU Ori
objects and symbiotic stars \citep[][and references there in]{har96},
are known to experience a similar evolution, going through alternating
accretion outburst and post-outburst (quiescent) states.
\section{Summary}
\label{summ}

We have mapped with unprecedented angular resolution (down to
$\sim$20\,mas$\approx$\,30\,au) the molecular line and dust continuum
emission from the central regions of the wind-prominent pPN
\oh\ (hereafter \ohs). Here we present the results from the dust
continuum emission and a selection of molecular transitions observed in
the range $\sim$216-261\,GHz with the {\sl Atacama Large
  Millimeter/submillimeter Array} (ALMA). We spatially resolve the
close stellar environment around the central AGB star \qx\ (\cs) and
the compact bipolar outflow that emerges from it (\sso). A major
result from this work is the discovery of a rotating circumbinary disk
that is selectively traced by NaCl, KCl, and \water. This is the first
time that equatorial rotation is reported in this object and, more
generally, in pPNe with massive bipolar outflows.  We summary the main
results from, and points addressed in, this work as follows:

\begin{itemize}

\item[-] The continuum emission from the compact region around
  \qx\ (\cs) is spatially resolved in two main components: an extended
  disk-like component, elongated in the direction perpendicular to the
  bipolar nebula, and an unresolved component. The properties of the
  extended component are consistent with a circular dusty disk of
  radius of $\sim$40\,au inclined $\lsim$40\mydeg\ with respect to the
  line of sight. The point-like continuum emission is consistent with
  being due to the stellar photosphere/radiosphere of \qx\ and to hot
  ($\sim$1400\,K) dust within a few stellar radii
  (\rd$\sim$7.5\,au). We observe a small offset ($\sim$6.6\,mas)
  between the centroid of the extended disk and the position of the
  point-like source.

\item[-] The continuum flux from \cs\ follows a \snu2 frequency
  dependence, both for the extended disk-like and the point-like
  component, probably due to emission by large
  (mm-sized) dust grains with flat emissivity at mm-wavelengths. Under
  this assumption, the dust mass is about
  \md$\sim$1.5\ex{-5}-1.5\ex{-4}\,\msun\ in the extended disk and
  \md$\sim$10$^{-6}$-10$^{-5}$\,\msun\ in the hot-dust region around
  \qx.

  
\item[-] We have detected a total of 8 different transitions of NaCl
  in different $\upsilonup$=0, 1, 2, and, tentatively 3, vibrational
  levels (Table\,\ref{tab:mols}), which have been combined to obtain a
  NaCl line-stacked emission cube. The NaCl emission is found to arise
  from the surface layers of the extended dust disk. These layers of
  the disk are in rotation, with the east (west) side receding from
  (approaching to) us.  The rotation velocity deduced from the NaCl
  maps is about \vrot$\sim$4\,\kms\ at a mean radial distance of
  $\sim$40\,au. In addition to rotation, there are also expansive
  motions with a strikingly low expansion speed of \vexp$\sim$3\,\kms.

 \item[-] The spatio-kinematics and physical conditions (density and
   temperature) of the circumbinary disk have been estimated from a
   rotational diagram analysis of the individual NaCl transitions
   (\S\,\ref{s-rd}) and from the comparison of the ALMA NaCl maps with
   the predictions of an LTE radiative transfer model
   (\S\,\ref{s-model}). The best-fit model parameters and synthetic
   maps are given in Table\,\ref{t-model} and
   Fig.\,\ref{f-naclmod}. We deduce an average rotational temperature
   of \trot$\sim$400-500\,K and H$_2$ densities of
   $\approx$10$^{9}$\,\cm3, resulting in a total mass of the disk's
   surface layers of about 2\ex{-3}\,\msun\ (for a fractional
   NaCl-to-H$_2$ abundance of $\sim$5\ex{-9}). The line-of-sight
   inclination of the disk's plane is constrained to values of
   $\sim$30-35\degr.
     
\item[-] In addition to NaCl, we have found two other molecular
  species that selectively trace the rotating equatorial structure at
  the core of \ohs, namely, potassium chloride (KCl, i.e.\,another
  salt) and water (\water). KCl is a new detection in this object and
  also represents the first detection of this molecule in an O-rich
  AGB CSE.

\item[-] The orbital separation of the central binary system of
  \ohs\ have been estimated to be $a$$\sim$20\,au, given the relative
  offset between the locus of \qx\ and the centroid of the extended
  dust disk, under the plausible hypothesis that the disk is
  circumbinary. We derive an orbital period of $P$$\sim$55\,yr.  From
  the, probably sub-Keplerian, rotation detected in the disk, we
  derive a lower limit to the present central mass of the binary
  system of $\sim$0.7\,\msun.
  
\item[-] The \sso, a compact ($\sim$1\s$\times$4\s) bipolar outflow that emerges from \cs, is traced by several rotational transitions in the
  $\upsilonup$=0 and $\upsilonup$=1 vibrational states of SiO and SiS
  (including some isotopologues). Here, we focus our study on the ALMA maps of the \sisdo\ and \siov\ transitions.  

\item[-] The lobes of the \sso\ have a conical geometry at their base,
  consistent with a wide opening angle ($\theta$$\sim$90\mydeg) wind,
  and a more rounded morphology at their tips. The \sso\ emerges from
  the surface layers of the rotating equatorial disk, where the
  \sisdo\ and NaCl emission partially overlap. We find a gradual
  outward acceleration of the gas along the lobes up to a terminal
  expansion speed of about $\sim$16\,\kms, reached at a radial
  distance of $\sim$350\,au. The radial expansion continues at
  constant velocity beyond this point. We have constrained the
  inclination of the \sso\ to values around $i$$\sim$30\mydeg\ with
  respect to the plane of the sky.

\item[-] The \siov\ 
  transition is selective probing the high-excitation regions of
  the \sso\ closer to the center (within $\sim$100\,au). The
  kinematics at the base of the \sso\ is predominantly expansive but
  the signature of rotation is also present, particularly in regions
  close to the equator (within $\pm$30\,au).

\item[-]
  Based on the presence of gas-phase SiO, a well known shock tracer,
  at the base of the \sso\ and partially overlapping with the
  NaCl-emitting regions, we believe that shocks are the main agents
  efficiently extracting both NaCl and SiO (and probably other
  refractory/ice species) from dust grains and returning it to the gas
  phase.

\item[-] The expansion velocity (\vexp$\sim$3\,\kms) at
  the surface layers of the rotating disk at
  $\sim$40\,au$\sim$6\ex{14}cm$\sim$20\,\rs\ is unexpectedly low.
  The reason for such an
  inefficient wind acceleration in \ohs\ is unknown.

\item[-] The circumbinary disk in \ohs\ probably results
  from wind Roche Lobe OverFlow (WRLOF). Indeed, \ohs\ brings
  together several favorable conditions for the formation of a dense
  equatorial structure under this scenario, namely: an extremely low
  expansion velocity (\vexp$\sim$3\,\kms), a very massive companion
  (q=$m_{\rm 1}$/$m_{2}$$\sim$0.4$<$1) and comparable sizes of the
  dust condensation radius (\rd$\sim$7.5\,au) and the Roche lobe
  radius (r$_{\rm L}$$\sim$6\,au).

\item[-] The angular momentum of the NaCl-emitting surface layers of
  the circumbinary disk of \ohs\ is found to be J$_{\rm
    disk}$$\sim$0.24\,\msun\,au\,\kms\,(7.2\ex{42} kg\,m/s).  The
  angular momentum of the circumbinary disk in its entreaty is
  probably at most J$_{\rm disk}$$\sim$few$\times$1\,\msun\,au\,\kms,
  which is two orders of magnitude lower than the angular momentum
  contained at present within the orbit of the central binary. This implies that a few per cent decrease in
  the orbital separation can account for the angular momentum of the circumbinary disk.
    
\item[-] The origin of the \sso\ is unclear.
  The \sso\ could represent the on-going companion-perturbed
  AGB wind escaping through the low-density poles of the circumbinary
  disk and running into the surrounding material.
  The \vexp$\propto$$r$ kinematic pattern observed may indicate that the
  \sso\ is undergoing acceleration (by a yet unknown mechanism) at linear scales of $\lsim$350\,au.

\item[-] The age of the \sso\ is between $\sim$100\,yr (assuming
  that is ballistically expanding) and $\sim$200\,yr (in the case of
  constant acceleration), to be compared with the $\sim$800\,yr age of
  the large-scale CO outflow.
  A lower  limit to the age of the rotating circumbinary disk of $\sim$200\,yr
  is deduced.

\item[-] Contrary to what probably happened $\sim$800 years ago, when
  the fast ($\approx$100\,\kms) large-scale bipolar lobes of
  \ohs\ where shaped, high-rate accretion and wind launching by a
  compact object is most likely not taking place at
  present. Therefore, the binary interaction mode and/or wind shaping
  process seem to have changed over the course of the evolution of
  this object.

\end{itemize}

\begin{acknowledgements}
We thank the anonymous referee for very useful suggestions.  This
paper makes use of the following ALMA data:
ADS/JAO.ALMA\#2017.1.00706.  ALMA is a partnership of ESO
(representing its member states), NSF (USA) and NINS (Japan), together
with NRC (Canada), NSC and ASIAA (Taiwan), and KASI (Republic of
Korea), in cooperation with the Republic of Chile. The Joint ALMA
Observatory is operated by ESO, auI/NRAO and NAOJ.  The data here
presented have been reduced by CASA (ALMA default calibration
software; {\tt https://casa.nrao.edu}); data analysis was made using
the GILDAS software ({\tt http://www.iram.fr/IRAMFR/GILDAS)}. This
work is part of the I+D+i projects PID2019-105203GB-C22,
PID2019-105203GB-C21, and PID2020-117034RJ-I00 funded by the Spanish
MCIN/AEI/10.13039/501100011033.  This research has made use of the JPL
Molecular Spectroscopy catalog, The Cologne Database for Molecular
Spectroscopy, the SIMBAD database operated at CDS (Strasbourg,
France), the NASA’s Astrophysics Data System and Aladin.

\end{acknowledgements}

%
%

\begin{appendix}
  \section{Additional figures}
  \label{extrafigs}
   \begin{figure*}[h!]
     \centering
     \includegraphics*[bb= 1 1 635 560, width=0.30\hsize]{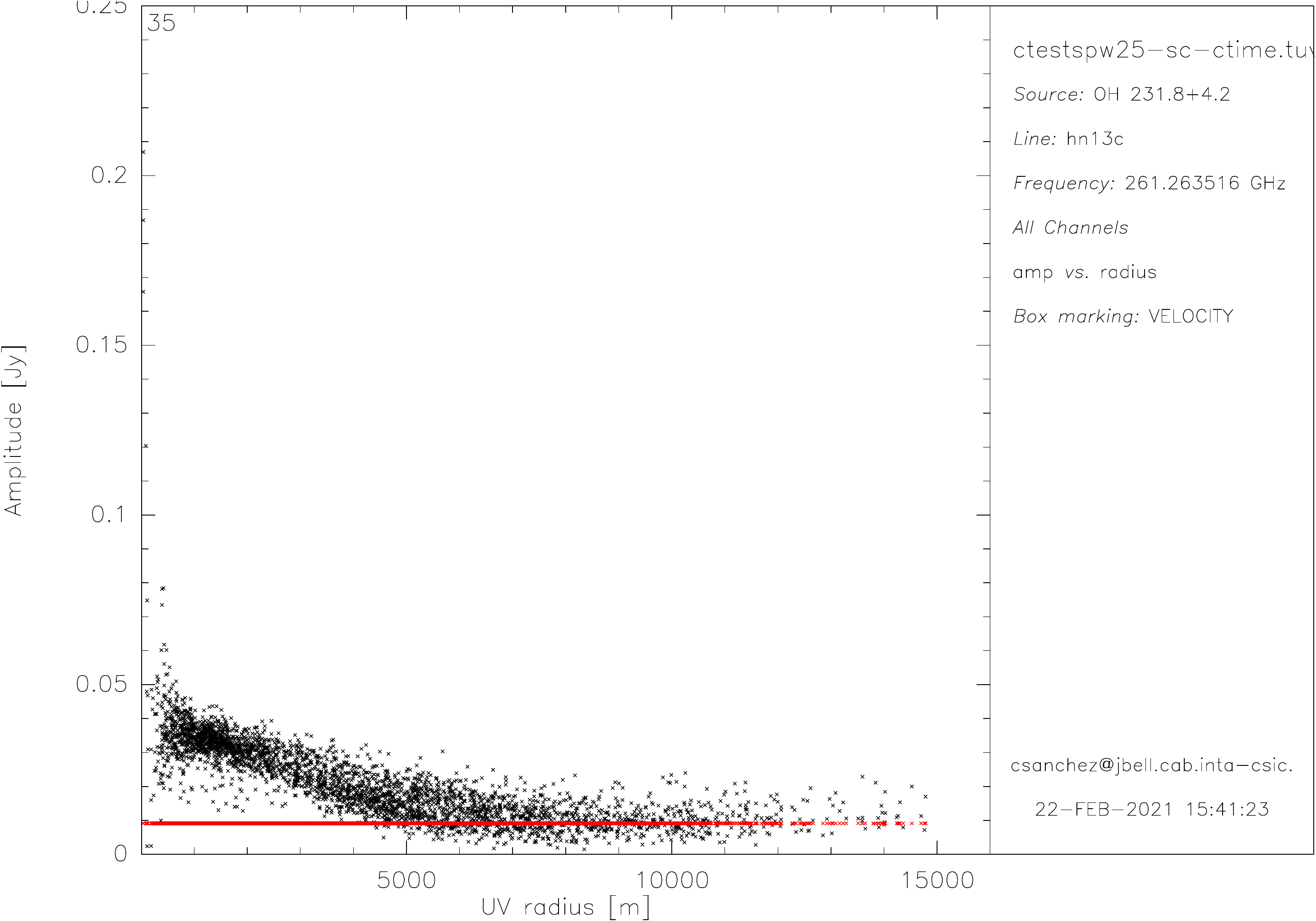}
     \includegraphics*[bb= 1 1 635 560, width=0.30\hsize]{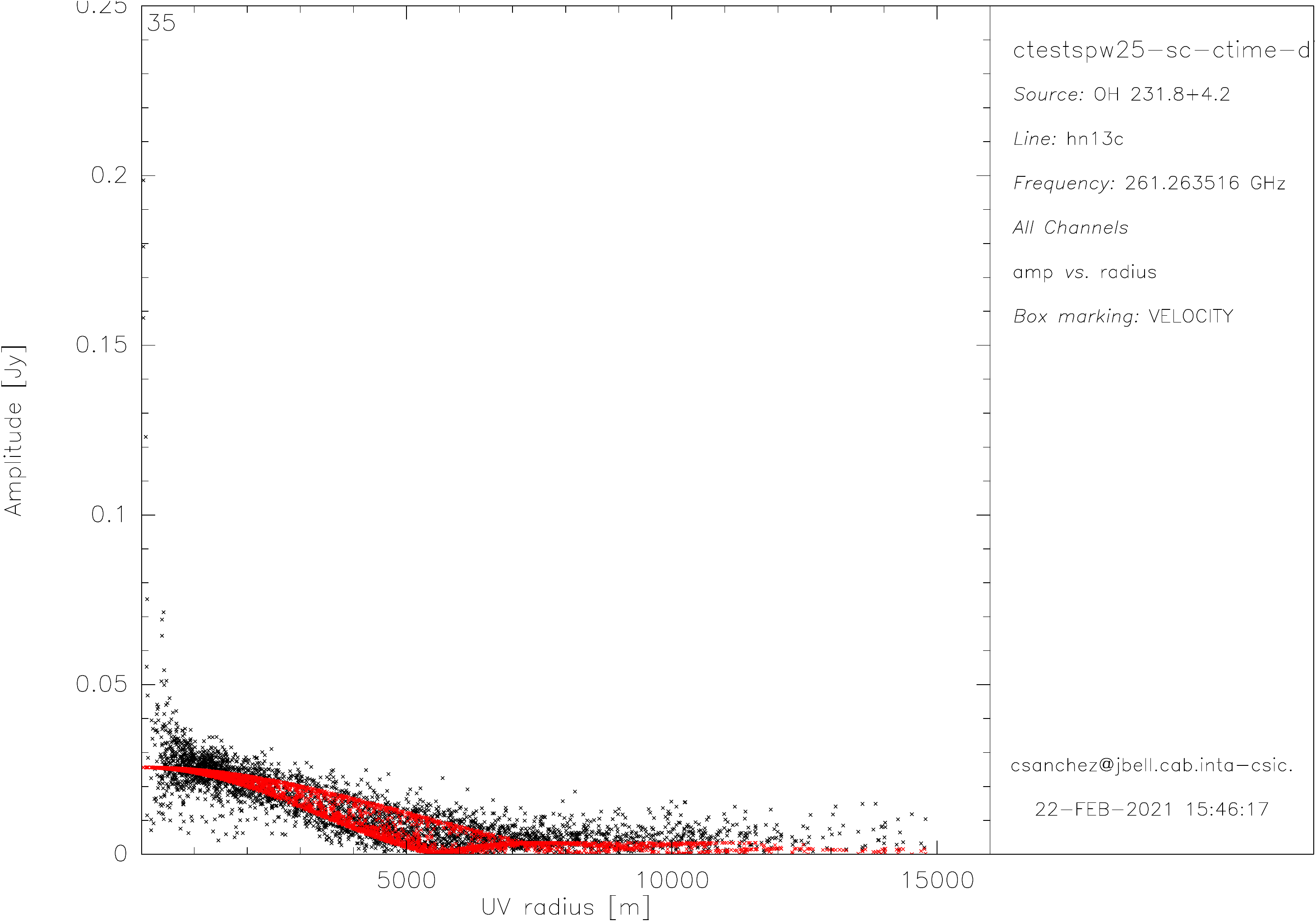}
     \includegraphics*[width=0.33\hsize]{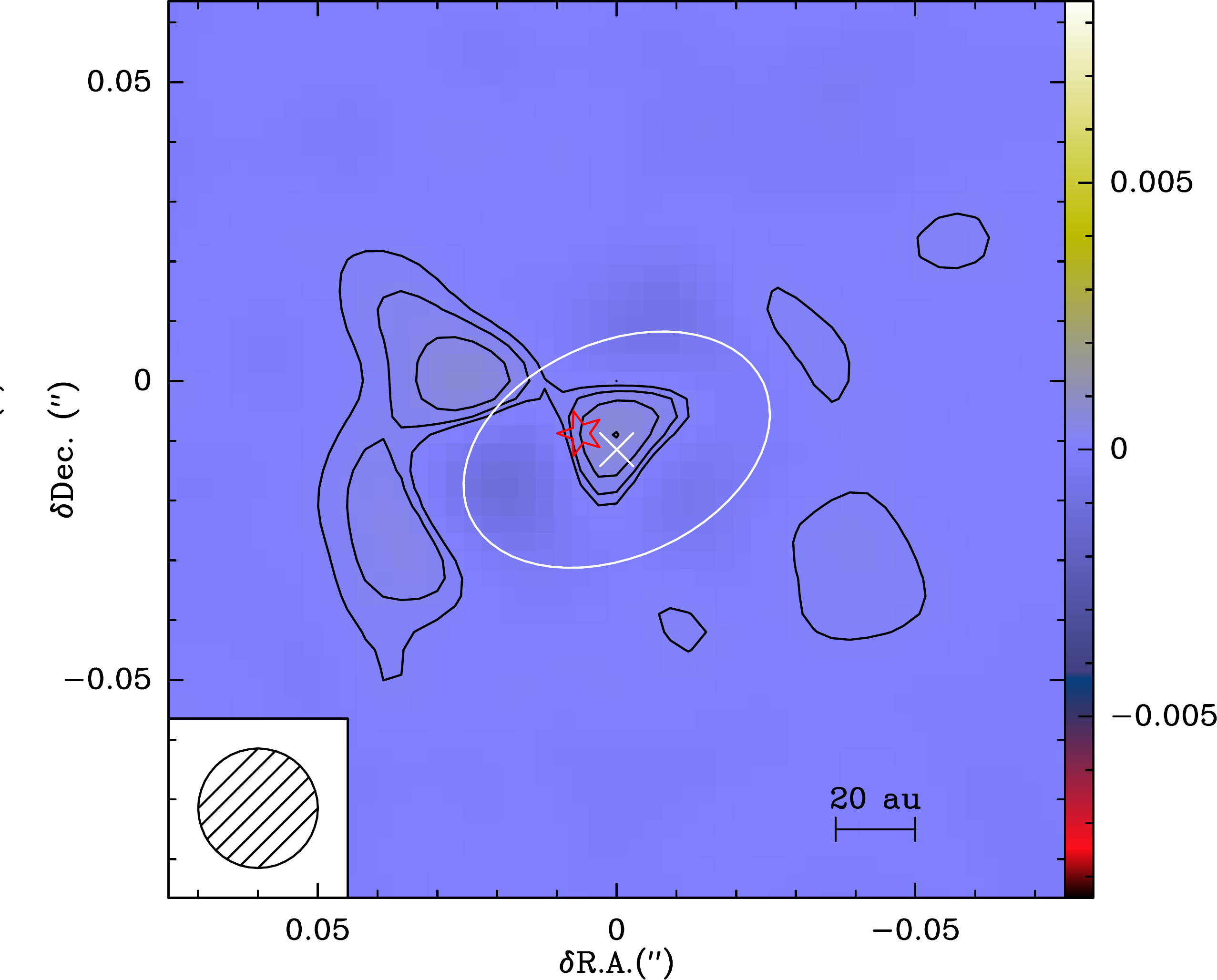}
  \caption{{\bf Left)} Visibity amplitude (Jy) vs.\ the antenna spacing in the $uv$ plane (m) of
    the continuum emission at 261.3\,GHz
    (selfcalibrated data)
    --- black --- and model fit of a point source (flux=
  9.1$\pm$0.1\,mJy) --- red.  {\bf Middle)} Residuals of the 
  continuum visibility amplitude after subtraction of the point source
  model in the top pannel --- black --- and new model fit of the residuals adopting
  a uniform elliptical disc model (flux= 25.6$\pm$0.2\,mJy) --- red. See
  details of the model fit parameters of both (point and elliptical
  disc) components in Sect.\,\ref{res-cont}. In both panels, the
  $uv$-data have been averaged over a time interval of 4500\,s
  (approximately the total on-source integration time in each
  observing block) to reduce the noise in these plots. {\bf Right)} Cleaned map of the residual after subtracting the point source and extended disk model to the observed 261\,GHz-continuum map shown in Fig.\,\ref{f-cont} (left), using the same contour level step.}
         \label{f-uvfit}
   \end{figure*}
%
\FloatBarrier

   \begin{figure*}[h!]
     \centering
     \includegraphics*[width=0.95\hsize]{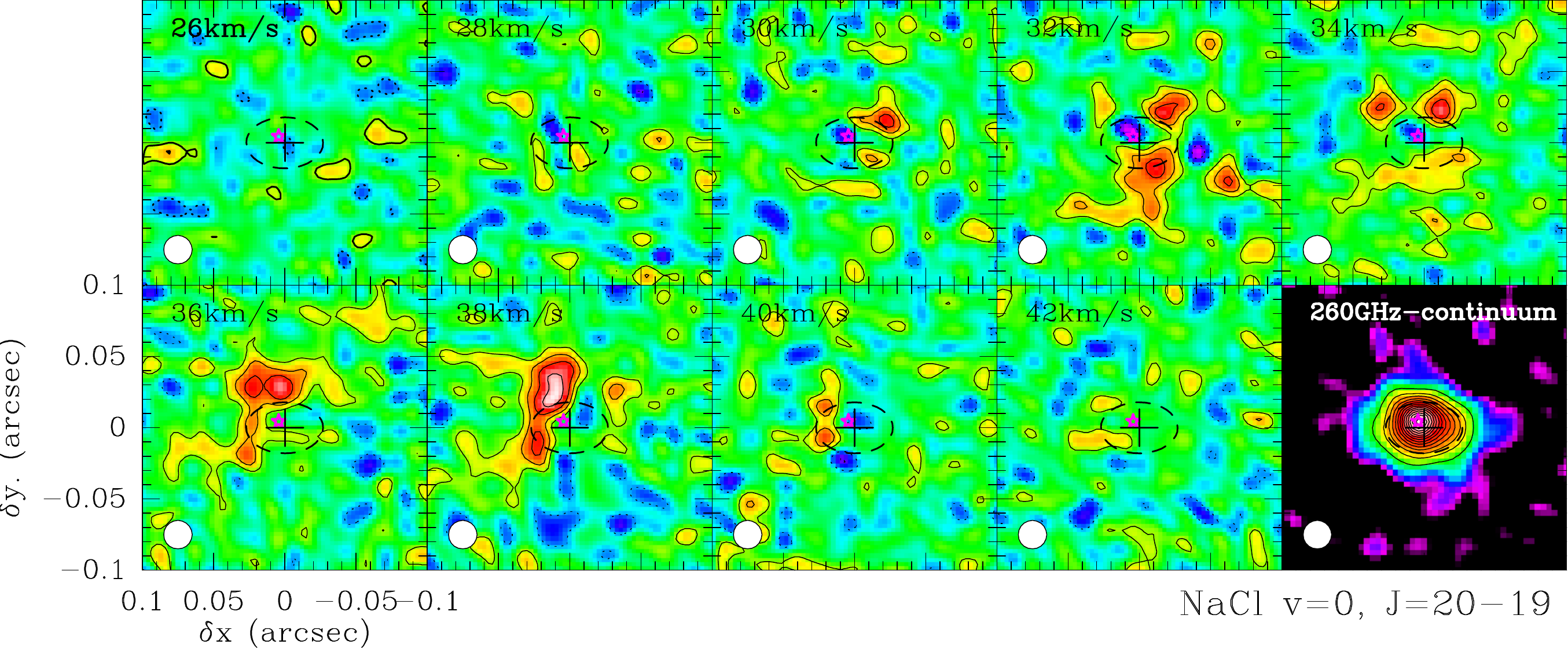}  
     \includegraphics*[width=0.95\hsize]{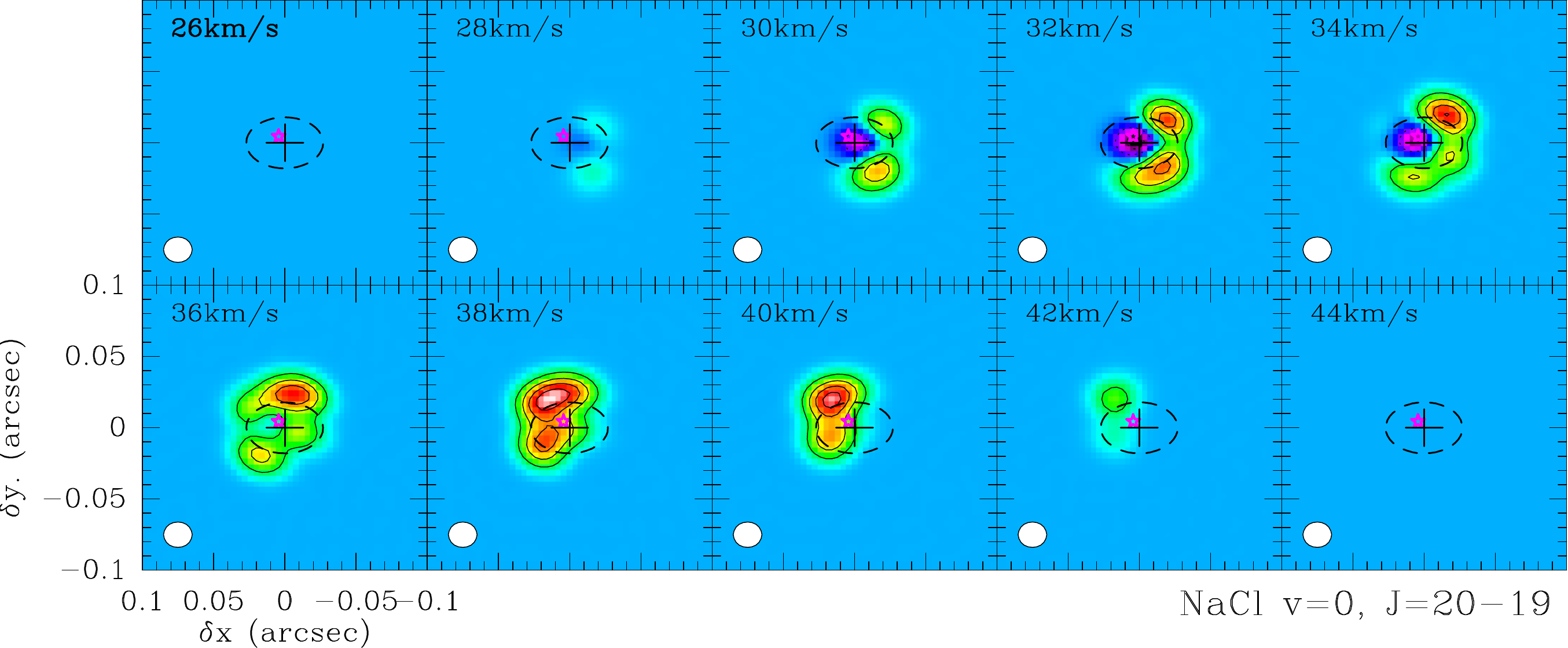}
     \caption{{\bf Top)} ALMA velocity-channel maps of the \naclvd\ transition; in the last panel, we show the 
       continuum ALMA maps obtained using the line-free
       channels of the
       spectral window that covers that
       transition. All maps are rotated by
       25\degr\ clockwise so the symmetry axis of the disk is
       vertical. Contours in the line and continuum maps are spaced
       every 1.1\,mJy/beam, equivalent to 50\,K in main-beam
       temperature units. 
       The clean beam
       (HPBW=0\farc02$\times$0\farc02) is plotted at the bottom-left
       corner of each panel. The dashed ellipse represents the dust
       disk model deduced from the analysis of the continuum maps
       presented in \S\,\ref{cmaps} and Fig.\,\ref{f-cont}. The position of \qx, coincident with the continuum emission peak, is indicated by the purple star-like symbol.
     {\bf Bottom)} Synthetic \naclvd\ velocity-channel maps from our model (\S\,\ref{s-model}) represented as in the top panel.}
       
     \label{f-nacl51}
   \end{figure*}
%
\FloatBarrier

   \begin{figure*}[h!]
     \centering
 \includegraphics[width=0.22\hsize]{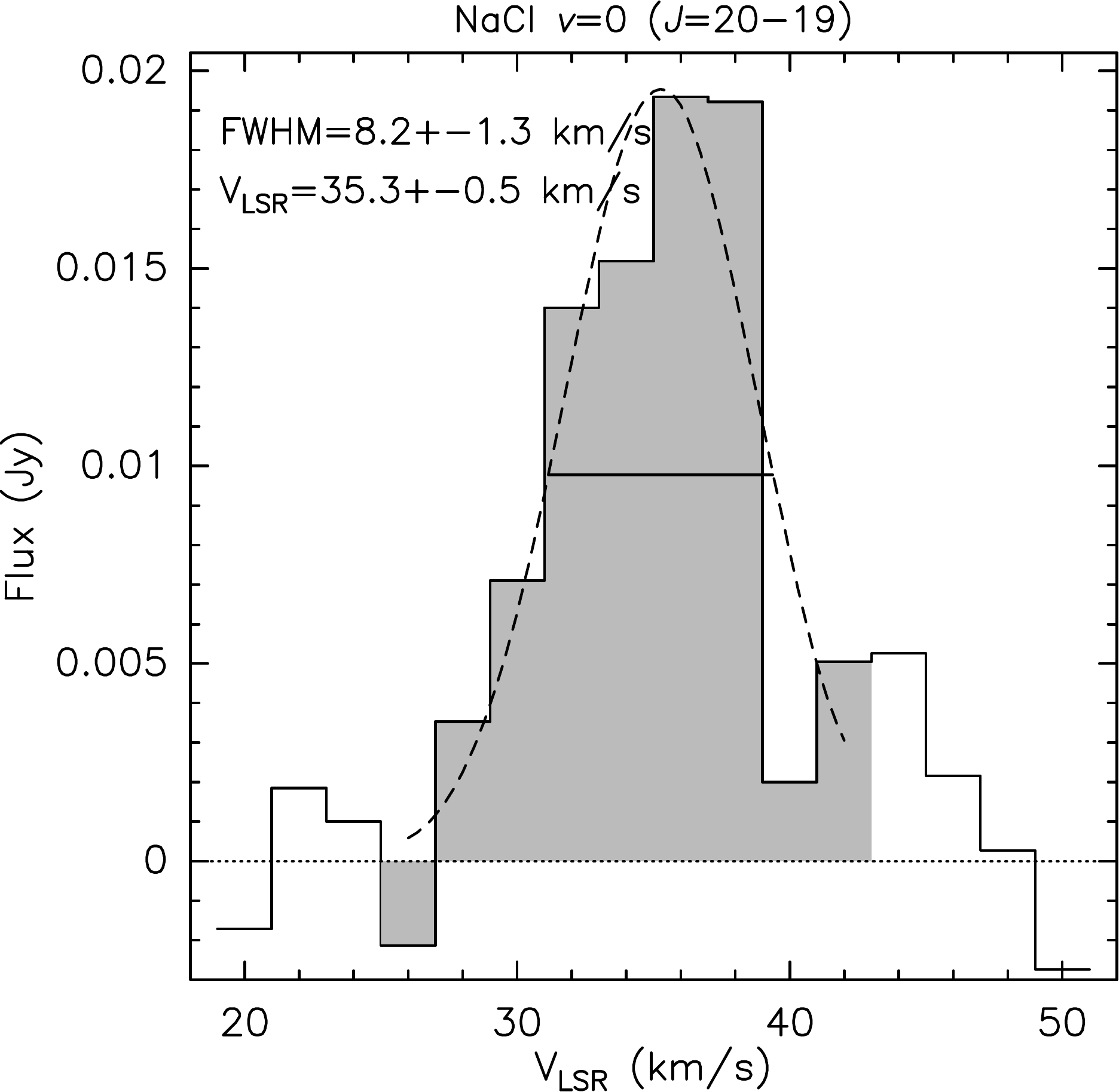}
 \includegraphics[width=0.22\hsize]{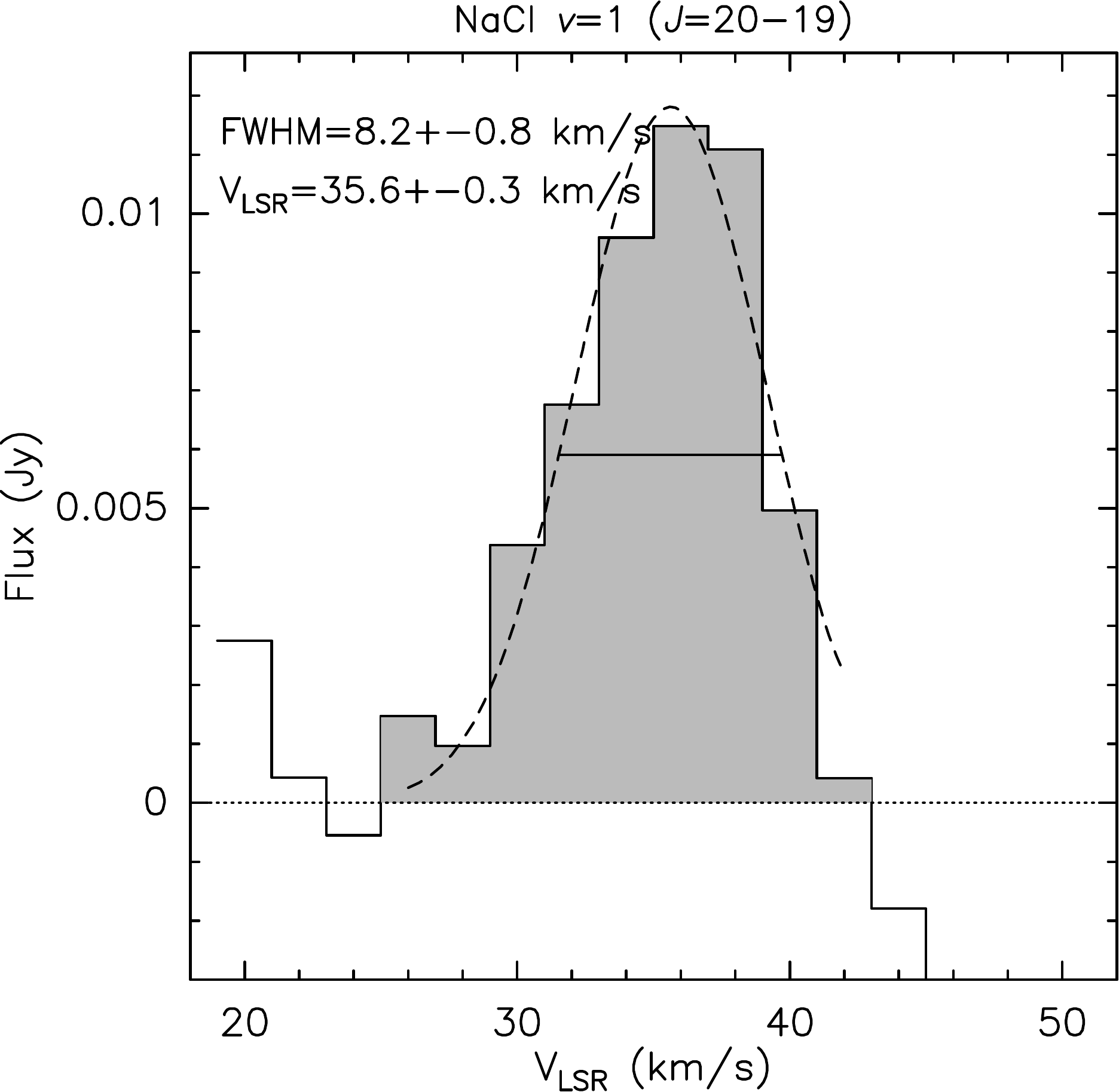}
 \includegraphics[width=0.23\hsize]{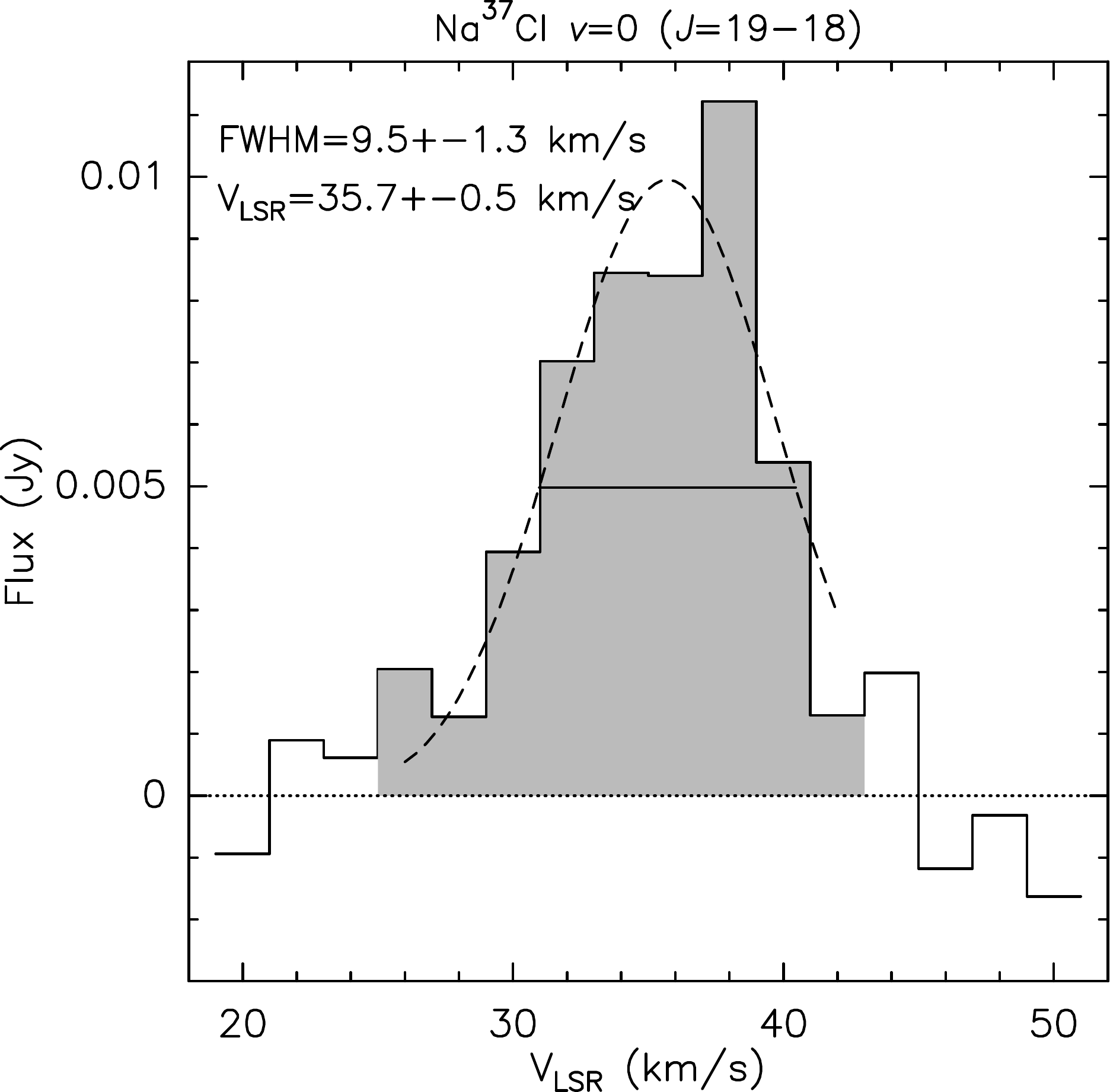}
 \includegraphics[width=0.23\hsize]{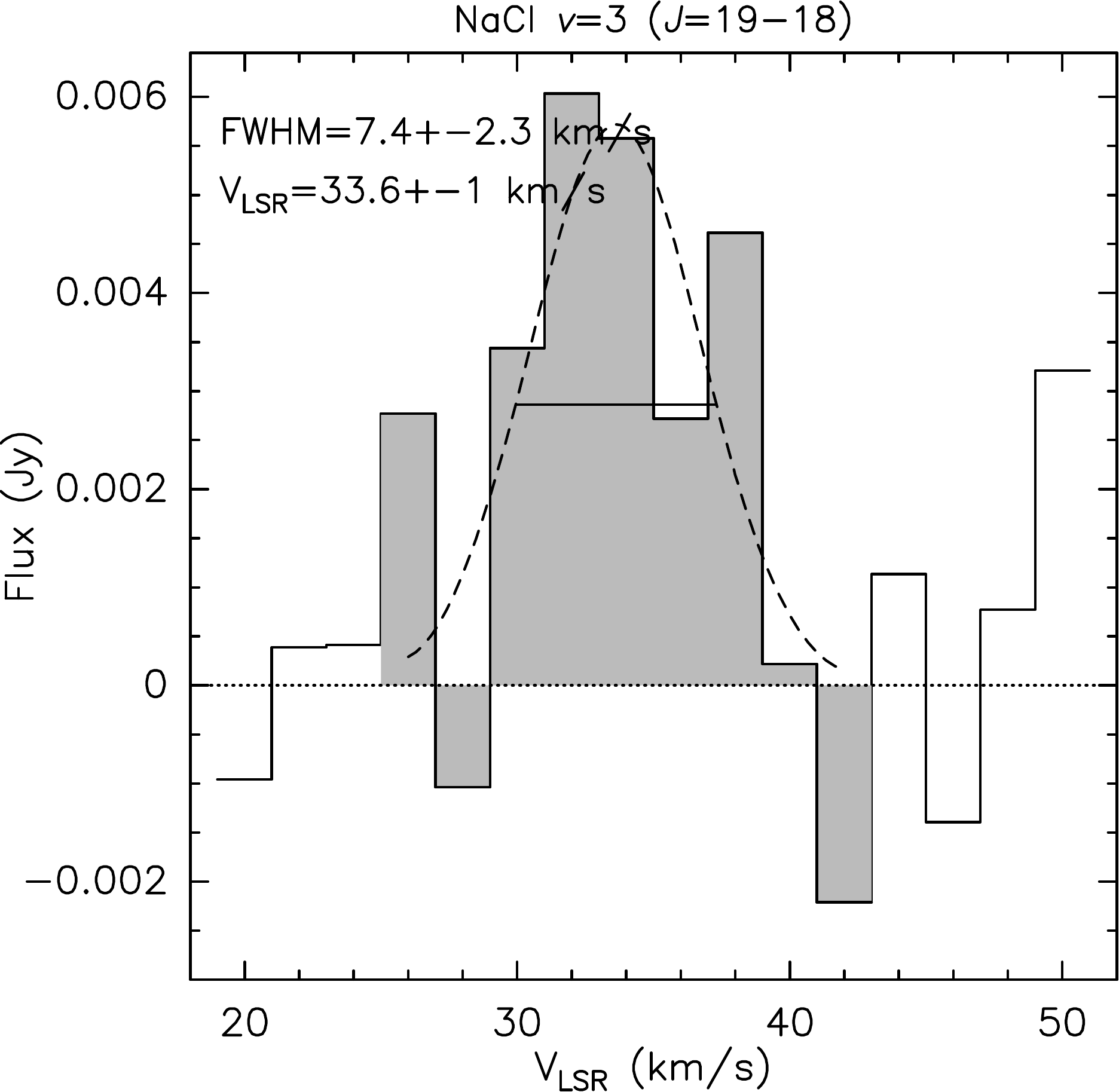}
 \vspace{0.25cm}
 
 \includegraphics[width=0.24\hsize]{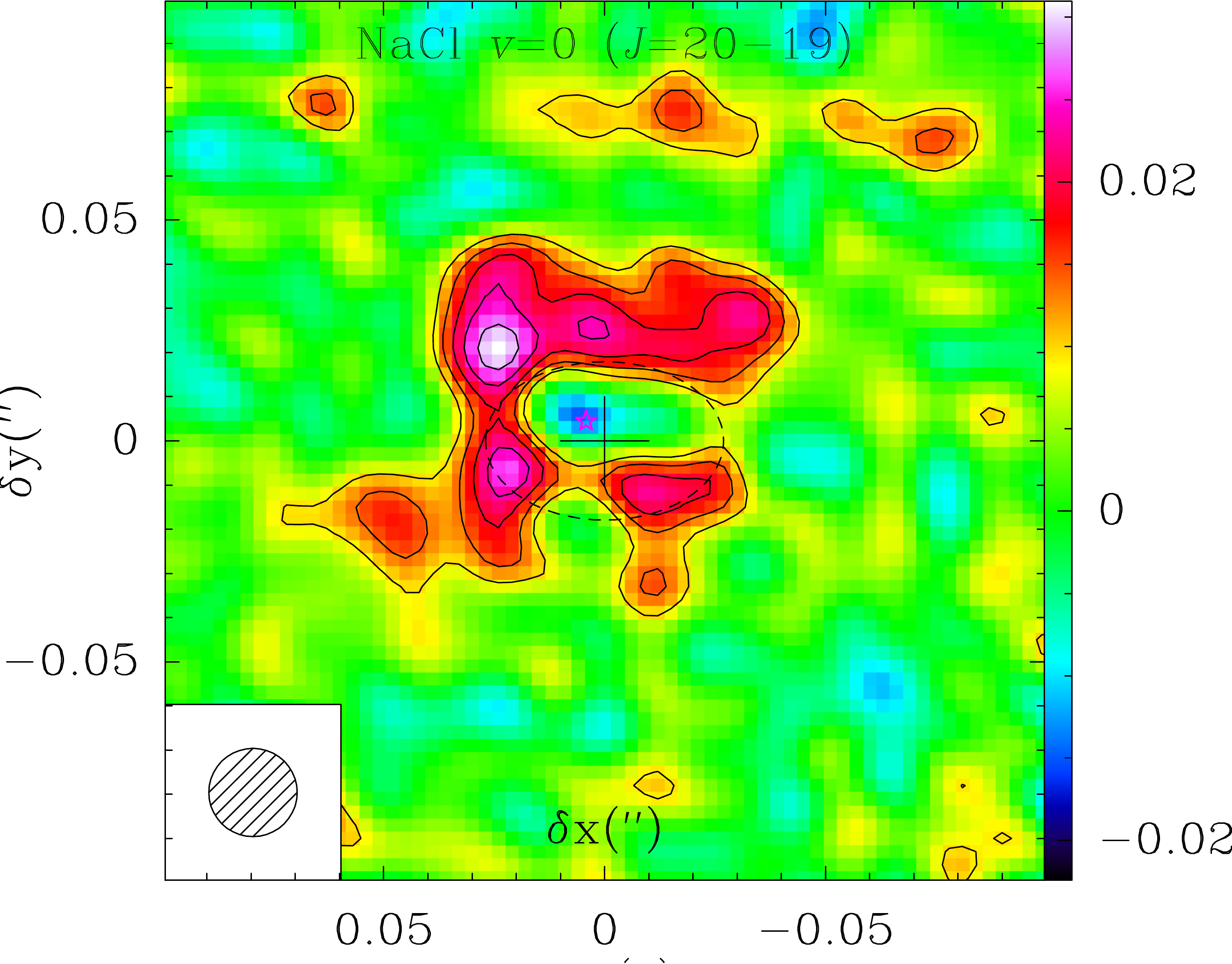}
 \includegraphics[width=0.24\hsize]{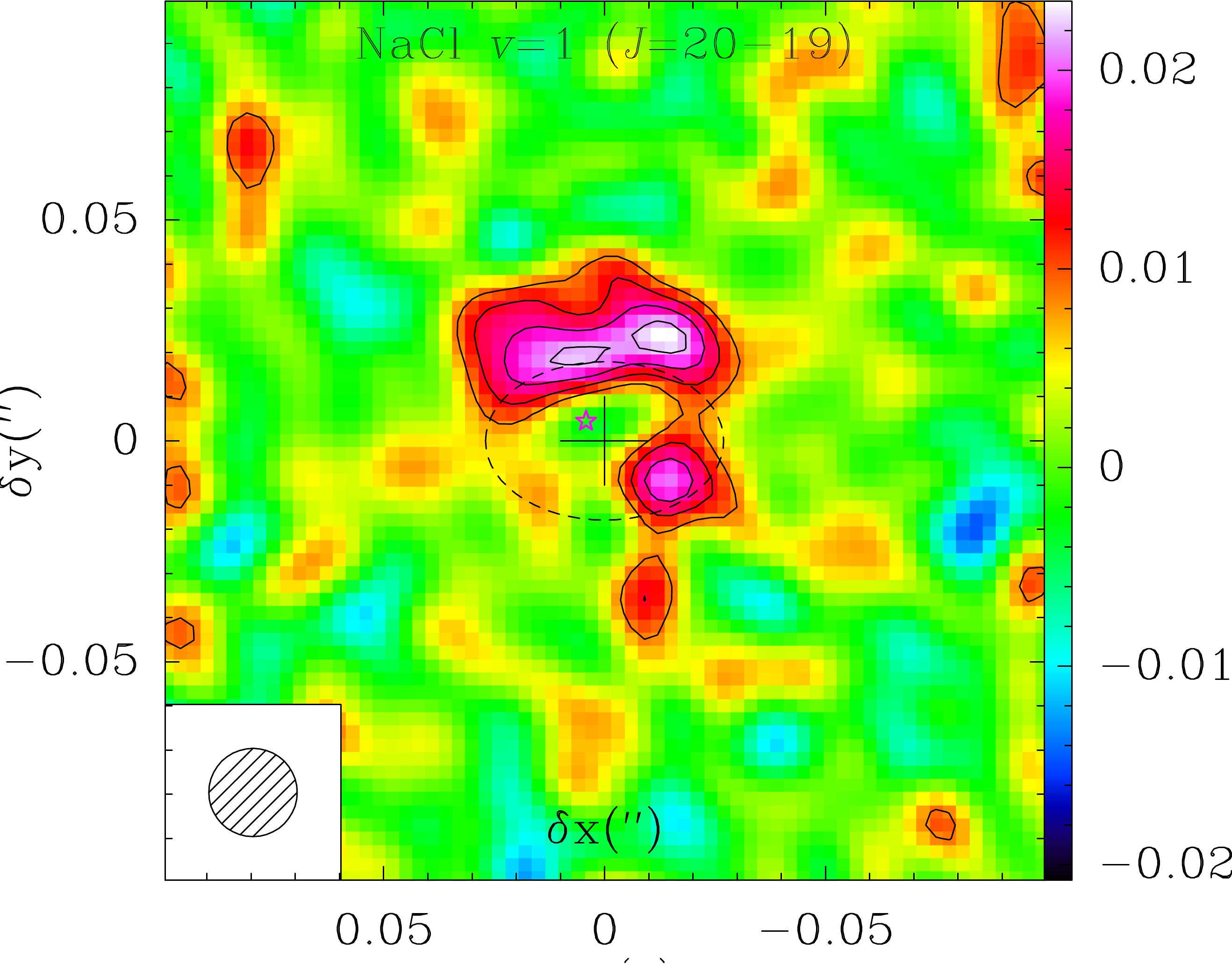}
 \includegraphics[width=0.24\hsize]{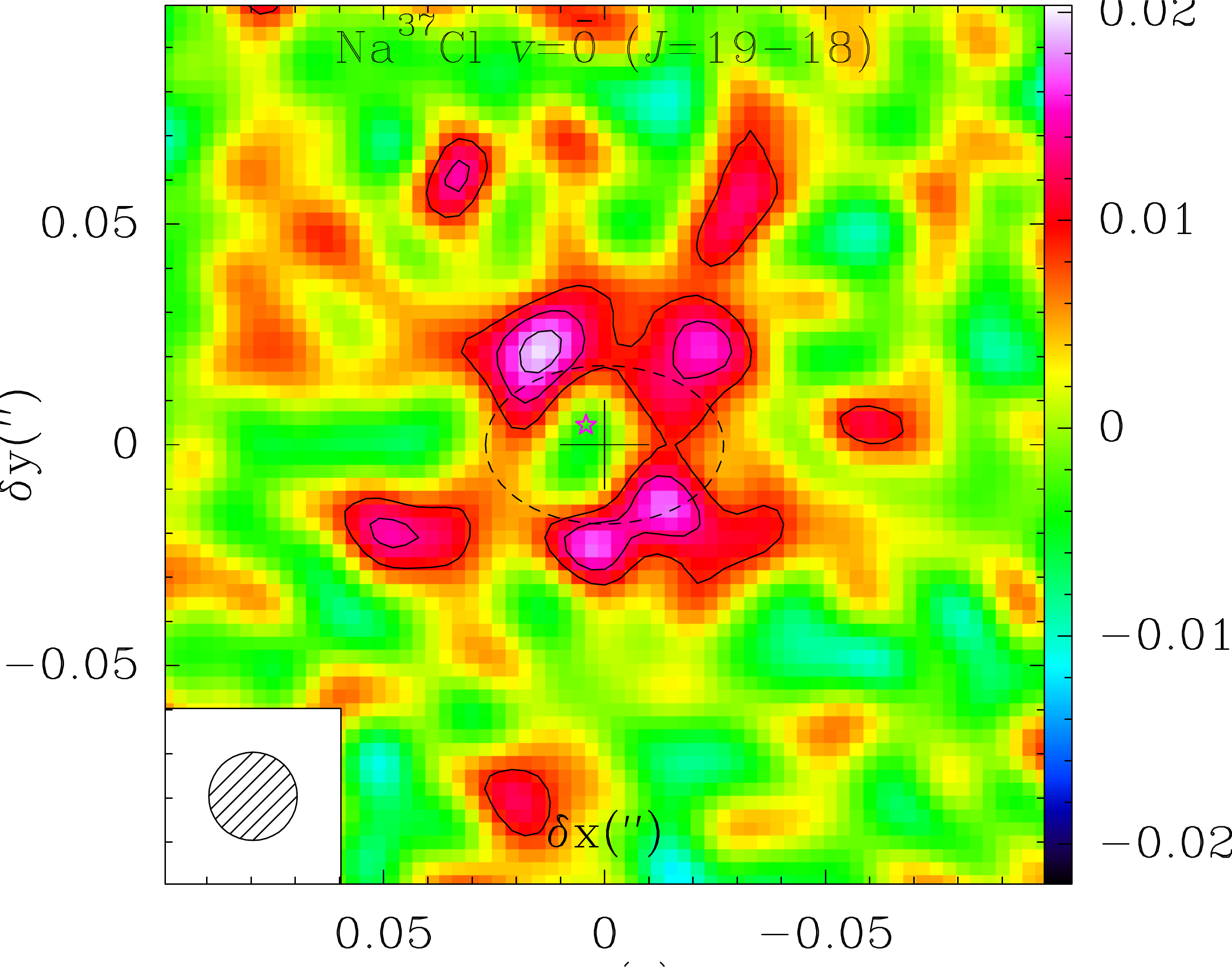}
 \includegraphics[width=0.24\hsize]{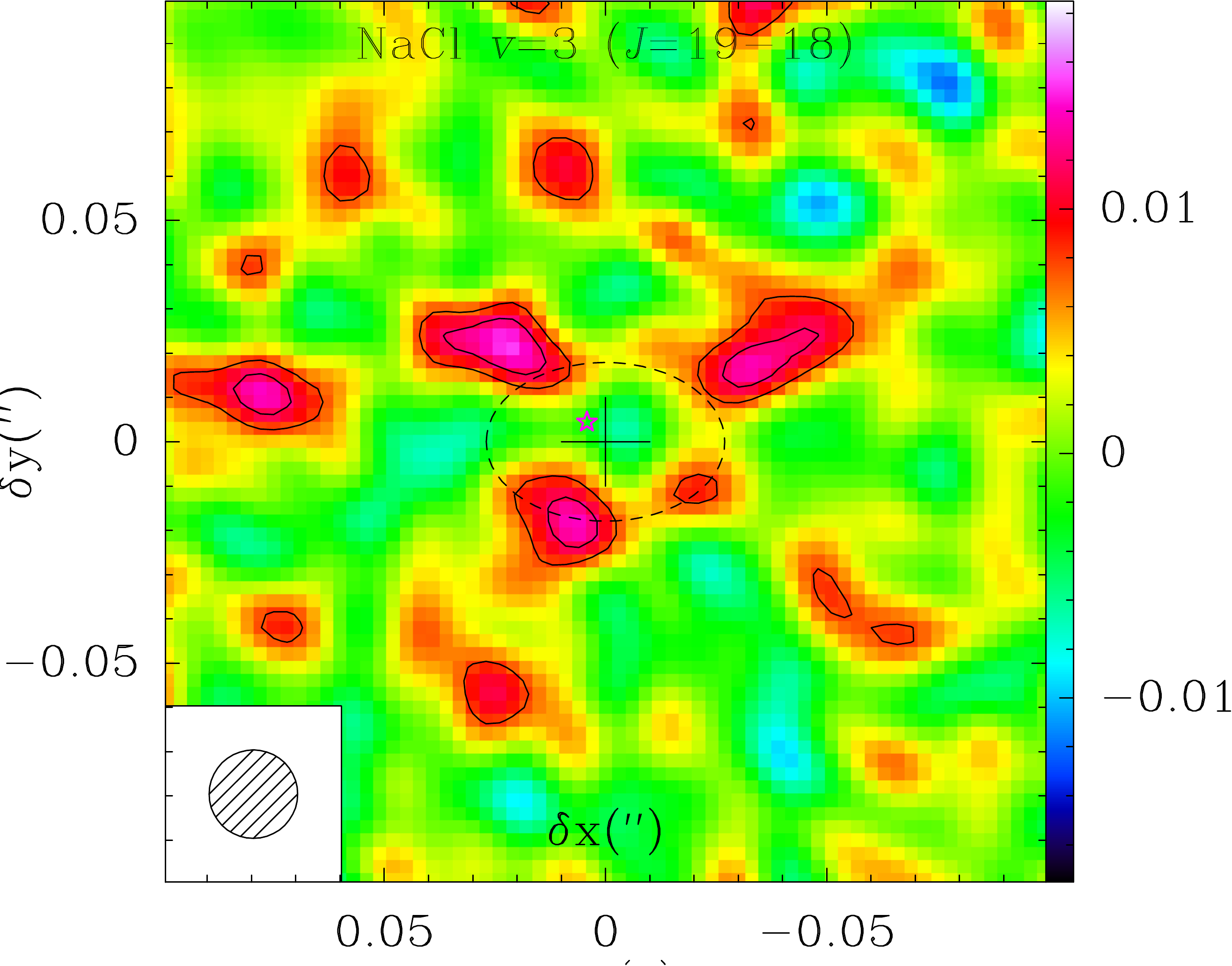} \\

 \vspace{1cm}

 \includegraphics[width=0.23\hsize]{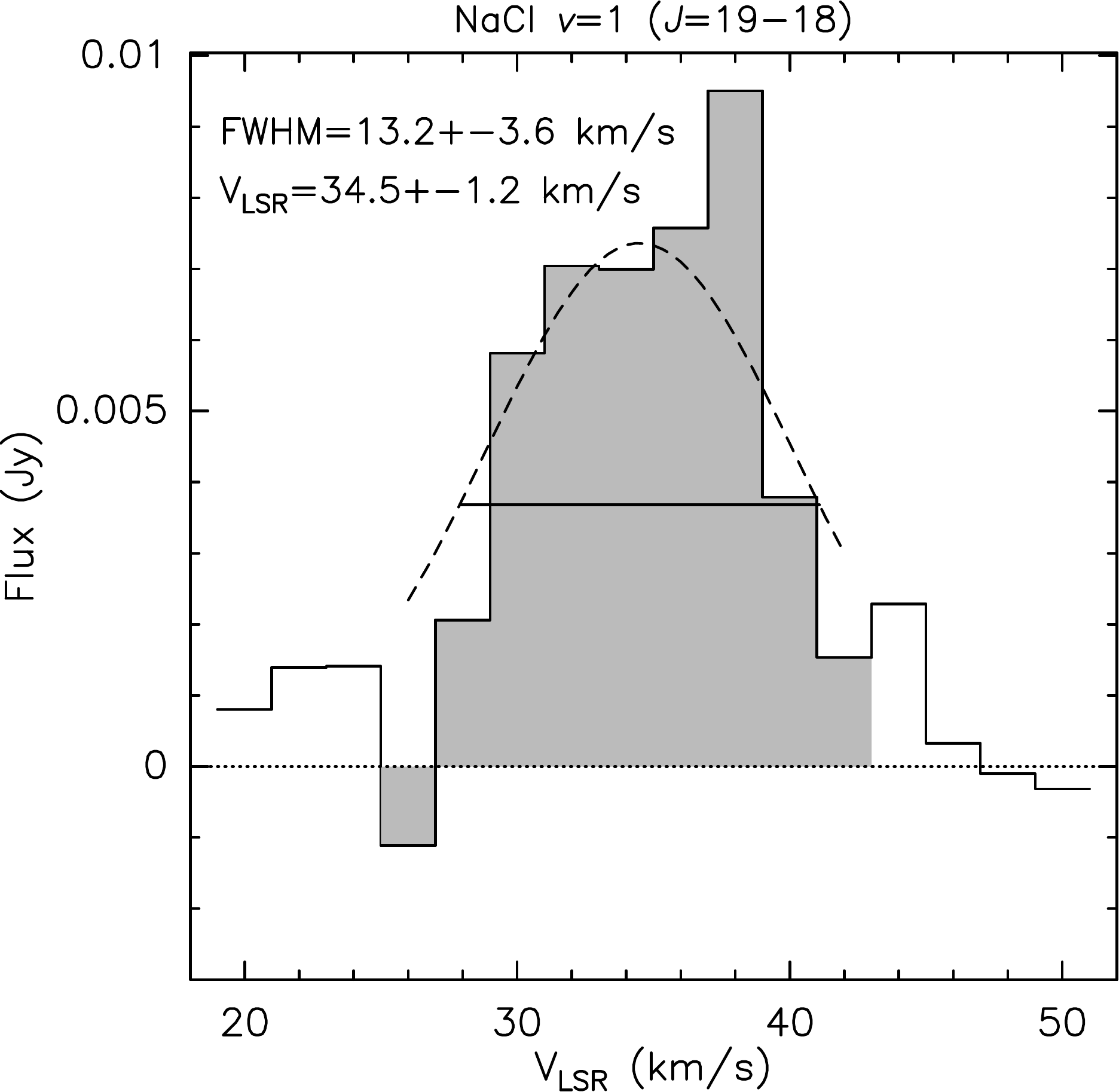}
 \includegraphics[width=0.23\hsize]{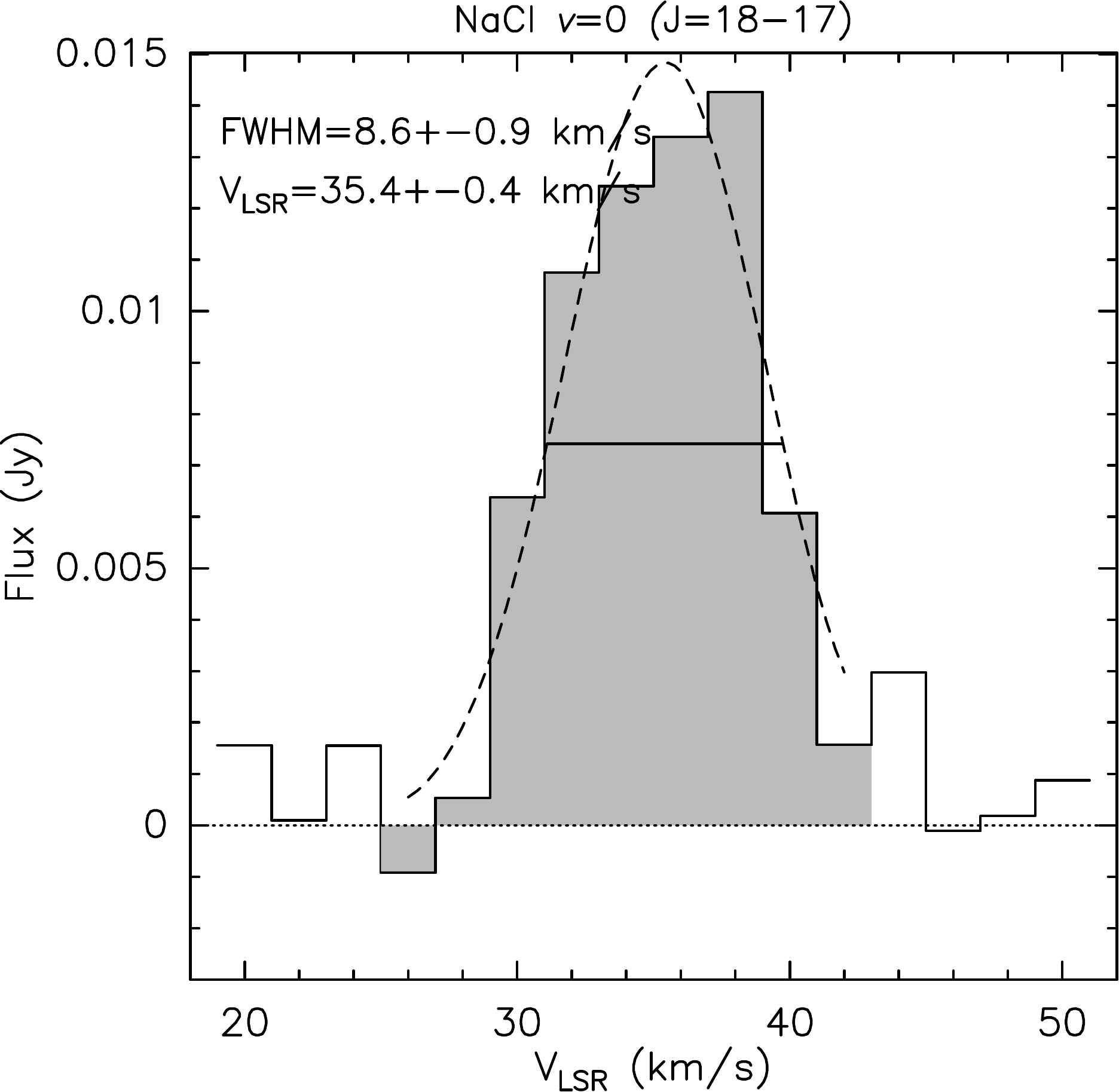}
 \includegraphics[width=0.23\hsize]{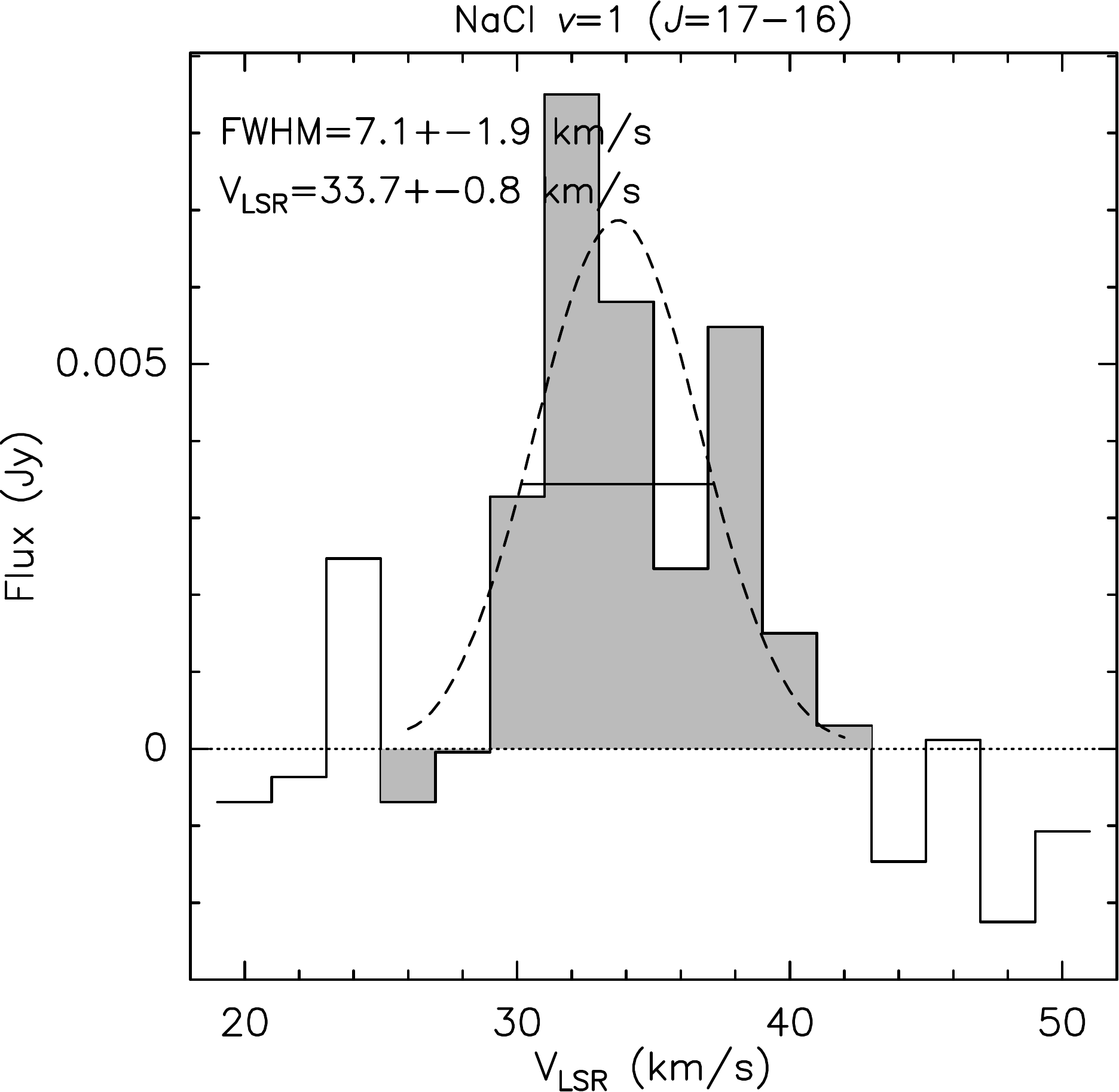}
 \includegraphics[width=0.23\hsize]{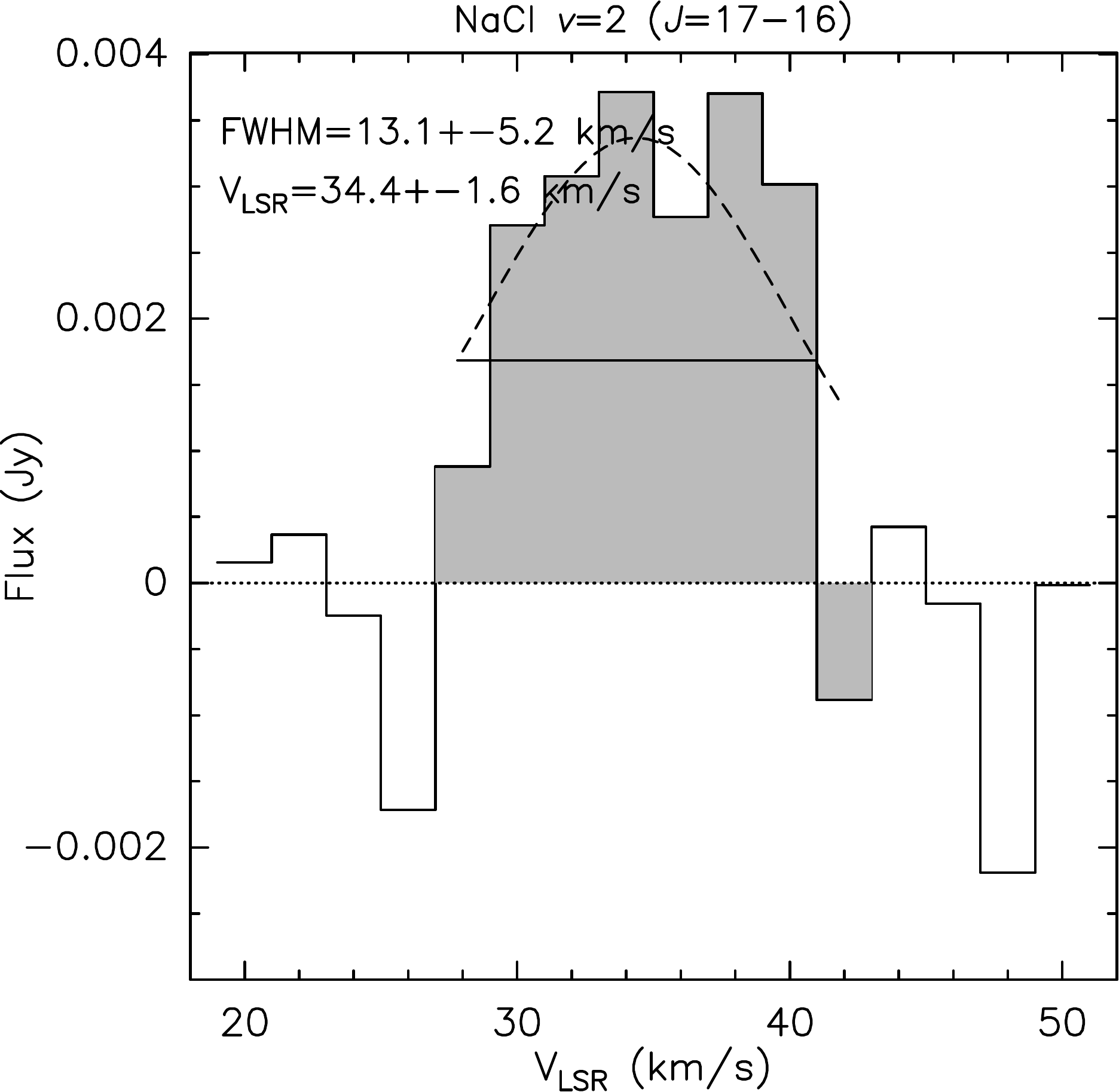}
 \vspace{0.25cm}
 
 \includegraphics[width=0.24\hsize]{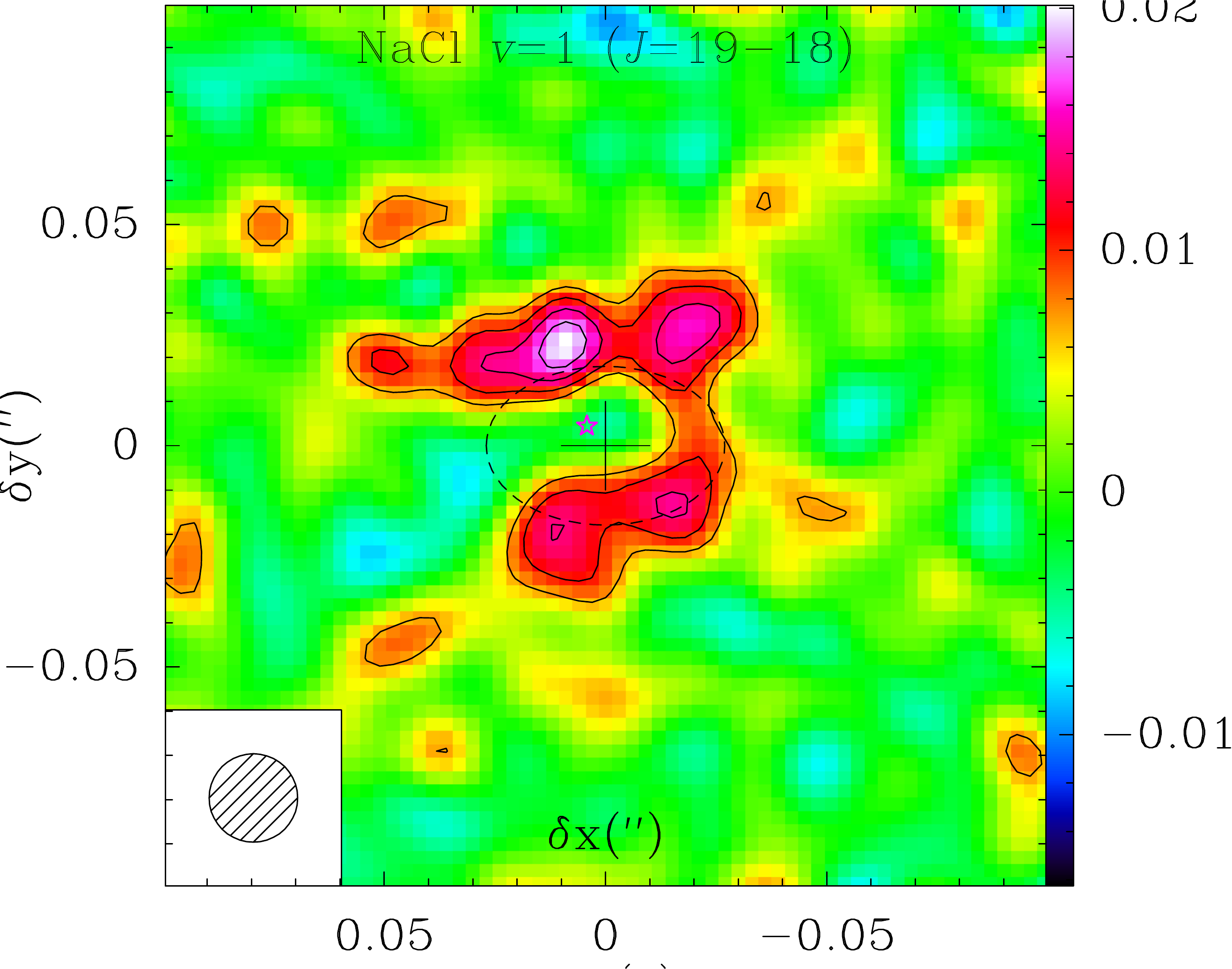}
 \includegraphics[width=0.24\hsize]{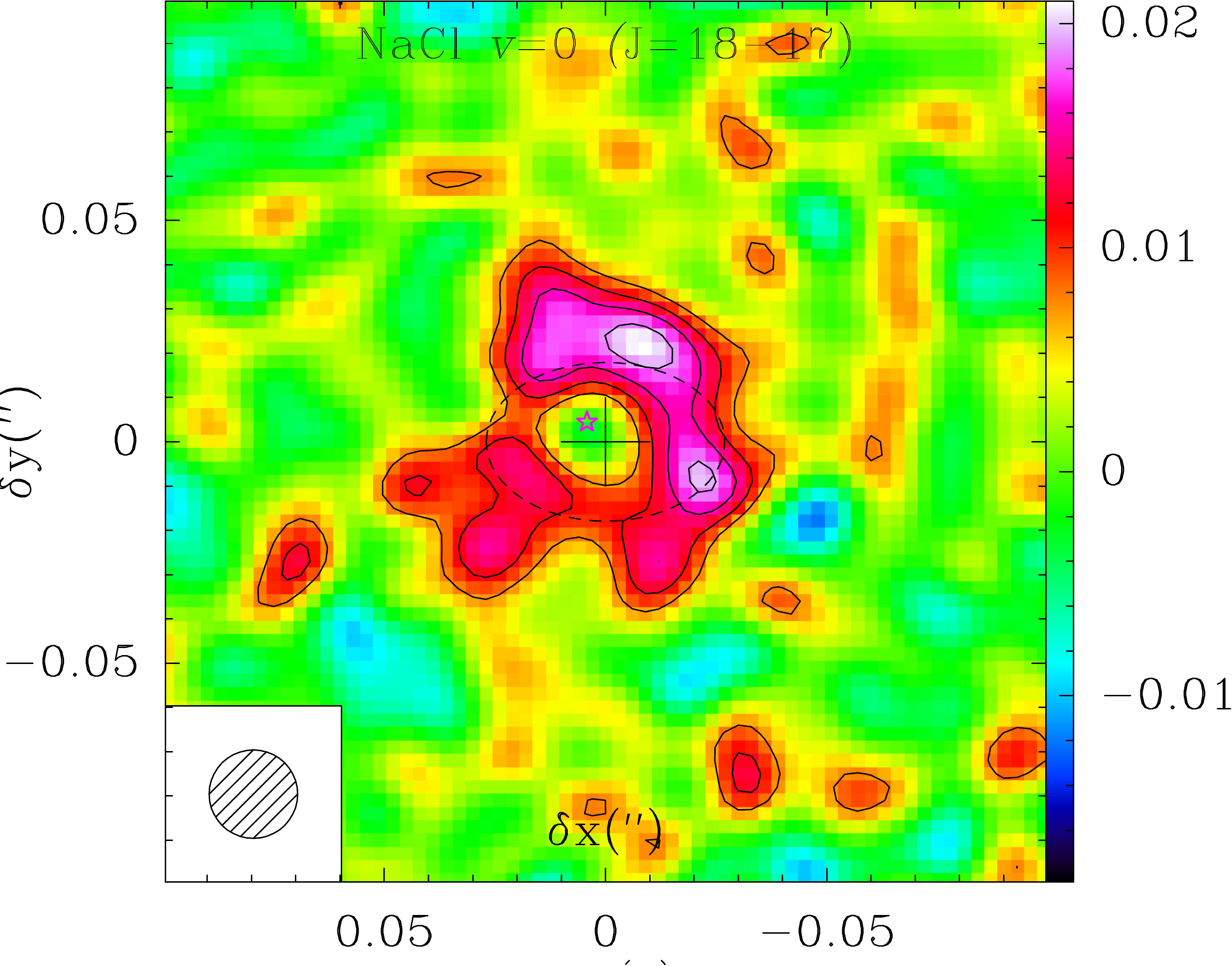}
 \includegraphics[width=0.24\hsize]{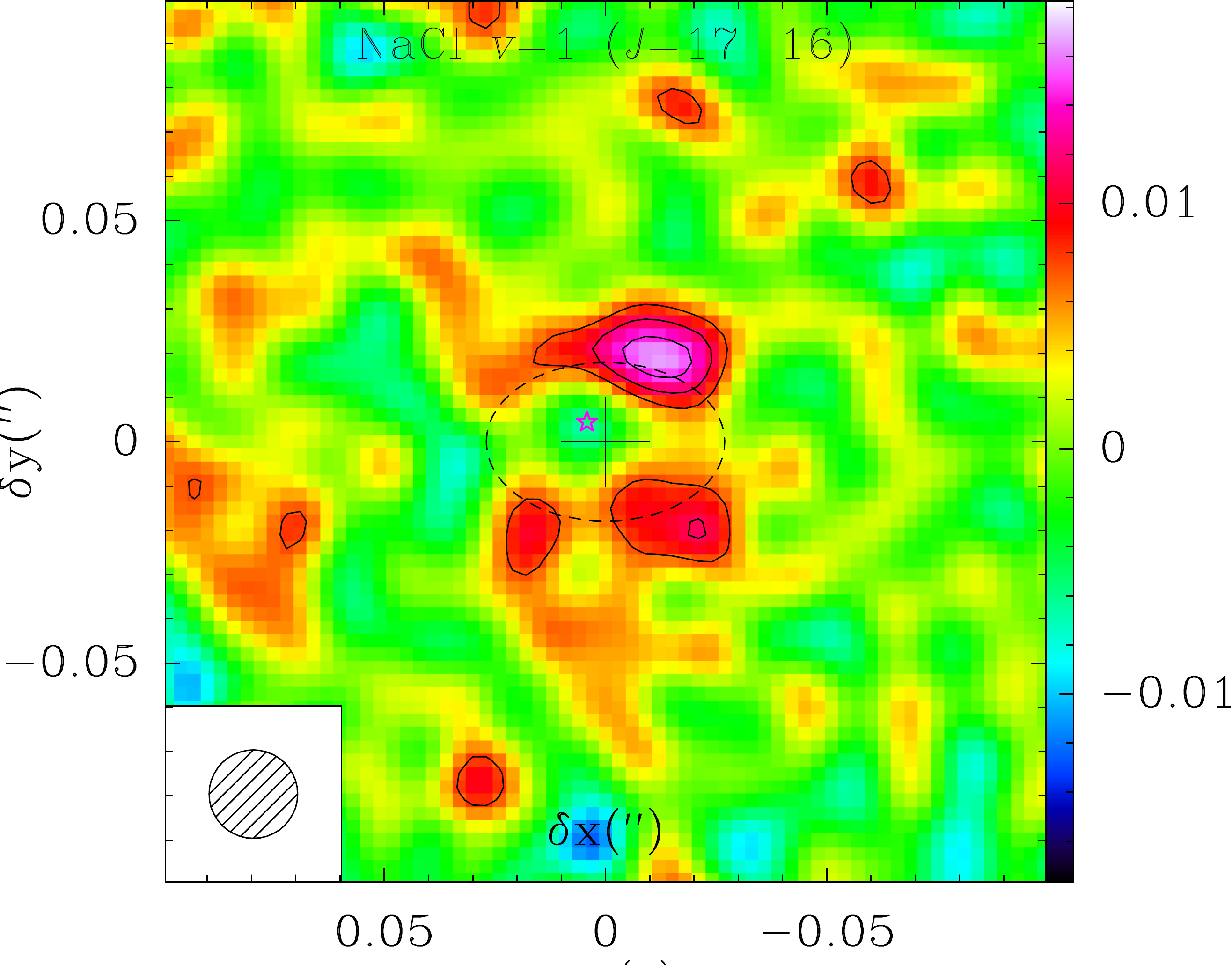}
 \includegraphics[width=0.24\hsize]{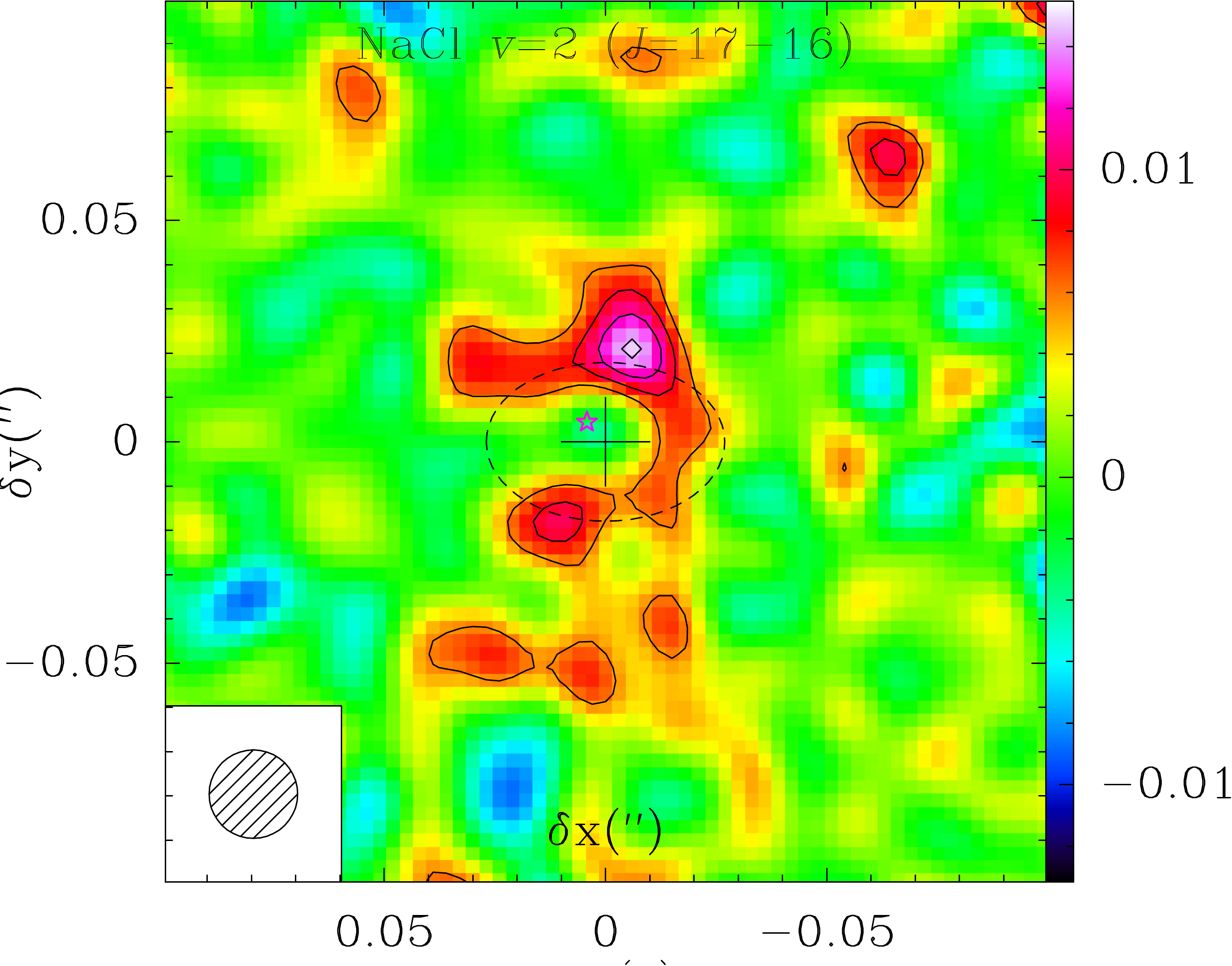}

 \caption{Total emission spectra and integrated intensity maps of the eigth NaCl transitions detected in this work (Table\,\ref{tab:mols}). Robust weigthing has been used to restore the emission maps with a half-power clean beam width of HPBW=0\farc02$\times$0\farc02. Contours are 2$\sigma$, 3$\sigma$, 4$\sigma$,... with $\sigma$=4.6, 4.3, 4.4, 3.9, 3.5, 3.8, 3.7, and 2.9\,mJy\,beam$^{-1}$, for the maps from left to right and top to bottom. Maps are
rotated by 25\degr\ clockwise so the symmetry axis of the nebula is vertical. 
 }
         \label{f-nacl-app}
   \end{figure*}
%

\FloatBarrier

   \begin{figure*}[h!]
     \centering

\includegraphics[width=0.22\hsize]{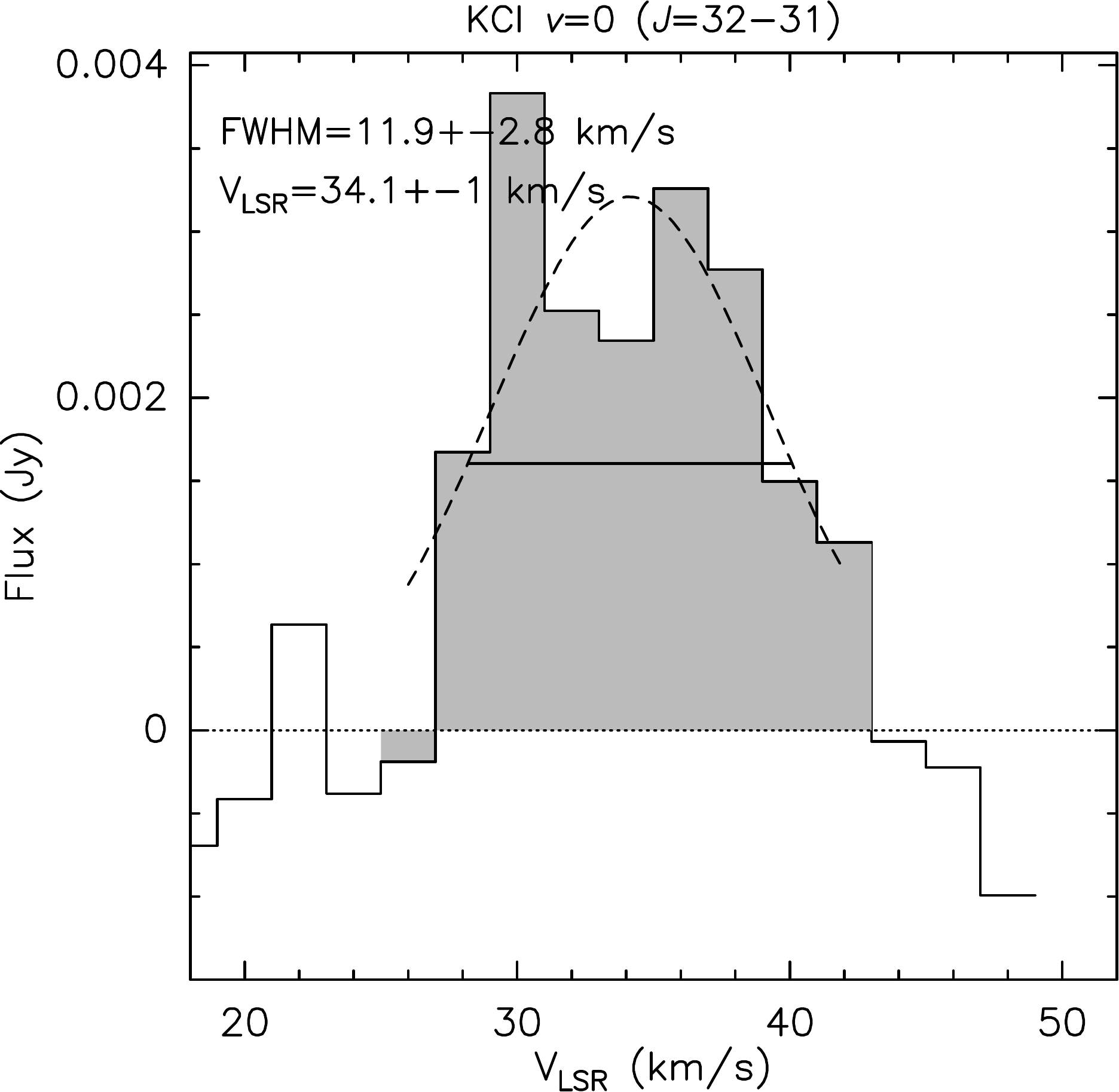}
 \includegraphics[width=0.22\hsize]{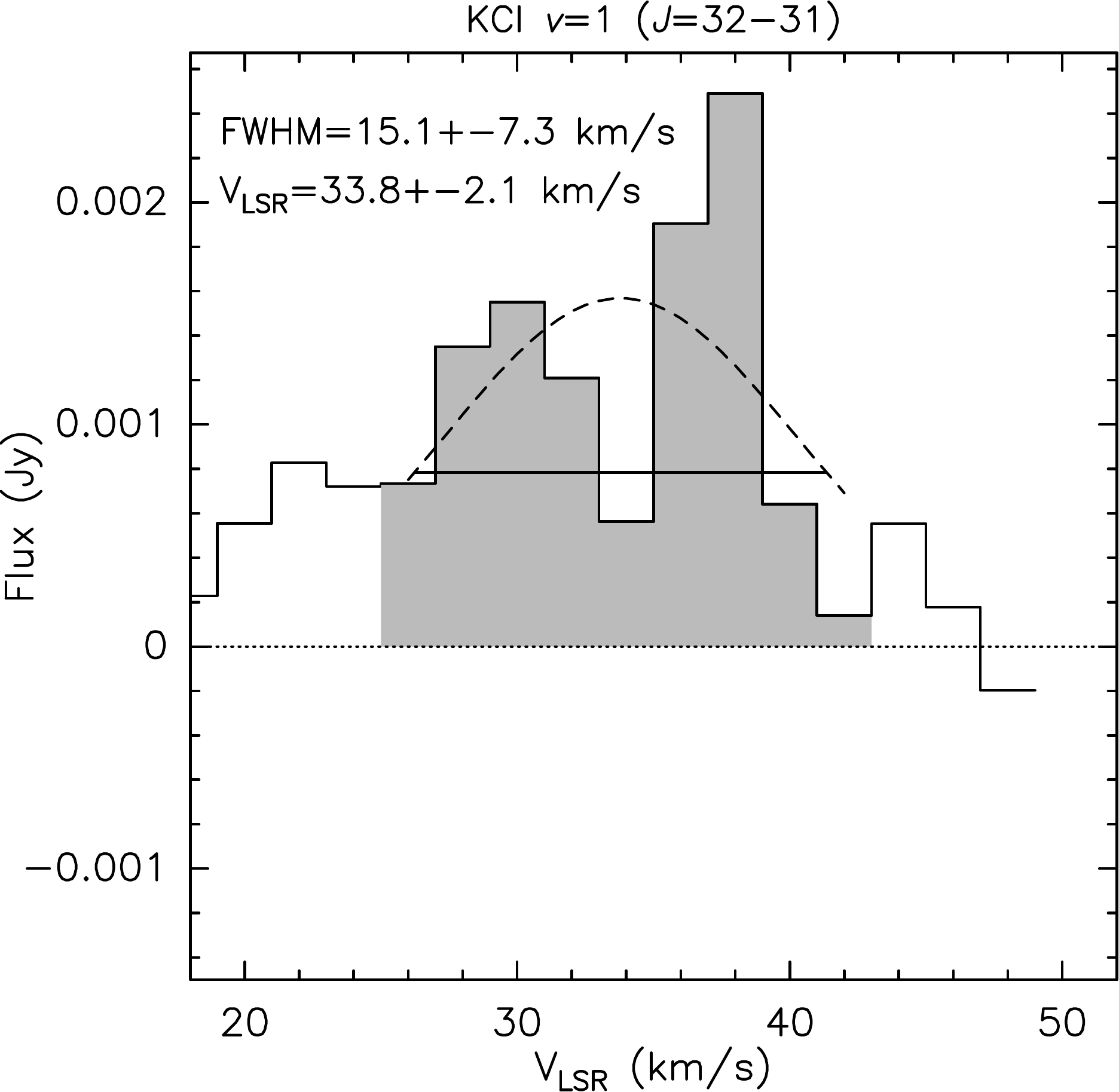}
 \includegraphics[width=0.23\hsize]{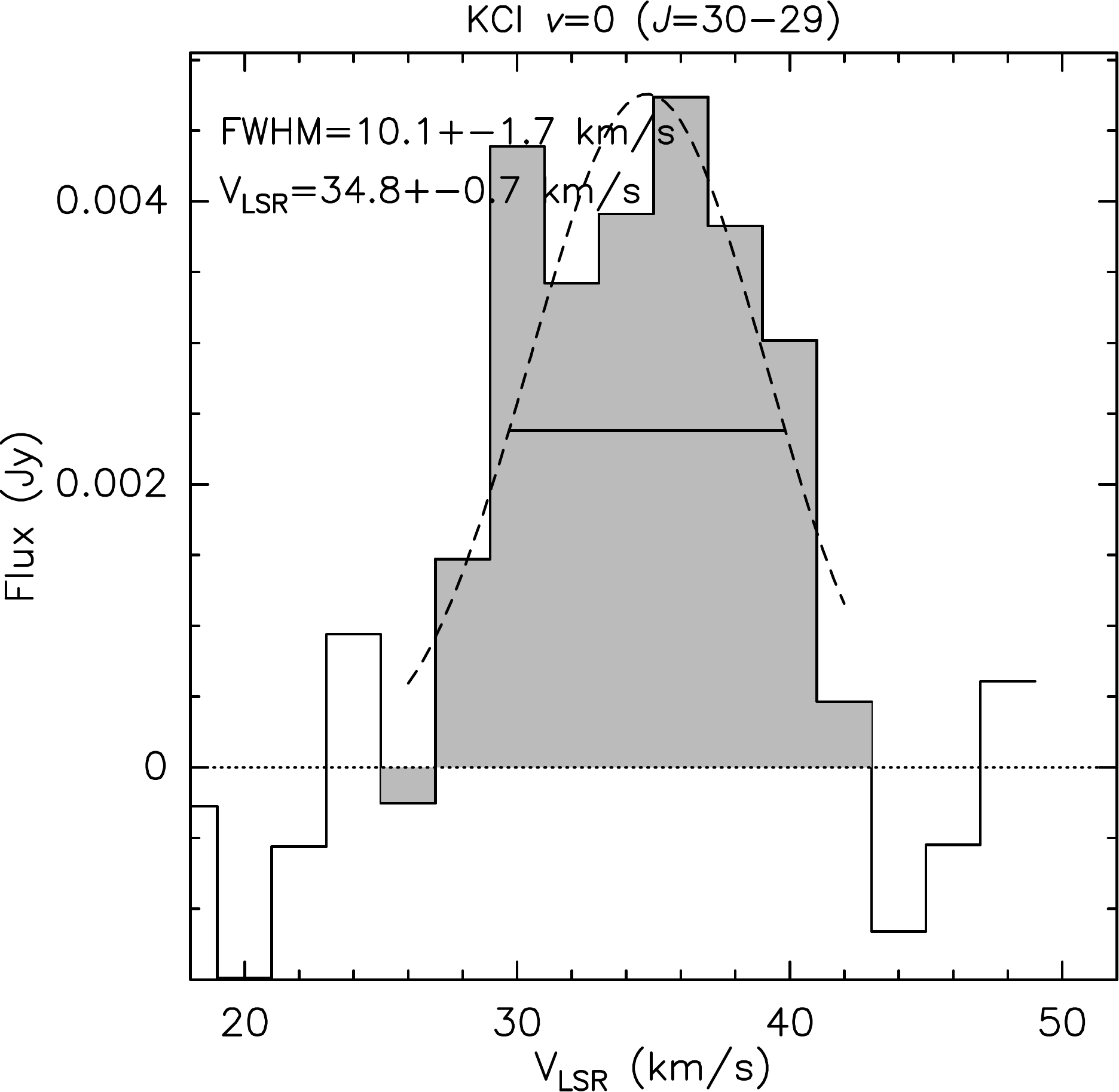} 
 \vspace{0.25cm}
 
 \includegraphics[width=0.24\hsize]{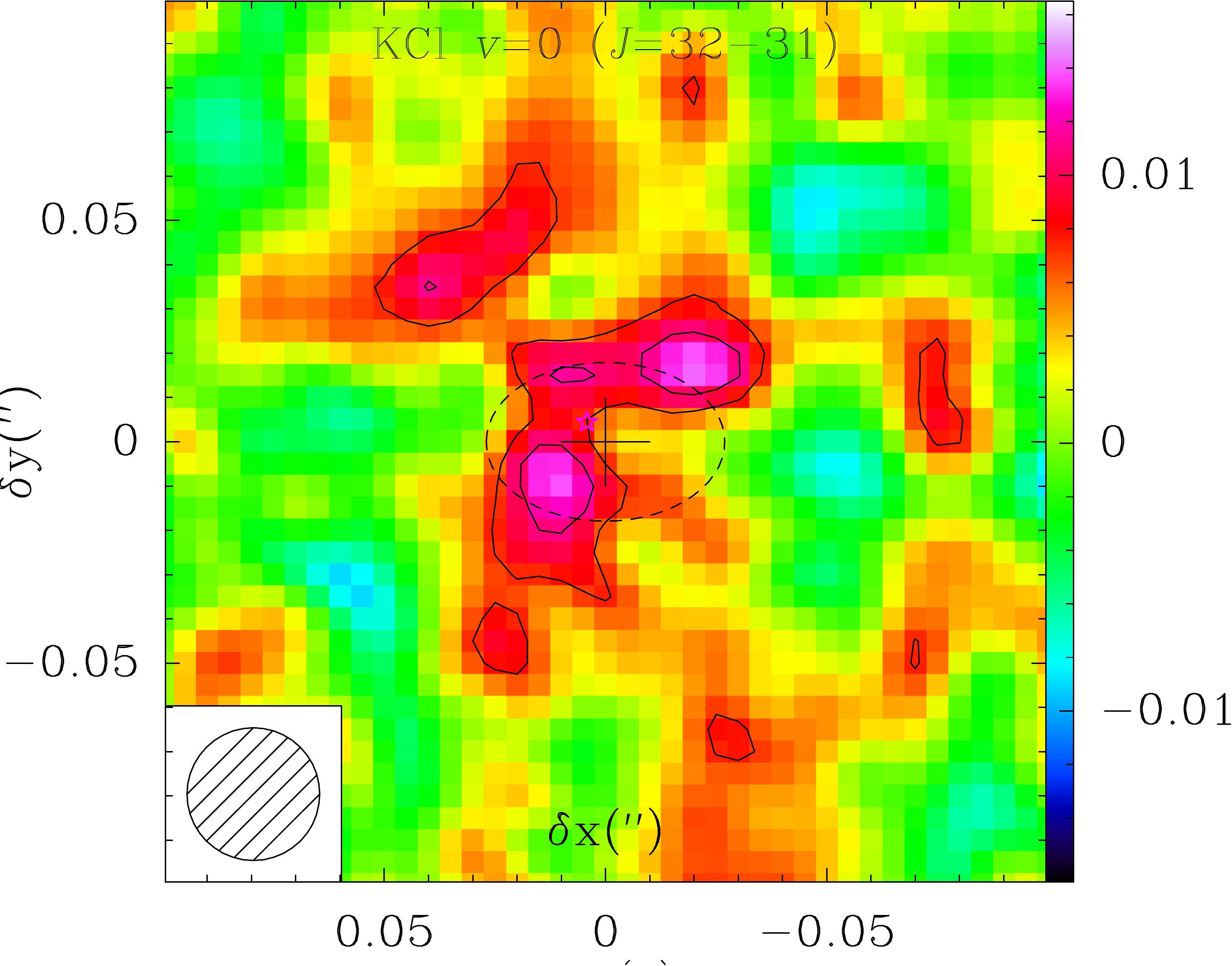}
 \includegraphics[width=0.24\hsize]{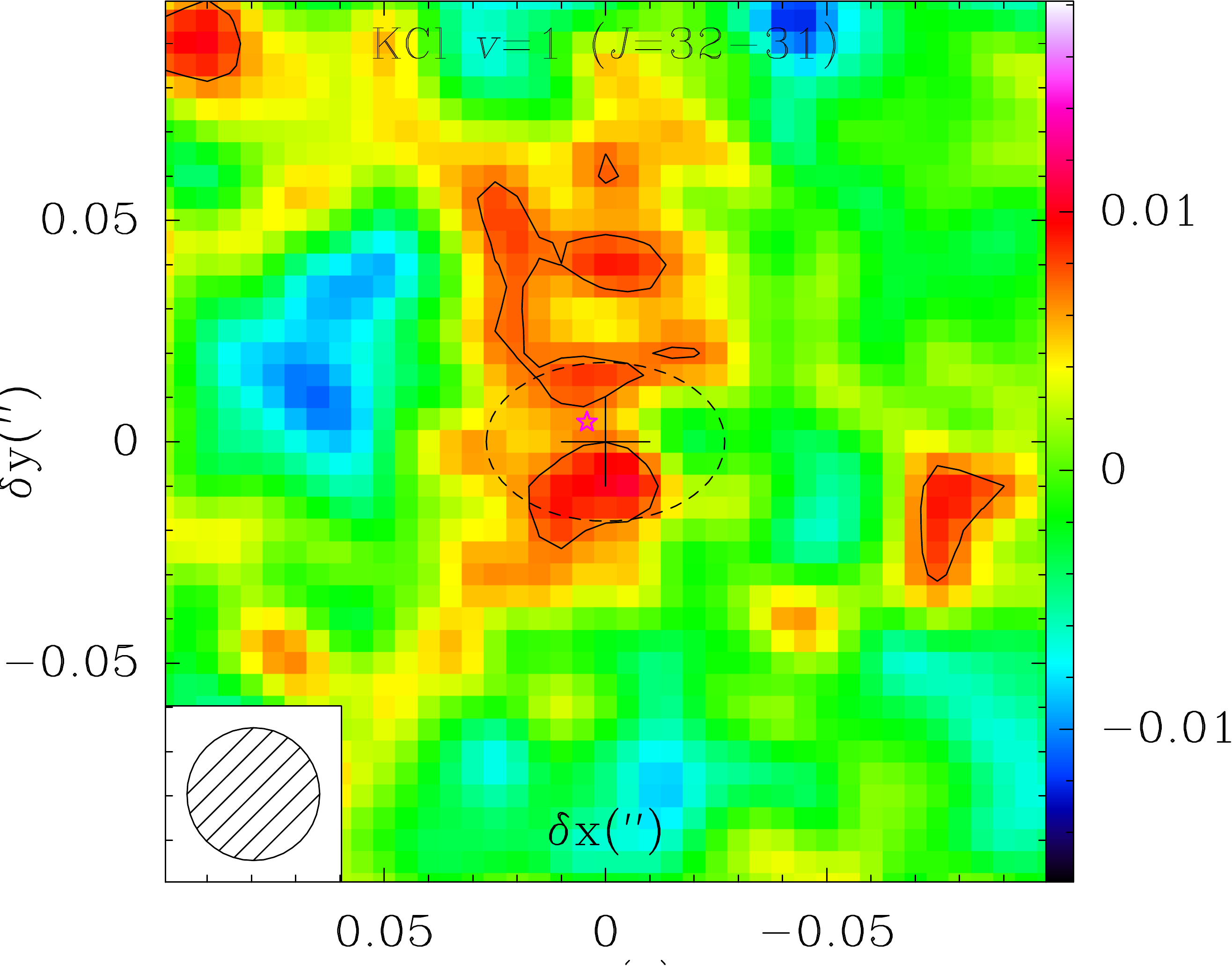}
 \includegraphics[width=0.24\hsize]{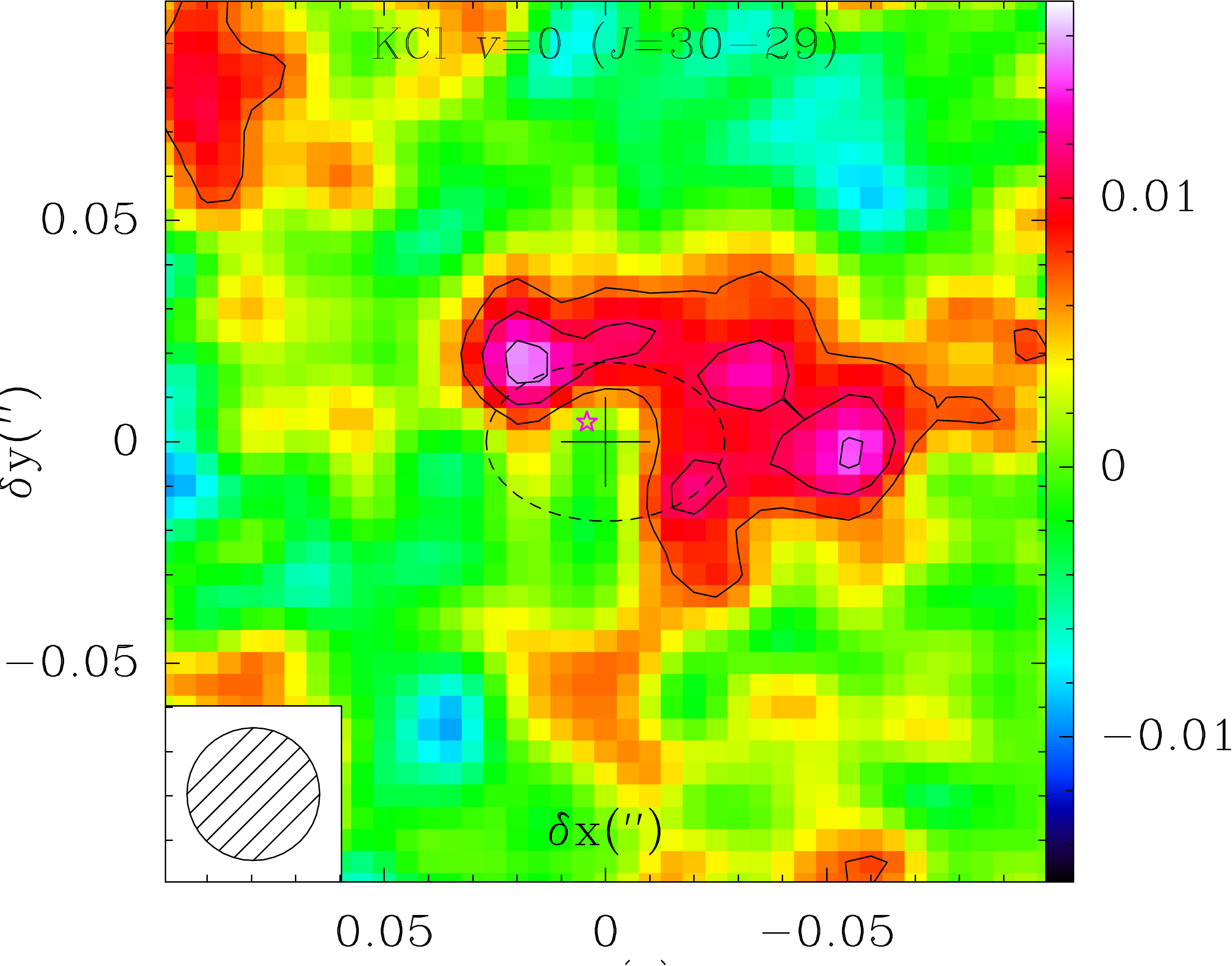}

 \caption{Same as in Fig.\,\ref{f-nacl-app} but for the three transtions of KCl detected in this work (Table\,\ref{tab:mols}). For these weak transitions, natural weigthing has been used to restore the emission maps with a half-power clean beam width of HPBW=0\farc03$\times$0\farc03. Contours are 2$\sigma$, 3$\sigma$, 4$\sigma$,... ($\sigma$=3.5\,mJy\,beam$^{-1}$).}
         \label{f-kcl-app}
   \end{figure*}
%
\FloatBarrier


%

\section{The current mass of \qx}
\label{mass-qx}

The mass of the Mira-type star \qx\ at the center of \ohs\ is
empirically poorly constrained. \cite{san02} obtained a rough
estimate of $\sim$1\,\msun\ from the analysis of the pulsation infall
motion and rotation of the SiO-masing regions at $\sim$6\,au from the
star.
The initial mass (in the
main-sequence) of \qx\ was estimated to be \mini$\sim$3\,\msun\ by
\cite{jur85}
given \ohs's probable membership to the Galactic
open cluster M\,46.  More recently, the age of the cluster has been
recomputed \citep{sha06,dav13}, resulting to be slightly younger, $\sim$225-250\,Myr, than
initially thought 
and hence suggesting a sightly larger initial mass of \mini$\sim$3.5\,\msun\ for \qx\ \citep{mil16}.
The initial mass of \qx\ is, then, confirmed to be higher than the mass of the
companion, A\,0V \citep{san04}, of about m$_{\rm
  2}$$\sim$2\,\msun, explaining the faster evolution of
\qx: note that for a \mini$\sim$3.5\,\msun\ star it takes $\sim$250\,Myr to leave the main-sequence,
while for a \mini$\sim$2\,\msun\ star it takes 4 times longer \citep{mil16}.

The empirically determined nebular mass of \ohs\ is 
$\sim$1\msun, including the mass in the large scale
CO-outflow (which accounts for most, 99\%, of the nebular mass), the NIR halo
surrounding the central parts of the nebula \citep[probably the relic of an
ancient wind ejected at $\sim$\dex{-6}\,\my,  $M_{\rm halo}$$\lsim$0.01\,\msun;][]{alc01}, and the ionized H$\alpha$-nebula \citep[$M_{\rm ion}$$\sim$5\ex{-4}\,\msun;][]{san00} -- see more details in \S\,\ref{intro}.
If this was the only mass lost by \qx\ up to date, then its stellar mass at present would be
\mini$-$1\,\msun$\sim$2.5\,\msun.
However, according to the stellar evolutionary models of low to intermediate mass stars, 
in the later stages (i.e.\,last cycles) of the TP-AGB phase, stars like \qx\ are expected to have already lost
about 80\% of the initial mass, which would have occurred during the last $\approx$10$^{5}$ years.
More specifically, conforming to theoretical AGB stellar tracks
\cite[][see their Fig.\,2]{blo95, ste98}
a \mini$\sim$3\,\msun\ star loses about 1\,\msun\ during the last
350-60\,kyr before the last thermal-pulse and another
$\sim$1\,\msun\ in the next 60-10\,kyr.  Therefore, in addition to the
$\sim$1\,\msun\ mass most recently lost and that is visible in
the large-scale $\sim$800\,yr-old nebula, \qx\ has probably lost in
much earlier times about $\sim$2\,\msun. However, for the most part,
this very ancient $\sim$2\,\msun\ ejecta would remain undetectable due to 
strong dilution and photodissociation effects after
being in expansion during $\approx$10-100\,kyr: note, that the most extended and oldest
haloes/envelopes around AGB stars or prePN/PN's central stars ever detected
are $\sim$20-50\,kyr \citep[e.g.][]{kwo07}.

In summary, current stellar evolutionary models predict that, at the tip of the
AGB, the mass of an AGB star with \mini$\sim$3.5\,\msun\ is
$m_{\rm 1}$$\sim$0.7\,\msun, which we then take as a probable value for the current mass of \qx.

\end{appendix}

\end{document}